\documentclass[useAMS,usenatbib,usegraphicx]{mn2e}

\usepackage{natbib}
\usepackage{amsmath}
\usepackage{amssymb}

\newcommand{\M}{Paper~\textrm{I}}
\newcommand{\HI}{H\texttt{I}~}
\newcommand{\HII}{H\texttt{II}~}

\title[Deep multiband surface photometry on $21$ ELGs]{Deep multiband surface photometry on star forming galaxies: \emph{II}. A volume limited sample of $21$ emission lines galaxies\thanks{Based in part on observations made with the Nordic Optical Telescope, operated on the island of La Palma jointly by Denmark, Finland, Iceland, Norway, and Sweden, in the Spanish Observatorio del Roque de los Muchachos of the Instituto de Astrofisica de Canarias.}\thanks{Based in part on observations collected at the European Organization for Astronomical Research in the Southern Hemisphere, Chile, (ESO ID 075.B--0220 and 077.B--0599).}}
\author[Micheva et al.]{Genoveva Micheva$^{1}$\thanks{E-mail: genoveva@astro.su.se (GM)}, G\"oran \"Ostlin$^{2}$, Erik Zackrisson$^{2}$, Nils Bergvall$^{3}$, Thomas 
\newauthor Marquart$^{1,3}$, Josefa Masegosa$^{4}$, Isabel Marquez$^{4}$, Robert Cumming$^{5}$, Florence Durret$^{6}$\\
$^{1}$Stockholm Observatory, Department of Astronomy, Stockholm University, 106\,91 Stockholm, Sweden\\
$^{2}$Oskar Klein Centre for Cosmoparticle Physics, Department of Astronomy, Stockholm University, 106\,91 Stockholm, Sweden\\
$^{3}$Division of Astronomy \& Space Physics, Uppsala university, 751\,20 Uppsala, Sweden\\
$^{4}$Instituto Astrofisica Andalucia (IAA), CSIC, Spain\\
$^{5}$Department of Earth and Space Sciences, Chalmers University of Technology, Onsala Observatory, SE-439 94 Onsala, Sweden\\
$^{6}$UPMC-CNRS, UMR7095, Institut d'Astrophysique de Paris, F-75014, Paris, France\\
}
\begin{document}

\date{Accepted .... Received ...; in original form ...}

\pagerange{\pageref{firstpage}--\pageref{lastpage}} \pubyear{2011}

\maketitle

\label{firstpage}

\begin{abstract}

\noindent We present deep surface photometry of a volume--limited sample of $21$ UM emission line galaxies in broadband optical $UBVRI$ and near infra-red (NIR) $HKs$ filters. The sample comprises $19$ blue compact galaxies (BCGs) and two spirals. For some targets the exposure times are the deepest to date. For the BCG UM462 we observe a previously undetected second disk component beyond a surface brightness level of $\mu_B=26$ mag arcsec${}^{-2}$. This is a true low surface brightness component with central surface brightness $\mu_0=24.1$ mag arcsec${}^{-2}$ and scale length $h_r=1.5$ kpc. All BCGs are dwarfs, with $M_B\ge-18$, and very compact, with an average scale length of $h_r\sim1$ kpc. We separate the burst and host populations for each galaxy and compare them to stellar evolutionary models with and without nebular emission contribution. We also measure the $A_{180}$ asymmetry in all filters and detect a shift from optical to NIR in the average asymmetry of the sample. This shift seems to be correlated with the morphological class of the BCGs. Using the color-asymmetry relation, we identify five BCGs in the sample as mergers, which is confirmed by their morphological class. Though clearly separated from normal galaxies in the concentration--asymmetry parameter space, we find that it is not possible to distinguish luminous starbursting BCGs from the merely star forming low luminosity BCGs.
\end{abstract}

\begin{keywords}
galaxies: dwarf - photometry - stellar content - halo, etc...
\end{keywords}

\section{Introduction}

\noindent Blue compact galaxies (BCGs) are low metallicity gas--rich galaxies at low redshifts, currently undergoing intense star formation (SF). Their star formation rates (SFR) are on average too high to be indefinitely sustained by the available gas supply. Their spectra are reminiscent of \HII regions, with strong emission lines superposed on a blue stellar continuum, which is why they are sometimes referred to as \HII galaxies. Deep optical and near infra--red (NIR) observations have revealed the presence of an old stellar population in these galaxies, often referred to as the ``host'', in which the starbursting regions are embedded. The original criteria of what constitutes a BCG~\citep{1981ApJ...247..823T} referred to compactness ($r_{25}\sim1$ kpc in diameter) on photographic plates, blue colors, and low total luminosity ($M_B\gtrsim-18$), however, with the discovery of an old and extended underlying host population in almost all BCGs~\citep[e.g.][]{1996A&AS..120..207P,1997MNRAS.286..183T,2001ApJS..133..321C,2001ApJS..136..393C,2002A&A...390..891B,2003ApJ...593..312C,2003A&A...410..481N}, these criteria have been relaxed to be more inclusive. Thus, BCGs comprise a heterogeneous group of galaxies, with varied morphologies, star formation histories, and total luminosities, but they all have \HII region emission line spectra, which is in practice their only unifying characteristic.\\
\subsection*{Sample selection}
\noindent This paper is part of a series and should be read as such. In~\citet[][hereafter \M]{Paper1} we presented and analysed $UBVRIHKs$ broadband imaging for a sample of 24 BCGs. That sample was hand--picked to contain interesting and representative cases of BCGs and is biased towards relatively luminous (median $M_B\sim-18$ mag) galaxies. The \M~ sample is defined in terms of galaxy class -- all galaxies are BCGs -- but the heterogeneous and hand--picked nature of the sample make it difficult to translate the properties of such an inherently mixed bag of BCGs to global properties of the galaxies in the local Universe. An inherent problem is that the BCG classification is somewhat ad hoc and based on criteria mainly relating to their appearance on photographic plates rather than their star forming properties, and most samples have ill determined completenesses. In an attempt to study a spatially well defined sample of BCGs complete in terms of luminosity we turned to emission line surveys. In magnitude limited surveys a galaxy's inclusion in the survey depends entirely on its apparent brightness, which introduces a bias against low luminosity systems despite the fact that those are the most common ones. Understandably, one would like to study the most common type of galaxy in the Universe which makes emission line surveys, with their small/no luminosity bias, a favorable place to look for a representative and abundant sample of such systems.~\citet{1989ApJS...70..447S} compiled a large sample of emission line galaxies (ELGs) from Lists $IV$ and $V$ of the University of Michigan (UM) objective-prism survey. The primary selection criteria for this survey are based on the strength and contrast of the [OIII] $\lambda5007$ emission line and it therefore contains a larger fraction of low luminosity dwarfs compared to magnitude limited surveys~\citep{1989ApJS...70..479S}. BCGs, being a subgroup of emission line galaxies, make up about two thirds of the UM survey~\citep{1989ApJS...70..479S}. Our approach in this paper is to take a volume of space and study all emission line galaxies in it. We use~\citet{1989ApJS...70..447S} to select a volume limited sample defined by $11\le RA\le14$h and $v\le2100$ km s${}^{-1}$. This velocity cut--off ensures that we have good completeness at the faint end~\citep[][completeness $\gtrsim95\%$ for $v<2500$km s${}^{-1}$]{1989ApJ...347..152S}. Inside of this volume are $21$ UM ELGs, of which $19$ are BCG--like and two are giant spiral galaxies. Thus selected, this sample is representative of the star forming galaxy population in the local Universe. It consists predominantly of compact low--luminosity dwarfs of various (burst) metallicities -- from low ($Z\sim0.004$) to close to solar ($Z\sim0.02$), and with varying gas content. Throughout this paper we refer to the sample galaxies, with the exception of the two spirals, as BCGs.\\

\noindent These galaxies and the targets from \M~together constitute a sample of $46$ high and low luminosity BCGs. The observations presented here are a part of our ongoing effort to study representative numbers of such galaxies. Kinematic data exist for the majority and are about to be published (\"Ostlin et al. 2012 in prep., Marquart et al. 2012 in prep.).~\citet{1997MNRAS.288...78T} find that such galaxies readily divide into two major morphological types (roughly into regular and irregular), which indicates that they may have different progenitors. The deep optical and NIR imaging data in this paper and in \M~will allow us to study the difference in the faint old populations of these two groups and compare their structural parameters and photometric properties. Though we will frequently refer to the properties of the BCGs in \M~throughout this paper, the bulk of the analysis juxtaposing low and high luminosity BCGs will appear in a dedicated future paper~\citep{Paper3}. We have assumed $H_0=73~km^{-1}s^{-1}Mpc^{-1}$.\\

\noindent The layout of this paper is as follows: \S~\ref{data} introduces the data and the calibration, and provides a log of the observations. \S~\ref{methods} briefly summarizes the derived profiles and the measured quantities. \S~\ref{individ} gives brief notes on the characteristics of the galaxies, as well as a detailed summary of how stellar evolutionary models (SEMs) compare with the observed colors for each galaxy. Where possible, an indication of the age and metallicity for the different populations is given. Observed trends in the integrated colors, asymmetries, total luminosities, and other galaxy properties are discussed in \S~\ref{discuss}. We summarize our conclusions in \S~\ref{conclude}.
\section[]{Observations}\protect\label{data}
\noindent The data consist of optical and NIR broadband imaging, obtained during the period $2003$--$2007$ with ALFOSC (at the Nordic Optical Telescope, NOT), MOSCA (NOT), and EMMI (at the European Southern Observatory New Technology Telescope, ESO NTT) in the optical, and with NOTCAM (NOT) and SOFI (ESO NTT) in the NIR. \\

\noindent We have presented in detail the reduction pipelines and the calibration of the data in~\citet{2010MNRAS.405.1203M} and \M. We shall not repeat it here, except to give some brief notes on MOSCA reductions since \M~did not contain any such data.\\

\noindent \textbf{MOSCA} is a multi-chip instrument (4 CCDs). Each CCD had its own illumination gradient, which was not aligned in concert with the rest. Since our pipeline fits and subtracts a sky from every reduced frame before stacking, it became necessary to adapt it to fit 4 separate skies and subtract those from the individual CCDs on each reduced frame, instead of fitting a single sky on the mosaiced (raw) frames. Dark current (DC) frames were available, however after subtracting the masterbias from the masterdark we found the remaining DC to be negligible for our longest exposure of 10 minutes, hence we did not use the DC frames in the reduction. The MOSCA bias level can occasionally fluctuate throughout the night on some of the CCDs, but again, after examination of the bias frames taken on five separate occasions throughout the night we found it to be very stable for all 4 CCDs. This is not universally the case with MOSCA data, so care must be taken to check the behavior of the dark and bias levels in the four chips for each individual night. The orientation of the four chips is slightly misaligned, which we have corrected for before making the final stacked images. \\

\noindent We should further mention that during the reductions of this sample we again made extensive use of the \textit{astrometry.net} software~\citep{2010AJ....139.1782L} to add a world coordinate system (WCS) to the headers of most of the NOTCAM data. Possibly useful for the community tips, derived from our experience with this software, can be found in \M.\\

\noindent Tables~\ref{exptable} and~\ref{nedtable} summarize the individual exposure times for each filter and the observation log for these data. The heliocentric redshift and distance in Mpc, both taken from NED\footnote{\footnotesize NASA/IPAC Extragalactic Database, http://ned.ipac.caltech.edu/}, are also provided. The filter number at the respective observatory is given for convenience. The sample is volume--limited, with $11\le RA\le 14$h and $v\le2100$ km s${}^{-1}$.\\
\setcounter{table}{0}
\begin{table}
  \begin{minipage}{70mm}
    \caption{Total integration times for the sample. All times are given in minutes and converted to the framework of a 2.56 meter telescope where needed. The values are for observations in a single filter, e.g. only $SOFI~Ks$, and not $SOFI~Ks+NOTCAM~Ks$. }\protect\label{exptable}
    \begin{tabular}{|lccccccc|}
      \hline
      &U&B&V&R&I&H&Ks\\\hline\hline
      UM422&20&60&30&$30^\dagger$&58&23&$125^\ddagger$\\\hline
      UM439&60&40&40&9&58&&249\\\hline
      UM446&&40&40&9&116&21&51\\\hline
      UM452&70&60&20&7&46&37&64\\\hline
      UM456&40&40&40&7&38&30&32\\\hline
      UM461&60&50&40&6&38&32&82\\\hline
      UM462&30&40&40&6&38&32&48\\\hline
      UM463&30&40&40&&38&30&73\\\hline
      UM465&40&40&50&&&&121\\\hline
      UM477&40&20&20&6&38&&62\\\hline
      UM483&20&35&30&9&48&&249\\\hline
      UM491&60&40&40&9&58&&249\\\hline
      UM499&40&36&36&&19&&62\\\hline
      UM500&60&60&30&$30^\dagger$&38&61&123\\\hline
      UM501&40&40&40&6&38&32&121\\\hline
      UM504&30&40&40&9&9&31&121\\\hline
      UM523&10&40&40&6&38&&62\\\hline
      UM533&20&40&40&6&38&&80\\\hline
      UM538&30&30&20&9&58&&145\\\hline
      UM559&60&40&40&9&12&7&123\\\hline
      \hline
    \end{tabular}
    \medskip
    ~\\
    $\dagger$ -- ALFOSC, $\ddagger$ -- SOFI
  \end{minipage}
\end{table}
\subsection[]{Photometric calibration}\protect\label{photcal}
\noindent All data were calibrated in the Vega photometric system. We remind the reader that the calibration in the optical was carried out with Landolt standard stars, while in the NIR we used 2MASS to calibrate against the mean zero point of field stars found in each individual frame, which makes the NIR calibration less dependent on photometric conditions. In the optical we compared the photometry of stars in our calibrated frames with SDSS photometry in the same fields. Any offset larger than 0.05 mag detected between our photometry and the SDSS photometry was then applied to our frames. For both wavelength regimes we estimated the zero point uncertainty, $\sigma_{zp}$, for each final frame as the average residual difference in magnitudes between SDSS/2MASS and our own measurements for different stars around each target (after any existing clear offset has been corrected). If a galaxy was observed on several nights in the same filter we added the uncertainties in quadrature to obtain a total $\sigma_{zp}$ for that galaxy and filter. We have further compared the photometry of our field stars to values from the Pickles stellar library in both optical and NIR, and found no significant offsets.\\
\section[]{Methods}\protect\label{methods}
\noindent The methods used in obtaining surface brightness and color profiles, structural parameters and other quantities of interest were presented in detail in \M. Here we will provide a brief outline of the major steps but we refer the interested reader to \M~for a more in--depth description of the procedures, the individual sources of uncertainty, motivation for the error composition, systematic errors consideration, etc. For the sake of brevity hereafter we will refer to $Ks$ as simply $K$ (except in the conclusions).
\subsection[]{Contour plots and RGB images}
\begin{figure}
\centering
\includegraphics[width=8cm,height=4.cm]{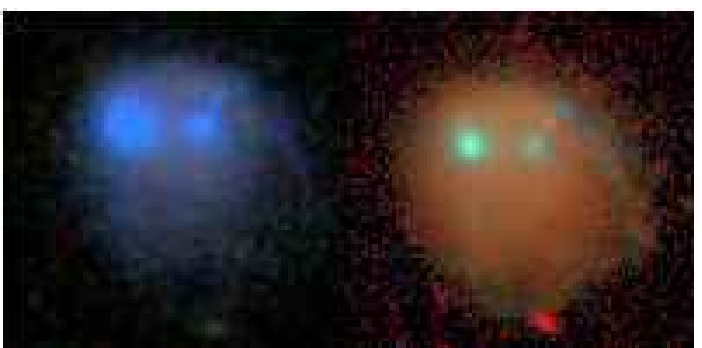}
\includegraphics[width=8cm,height=4.cm]{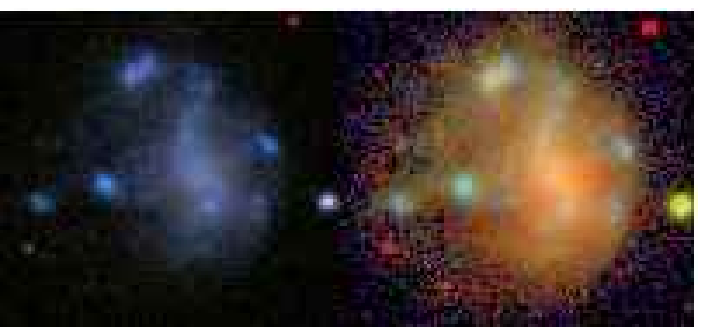}
\caption{\textbf{UM461} (top) and \textbf{UM500} (bottom) in SDSS (left) and our (right) RGB color schemes. }
\protect\label{sdssRGB}
\end{figure}

\noindent Contour plots were obtained with a combination of the python \emph{astLib} package and the built--in \emph{pylab} function \emph{contour}. The isophotal bin is $0.5$ mag for all galaxies. To reduce the noise in the fainter isophotes the images were partially smoothed with the boxcar median filter. The RGB images for each galaxy were made with our own implementation of the~\citet{2004PASP..116..133L} algorithm, where the same scaling and stretch factors were applied to all galaxies in order to facilitate direct comparison. The contour plots and RGB images are both oriented so that North is up and East is to the left. To illustrate the difference in color schemes between the \emph{SDSS} and our own RGB image we show such a comparison in Figure~\ref{sdssRGB} for two random galaxies from our sample. Figure~\ref{datafig} contains the contour and RGB plots for each galaxy. The individual boxcar median filter width and the isophotal level at which it was applied is also indicated in this figure.
\subsection[]{Surface brightness profiles}
\noindent We obtained isophotal and elliptical integration surface brightness profiles for the galaxies in the sample. In the former case we used a constant magnitude bin size of $0.5^{m}$ for all galaxies and all filters, and the deepest image (usually the $B$ band) to define the area of integration at each step. This area was then applied to the rest of the filters. In the case of elliptical integration the radial bin size is $1$ arcsec for all galaxies and all filters, and we again used the $B$ band to define the parameters of the integrating ellipse and applied those to the rest of the filters. In other words, the same physical area is sampled at each magnitude or radial bin in all filters. All foreground and background sources, except the target galaxy, were masked out prior to performing surface photometry on the images, where the mask size is usually a factor of $2.5$ larger than what is returned by \emph{SExtractor}. Though the source detection and masking procedures are automatic, all masks were visually inspected and modified if it was deemed necessary. \\

\noindent The elliptical integration errors include the zero point uncertainty $\sigma_{zp}$, the uncertainty in the sky $\sigma_{sky}$, and the uncertainty in the mean flux level, represented by the standard deviation of the mean flux in each elliptical ring, $\sigma_{sdom}$. The isophotal errors are similarly obtained but of course exclude $\sigma_{sdom}$. The details of the error estimation and the integration procedures are described in \M. Figure~\ref{datafig} shows the isophotal and elliptical surface brightness profiles as well as the resulting radial color profiles for all galaxies. 

\subsection[]{Integrated surface photometry}\protect\label{integsurfphot}
\noindent Similar to \M, in Tables~\ref{pa_holm_tbl} and~\ref{totlumtbl} we present the parameters derived from the surface photometry, including position angle and ellipticity, the Holmberg radius $r_H$, and the total apparent and absolute magnitudes for each galaxy measured down to $r_H$. The error on the total luminosity was obtained by varying the position angle and ellipticity parameters by $\pm 5^\circ$, respectively $\pm0.1$. In Table~\ref{morph_tab} we summarize some general information for each target, such as oxygen--based metallicities and $H_\beta$ equivalent widths, as well as the morphological class obtained either from the literature or through our own analysis of the morphology, where such classification was missing. We also split the underlying host galaxy in two regions, one between $\mu_B\sim24$--$26$ and one between $\mu_B\sim26$--$28$ mag arcsec${}^{-2}$, and calculate the total color over these regions (Table~\ref{totclrtbl}). In the same table we also provide the color of the central region, from the center of integration down to $\mu_B\sim24$ mag arcsec${}^{-2}$, which contains contributions from both the star forming regions and the underlying host galaxy.\\

\noindent Most of the host galaxies in this sample are well approximated by a disk, so it is meaningful to estimate the scale length $h_r$ and the central surface brightness $\mu_0$ (Table~\ref{scalelentbl}) for the sample, which we do in the same way as in \M. The assumption of knowing the exact shape of the underlying host enables us to give an estimate of the burst luminosity, i.e. the excess light above the exponential disk (Table~\ref{burstclrtbl}). The burst errors include the fitting errors of the exponential disk, scaled to units of the profile errors, and the zero point uncertainty. Thus they may be underestimated since there is no measure of the uncertainty in the exact flux level of the burst region. The latter is not included since the burst region is never explicitly defined in $2D$. All color measurements are always performed over identical physical ranges for all filters, taking the $B$ band as reference for defining the respective regions and then applying these regions to the rest of the filters. The errors of the colors are the composite of the individual errors in the two filters, which in turn contain contributions from all three relevant sources of uncertainty -- $\sigma_{zp}$, $\sigma_{sky}$, and $\sigma_{sdom}$. The structural parameters errors, $\sigma(h_r)$ and $\sigma(\mu_0)$, are the propagated errors of the fitted slope, and a composite of the fit error and zero point uncertainty, respectively.
\subsection[]{Asymmetry and concentration}\protect\label{casparam}
\noindent Table~\ref{asymtbl} shows the individual minimum Petrosian asymmetry ($A_P$) for each galaxy. These are calculated following~\citet[][]{2000ApJ...529..886C} as $A=\frac{\sum\lvert I_0-I_\phi\rvert}{2\sum\lvert I_0\rvert}$, where $\phi=180$ degrees. The measurements were performed over the area included in the Petrosian radius $r[\eta(0.2)]$, where all pixels below the corresponding flux level are masked out. We use the inverted $\eta$, defined as the ratio between the local surface brightness at some radius and the average surface brightness inside that radius~\citep[see][and references therein]{2000AJ....119.2645B}. The individual Petrosian radii are also presented in Table~\ref{asymtbl}, since it can be informative to know how large the enclosed area is.\\

\noindent Alternative measures of asymmetry are shown in Table~\ref{tab:altasym}, namely the Holmberg ($A_H^\prime$) and the dynamical ($A_{dyn}$) asymmetry. $A_H^\prime$ is calculated over images smoothed by a boxcar average filter of $1\times1$ kpc from the area enclosed by the Holmberg radius at $\mu=26.5$ mag arcsec${}^{-2}$ in the optical and $R_{23}$ at $\mu=23$ mag arcsec${}^{-2}$ in the NIR. $A_{dyn}$, the dynamical asymmetry, is also calculated over smoothed images, but all pixels brighter than $\mu=25$, $\mu=21$ mag arcsec${}^{-2}$ are set to the constant value of $25,~21$ mag arcsec${}^{-2}$ in the optical, respectively the NIR. This means that all star forming regions contribute nothing to the total asymmetry, allowing $A_{dyn}$ to give more weight to the faint dynamical structures. The faintest isophote in $A_{dyn}$ is $27$ mag arcsec${}^{-2}$ in the optical and $23$ mag arcsec${}^{-2}$ in the NIR.\\

\noindent The concentration index~\citep[e.g.][]{2000AJ....119.2645B} was calculated from $C=5\times\log{\frac{r_{80\%}}{r_{20\%}}}$ where $r_{20}$,$r_{80}$ are the radii at $20\%$, respectively $80\%$ light over an area inside the $1.5\times r[\eta(0.2)]$ radius. These values are listed in Table~\ref{tab:conc}.
\section{Characteristics of individual galaxies}\protect\label{individ}
\noindent In what follows we provide brief notes on the characteristics and relevant information from the literature on the individual galaxies. We have also analyzed color--color diagrams of all combinations of our five primary colors ($U-B$, $B-V$, $V-R$, $V-I$, $V-K$, and $H-K$) for each galaxy and compared them to two stellar evolutionary models -- one with nebular emission contribution and assumed instantaneous burst at zero redshift, and one with a pure stellar population without gas and an e--folding time of $10^9$ yrs, also at zero redshift. The model tracks with nebular emission are based on the \emph{Yggdrasil} spectral synthesis code \citep{2011ApJ...740...13Z}, whereas the pure stellar population tracks are based on \citet{2008A&A...482..883M} isochrones. \M~can be consulted for further details on these models. In each such diagram we plotted the total galaxy color, the central color down to $\mu_B\sim24$ mag arcsec${}^{-2}$, the colors between $24\lesssim\mu_B\lesssim26$ and $26\lesssim\mu_B\lesssim28$ mag arcsec${}^{-2}$, and the burst color estimate. Since the number of plots grew to over $400$, we have not included them in this paper, but they are available on demand\footnote{\footnotesize\thanks{E-mail: genoveva@astro.su.se}}. Any statement we make that indicates the possible metallicity or age of the young or old populations is based on the individual analysis of these plots.\\

\noindent We further provide the morphological class of the galaxy if it is available in the literature, and in the cases where it is missing we assign such a class based on our analysis of the contour and RGB plots for each galaxy. We will adhere to the morphological classification of~\citet[][]{1986sfdg.conf...73L}, namely \emph{iE} (irregular inner and elliptical outer isophotes), \emph{nE} (central nucleus in an elliptical host), \emph{iI,C} (off--center nucleus in a cometary host), and \emph{iI,M} (off--center nucleus in an apparent merger). We further reiterate the~\citet{1989ApJS...70..479S} classification for each galaxy based on spectral features, namely \emph{DANS} (dwarf amorphous nuclear starburst galaxies), \emph{Starburst nucleus galaxies}, \emph{H\texttt{II}H} (\HII hotspot galaxies), \emph{DH\texttt{II}H} (dwarf \HII hotspot galaxies), \emph{SS} (Sargent--Searle objects). These are summarized in Table~\ref{morph_tab}.
\subsection*{UM422}
\noindent This composite object is embedded in one of the most massive \HI envelopes of the sample and has an \HI companion~\citep{1995ApJS...99..427T}. UM422 is actually the blue knot visible to the North--West in the RGB image, in close proximity to an extended red irregular galaxy. This galaxy is so close to its much more massive neighbor that we are unable to separate it and extract its individual surface brightness or color profiles. Since the two galaxies are most likely merging, we present instead the surface brightness and color profiles for the composite object. Due to the irregular morphology of the merger our simplified burst estimation fails for the region $\mu_B=26$--$28$ mag arcsec${}^{-2}$. UM422 has a diameter of $\sim3$ kpc measured down to the Holmberg radius, which makes it a true \emph{SS} object~\citep{1989ApJS...70..479S}. We have measured total colors down to the Holmberg radius, as well as integrated the colors in the two $\mu_B=24$--$26$ and $\mu_B=26$--$28$ mag arcsec${}^{-2}$ regions for the composite object ($UM422~+~$neighbor), but we acknowledge that those will represent the colors of the neighbor more than the colors of $UM422$ itself. All colors indicate the dominant presence of an old stellar component. Specifically, the $B-V$ vs. $V-R$, $V-I$, or $V-K$ colors of the composite object show very little nebular emission contamination and are well--fitted by a  stellar population with age $\gtrsim3$ Gyr. They are also well--fitted by model tracks including nebular emission but also for an old age $\gtrsim1$ Gyr. Since both models indicate the sampled population is old, we must place higher weight on the pure stellar population model, since it is better suited to model old populations than \emph{Yggdrasil}. 
\subsection*{UM439}
\noindent This is a Wolf--Rayet galaxy~\citep{1999A&AS..136...35S} classified as \emph{iE} BCD by~\citet{2003ApJS..147...29G}, and a~\emph{DH\texttt{II}H} by~\citet{1989ApJS...70..479S}. This galaxy has an \HI distribution asymmetric in the North--East, which might indicate an interaction companion or a tidal feature~\citep{1995ApJS...99..427T}, however, the galaxy appears isolated in the sense that there are no detected companions within $1$ Mpc~\citep{1991A&A...241..358C,1993AJ....106.1784C,1995ApJS...99..427T}. The strongest star forming region is located at the highest \HI column density, with the weaker SF regions being remnants of recent burst activity which seems to be dying out~\citep{1998AJ....116.1186V}. Based on its IRAS $f_{25}/f_{100}$ and $f_{60}/f_{100}$ color indices the starburst is still the dominant property of this galaxy, with relatively low extinction based on $H_\alpha/H_\beta$~\citep{1991A&AS...91..285T}. The extinction is enough, however, to affect the $U-B$ and $B-V$ burst colors. The central colors are indicative of a young population $>10$ Myrs and nebular emission contribution. The latter drops dramatically for $24\lesssim\mu_B\lesssim26$ and $26\lesssim\mu_B\lesssim28$ mag arcsec${}^{-2}$ region colors, specifically $V-I>1$ and $V-R>0.6$ colors are observed for both regions, indicative of an old $\sim10$ Gyr stellar population of metallicity higher than $Z\sim0.004$. There is a bright strongly saturated star in the \emph{I} band to the East of the galaxy, which we have of course masked out when obtaining the surface brightness profiles or measuring burst or host integrated colors. 

\begin{center}
  \begin{table*}
    \begin{minipage}{150mm}
      \tiny
      \caption{Log of the observations. Heliocentric redshift and cosmology--corrected luminosity distances from \emph{NED}.}\protect\label{nedtable}
      \begin{tabular}{@{}|lllllll|@{}}
        \hline
        Name &Ra Dec (J2000)&Redshift&$\textrm{D}~[\textrm{Mpc}]$&Year&Instrument&Filters\\\hline
        UM422&11h20m14.6s&0.005360&27.2&2003&ALFOSC-FASU&U\#7, B\#74, V\#75, R\#76\\
        &+02d31m53s&&&2005&EMMI&R\#608, I\#610\\
        &&&&&SOFI&Ks\#13\\
        &&&&&NOTCAM&H\#204\\
        &&&&2007&NOTCAM&Ks\#207\\\hline
        UM439&11h36m36.8s&0.003666&20.2&2004&ALFOSC-FASU&B\#74, V\#75\\
        &+00d48m58s&&&2005&EMMI&R\#608, I\#610\\
        &&&&&ALFOSC-FASU&U\#7\\
        &&&&2006&SOFI&Ks\#13\\\hline
        UM446&11h41m45.6s&0.006032&30.0&2004&ALFOSC-FASU&B\#74, V\#75\\
        &-01d54m05s&&&2005&EMMI&R\#608, I\#610\\
        &&&&&NOTCAM&H\#204\\
        &&&&2006&NOTCAM&Ks\#207\\\hline
        UM452&11h47m00.7s&0.004931&25.4&2003&ALFOSC-FASU&B\#74, V\#75\\
        &-00d17m39s&&&2005&EMMI&R\#608, I\#610\\
        &&&&&NOTCAM&H\#204\\
        &&&&&ALFOSC-FASU&U\#7\\
        &&&&2007&NOTCAM&Ks\#207\\\hline
        UM456&11h50m36.3s&0.005940&29.6&2004&ALFOSC-FASU&B\#74, V\#75\\
        &-00d34m03s&&&2005&EMMI&R\#608, I\#610\\
        &&&&&NOTCAM&H\#204\\
        &&&&2006&MOSCA&U\#104\\
        &&&&&NOTCAM&Ks\#207\\\hline
        UM461&11h51m33.3s&0.003465&19.3&2004&ALFOSC-FASU&B\#74, V\#75\\
        &-02d22m22s&&&2005&EMMI&R\#608, I\#610\\
        &&&&&ALFOSC-FASU&U\#7\\
        &&&&&NOTCAM&H\#204\\
        &&&&2006&NOTCAM&Ks\#207\\\hline
        UM462&11h52m37.2s&0.003527&19.6&2004&ALFOSC-FASU&B\#74, V\#75\\
        &-02d28m10s&&&2005&ALFOSC-FASU&U\#7\\
        &&&&&EMMI&R\#608, I\#610\\
        &&&&&NOTCAM&H\#204\\
        &&&&2007&NOTCAM&Ks\#207\\\hline
        UM463&11h52m47.5s&0.004640&24.2&2004&ALFOSC-FASU&B\#74, V\#75\\
        &-00d40m08s&&&2005&NOTCAM&H\#204\\
        &&&&&EMMI&I\#610\\
        &&&&2006&MOSCA&U\#104\\
        &&&&2007&NOTCAM&Ks\#207\\\hline
        UM465&11h54m12.3s&0.003820&20.7&2004&ALFOSC-FASU&B\#74, V\#75\\
        &+00d08m12s&&&2006&SOFI&Ks\#13\\
        &&&&&MOSCA&U\#104\\\hline
        UM477&12h08m11.1s&0.004426&23.2&2005&EMMI&R\#608, I\#610\\
        &+02d52m42s&&&&SOFI&Ks\#13\\
        &&&&2006&ALFOSC-FASU&B\#74, V\#75\\
        &&&&&MOSCA&U\#104\\\hline
        UM483&12h12m14.7s&0.007792&37.2&2005&ALFOSC-FASU&U\#7\\
        &+00d04m20s&&&&EMMI&R\#608, I\#610\\
        &&&&2006&ALFOSC-FASU&B\#74, V\#75\\
        &&&&&SOFI&Ks\#13\\\hline
        UM491&12h19m53.2s&0.006665&32.4&2004&ALFOSC-FASU&B\#74, V\#75\\
        &+01d46m24s&&&2005&EMMI&R\#608, I\#610\\
        &&&&&ALFOSC-FASU&U\#7\\
        &&&&2006&SOFI&Ks\#13\\\hline
        UM499&12h25m42.8s&0.007138&34.3&2004&ALFOSC-FASU&B\#74, V\#75\\
        &+00d34m21s&&&2005&EMMI&I\#610\\
        &&&&&SOFI&Ks\#13\\
        &&&&2006&MOSCA&U\#104\\\hline
        UM500&12h26m12.8s&0.007000&33.8&2003&ALFOSC-FASU&B\#74, V\#75, R\#76\\
        &-01d18m16s&&&2005&EMMI&R\#608, I\#610\\
        &&&&&NOTCAM&H\#204\\
        &&&&&ALFOSC-FASU&U\#7\\
        &&&&2006&SOFI&Ks\#13\\\hline
        UM501&12h26m22.7s&0.006761&32.8&2004&ALFOSC-FASU&B\#74, V\#75\\
        &-01d15m12s&&&2005&EMMI&R\#608, I\#610\\
        &&&&&NOTCAM&H\#204\\
        &&&&2006&SOFI&Ks\#13\\
        &&&&&MOSCA&U\#104\\\hline
        UM504&12h32m23.6s&0.006800&32.9&2004&ALFOSC-FASU&B\#74, V\#75\\
        &-01d44m24s&&&2005&NOTCAM&H\#204\\
        &&&&2006&MOSCA&U\#104\\
        &&&&&SOFI&Ks\#13\\
        &&&&&ALFOSC-FASU&R\#76, I\#12\\\hline
        UM523&12h54m51.0s&0.003052&17.1&2004&ALFOSC-FASU&B\#74, V\#75\\
        &+02d39m15s&&&2005&SOFI&Ks\#13\\
        &&&&&EMMI&R\#608, I\#610\\
        &&&&2006&MOSCA&U\#104\\\hline
        UM533&12h59m58.1s&0.002957&16.7&2004&ALFOSC-FASU&B\#74, V\#75\\
        &+02d02m57s&&&2005&EMMI&R\#608, I\#610\\
        &&&&&SOFI&Ks\#13\\
        &&&&2006&MOSCA&U\#104\\\hline
        UM538&13h02m40.8s&0.003065&17.1&2005&EMMI&R\#608, I\#610\\
        &+01d04m27s&&&&SOFI&Ks\#13\\
        &&&&2006&MOSCA&U\#104\\
        &&&&&ALFOSC-FASU&B\#74, V\#75\\\hline
        UM559&13h17m42.8s&0.004153&21.5&2004&ALFOSC-FASU&B\#74, V\#75\\
        &-01d00m01s&&&2005&ALFOSC-FASU&U\#7\\
        &&&&&NOTCAM&H\#204\\
        &&&&2006&SOFI&Ks\#13\\
        &&&&&ALFOSC-FASU&R\#76, I\#12\\\hline
        \hline
      \end{tabular}
    \end{minipage}
  \end{table*}
\end{center}
\begin{figure*}
\begin{minipage}{150mm}
\centering
\includegraphics[width=15cm,height=18cm]{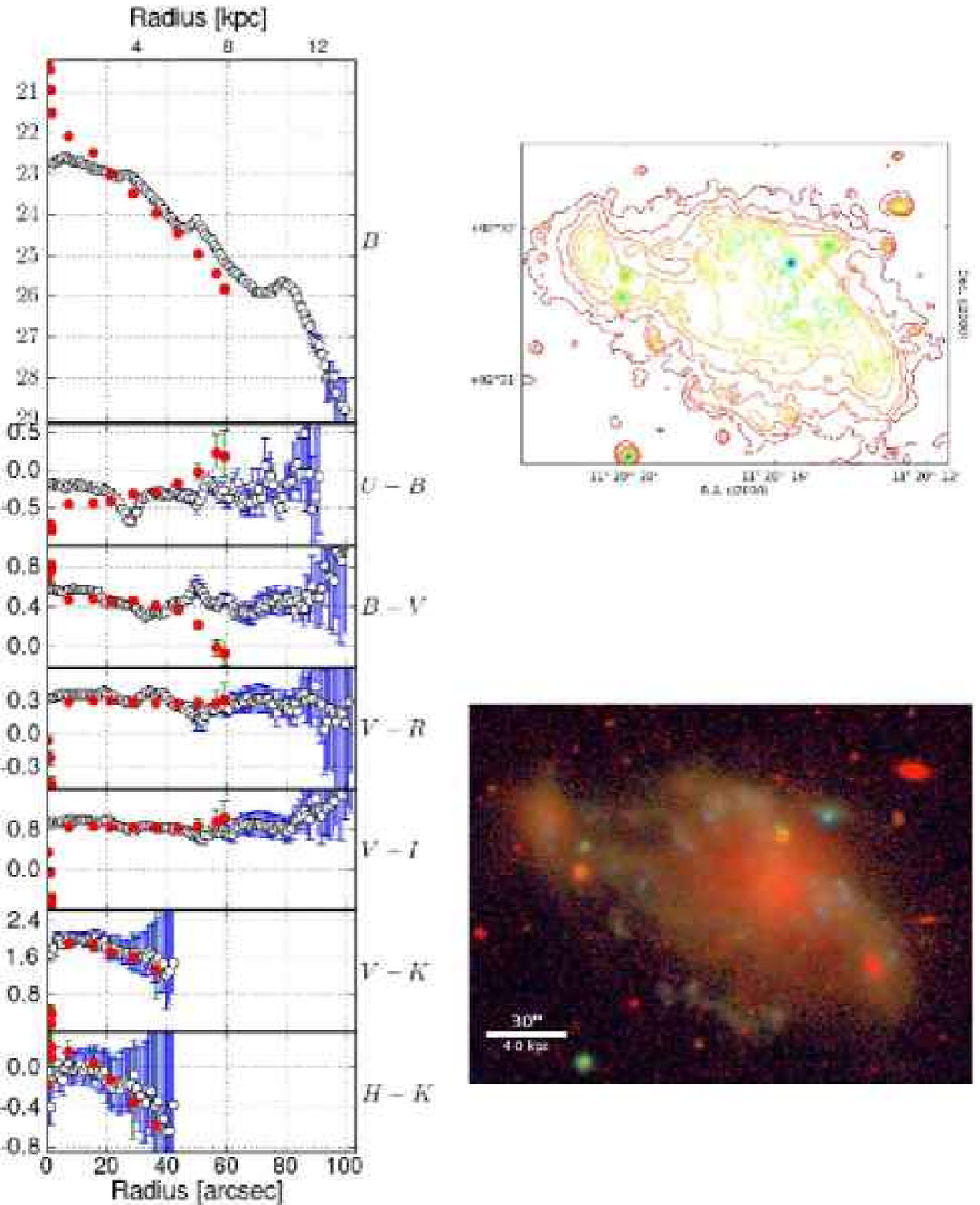}
\caption{\label{datafig}\textbf{UM422}. \textit{Left panel}: Surface brightness and color radial profiles for elliptical (open circles) and isophotal (red circles) integration in the Vega photometric system. \textit{Upper right panel}: contour plot based on the $B$ band. Isophotes fainter than $23.0$, $25.5$ are iteratively smoothed with a boxcar median filter of sizes $5$, $15$ pixels respectively. \textit{Lower right panel}: A true color RGB composite image using the U,B,I filters. Each channel has been corrected for Galactic extinction following~\citet{1998ApJ...500..525S} and converted to the AB photometric system. The RGB composite was created by adapting the~\citet{2004PASP..116..133L} algorithm.}
\end{minipage}
\end{figure*}
\clearpage
\begin{figure*}
\begin{minipage}{150mm}
\centering
\includegraphics[width=15cm,height=18cm]{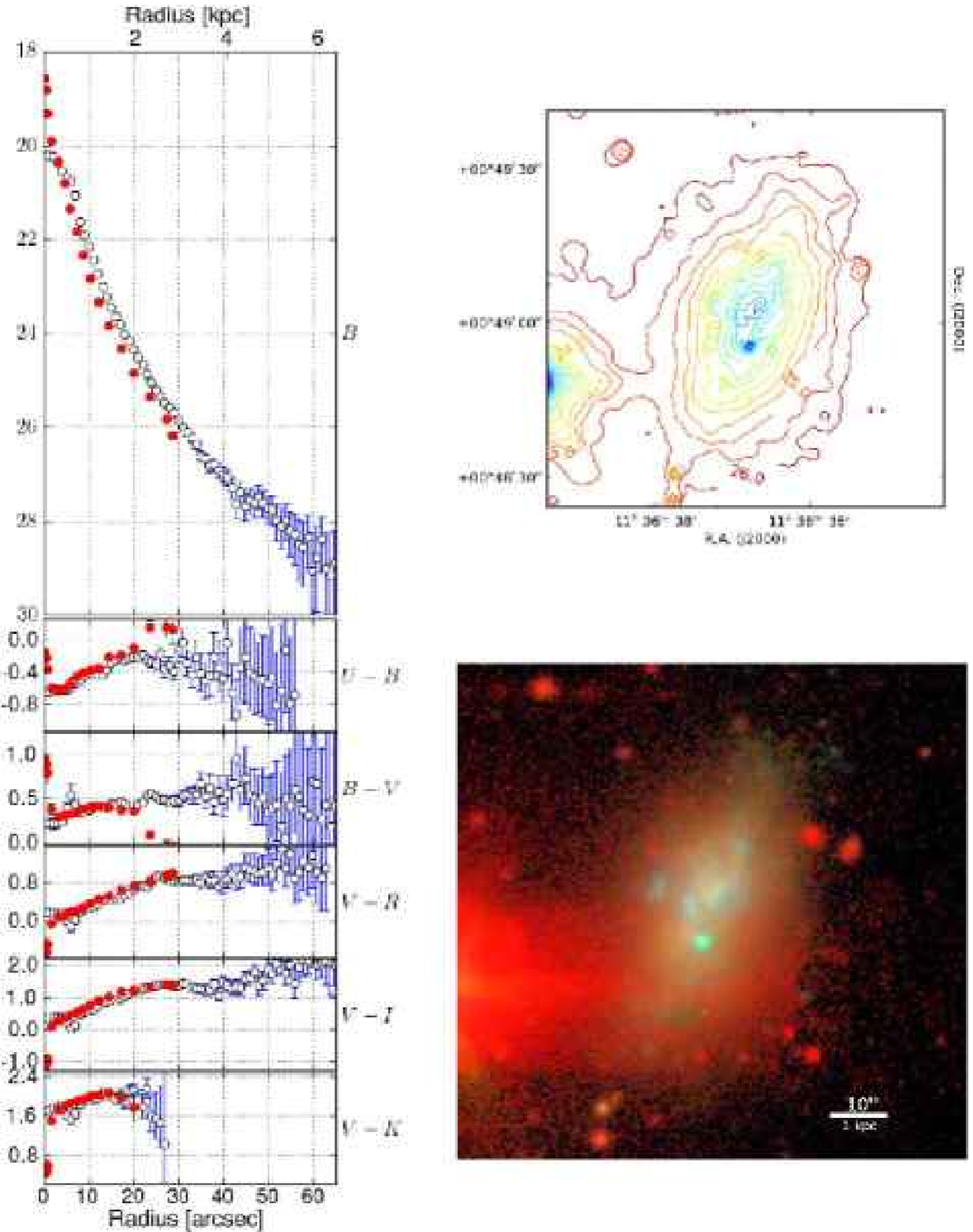}
\contcaption{\textbf{UM439}. \textit{Left panel}: Surface brightness and color radial profiles for elliptical (open circles) and isophotal (red circles) integration in the Vega photometric system. \textit{Upper right panel}: contour plot based on the $B$ band. Isophotes fainter than $23.0$, $25.5$ are iteratively smoothed with a boxcar median filter of sizes $5$, $15$ pixels respectively. \textit{Lower right panel}: A true color RGB composite image using the U,B,I filters. Each channel has been corrected for Galactic extinction following~\citet{1998ApJ...500..525S} and converted to the AB photometric system. The RGB composite was created by adapting the~\citet{2004PASP..116..133L} algorithm.}
\end{minipage}
\end{figure*}
\clearpage
\begin{figure*}
\begin{minipage}{150mm}
\centering
\includegraphics[width=15cm,height=18cm]{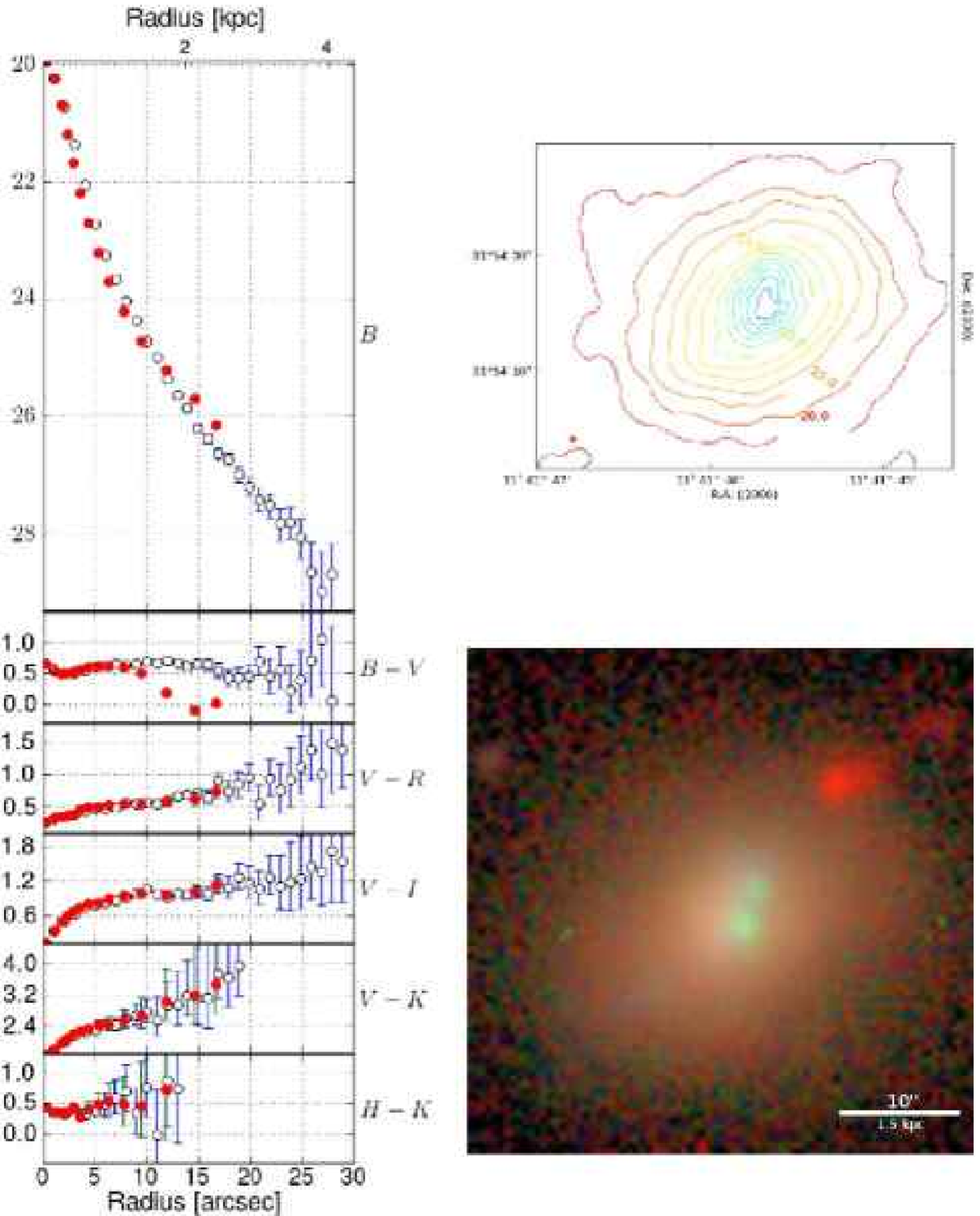}
\contcaption{\textbf{UM446}. \textit{Left panel}: Surface brightness and color radial profiles for elliptical (open circles) and isophotal (red circles) integration in the Vega photometric system. \textit{Upper right panel}: contour plot based on the $B$ band. Isophotes fainter than $23.0$, $25.5$ are iteratively smoothed with a boxcar median filter of sizes $5$, $15$ pixels respectively. \textit{Lower right panel}: A true color RGB composite image using the B,V,I filters. Each channel has been corrected for Galactic extinction following~\citet{1998ApJ...500..525S} and converted to the AB photometric system. The RGB composite was created by adapting the~\citet{2004PASP..116..133L} algorithm.}
\end{minipage}
\end{figure*}
\clearpage
\begin{figure*}
\begin{minipage}{150mm}
\centering
\includegraphics[width=15cm,height=18cm]{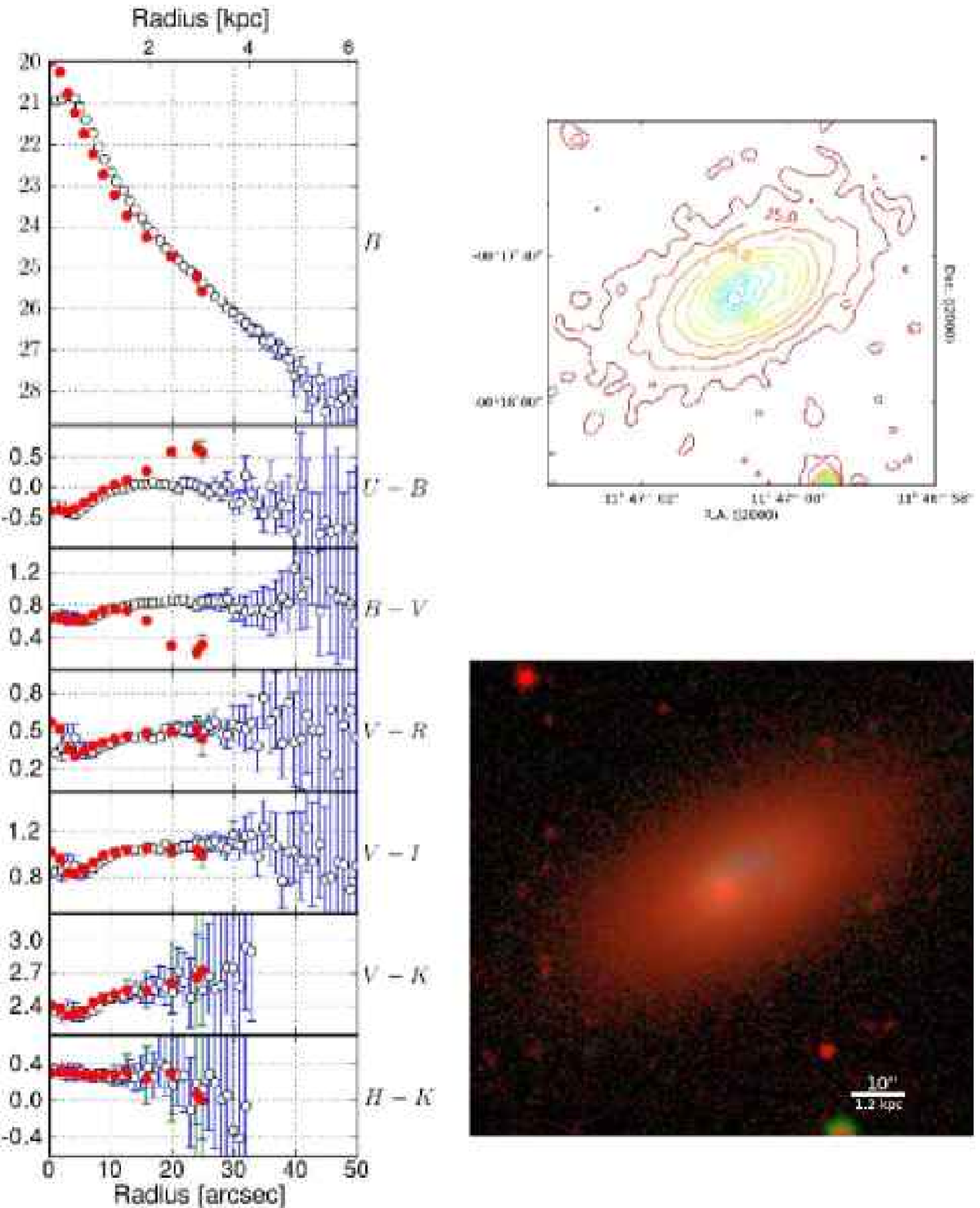}
\contcaption{\textbf{UM452}. \textit{Left panel}: Surface brightness and color radial profiles for elliptical (open circles) and isophotal (red circles) integration in the Vega photometric system. \textit{Upper right panel}: contour plot based on the $B$ band. Isophotes fainter than $22.5$ are smoothed with a boxcar median filter of size $5$ pixels. \textit{Lower right panel}: A true color RGB composite image using the U,B,I filters. Each channel has been corrected for Galactic extinction following~\citet{1998ApJ...500..525S} and converted to the AB photometric system. The RGB composite was created by adapting the~\citet{2004PASP..116..133L} algorithm.}
\end{minipage}
\end{figure*}
\clearpage
\begin{figure*}
\begin{minipage}{150mm}
\centering
\includegraphics[width=15cm,height=18cm]{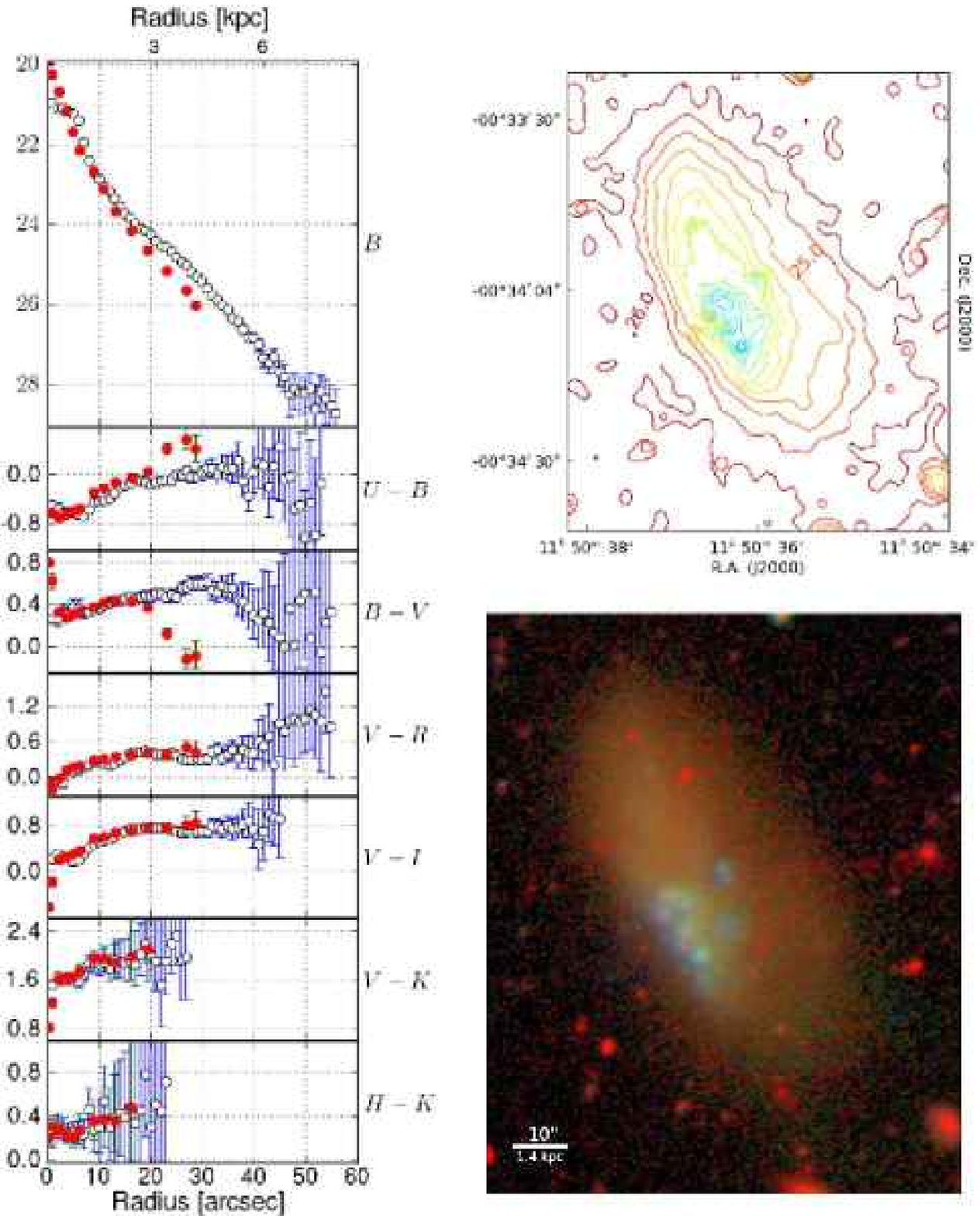}
\contcaption{\textbf{UM456}. \textit{Left panel}: Surface brightness and color radial profiles for elliptical (open circles) and isophotal (red circles) integration in the Vega photometric system. \textit{Upper right panel}: contour plot based on the $B$ band. Isophotes fainter than $23.0$ are smoothed with a boxcar median filter of size $5$ pixels. \textit{Lower right panel}: A true color RGB composite image using the U,B,I filters. Each channel has been corrected for Galactic extinction following~\citet{1998ApJ...500..525S} and converted to the AB photometric system. The RGB composite was created by adapting the~\citet{2004PASP..116..133L} algorithm.}
\end{minipage}
\end{figure*}
\clearpage
\begin{figure*}
\begin{minipage}{150mm}
\centering
\includegraphics[width=15cm,height=18cm]{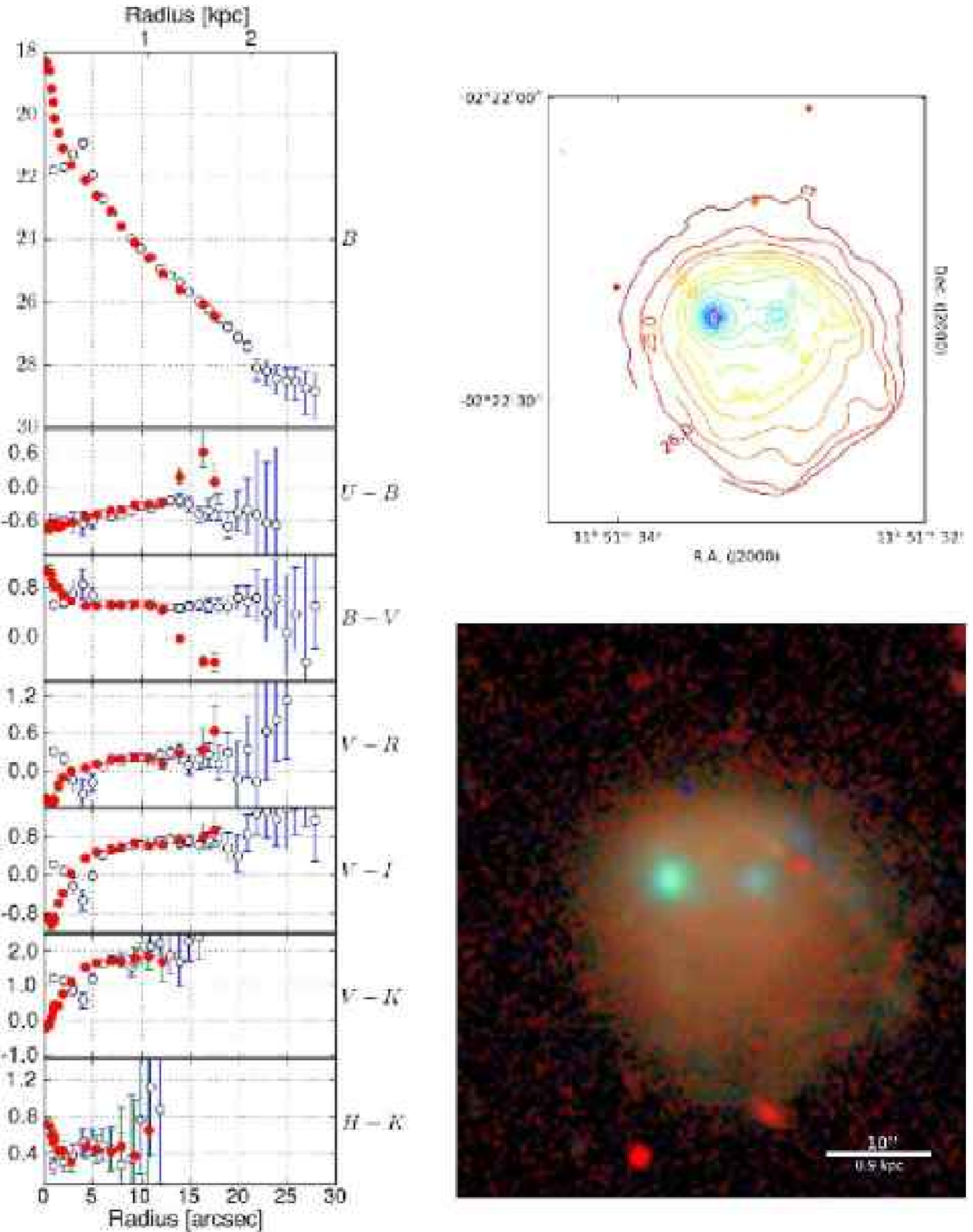}
\contcaption{\textbf{UM461}. \textit{Left panel}: Surface brightness and color radial profiles for elliptical (open circles) and isophotal (red circles) integration in the Vega photometric system. \textit{Upper right panel}: contour plot based on the $B$ band. Isophotes fainter than $23.5$, $25.5$ are iteratively smoothed with a boxcar median filter of sizes $5$, $15$ pixels respectively. \textit{Lower right panel}: A true color RGB composite image using the U,B,I filters. Each channel has been corrected for Galactic extinction following~\citet{1998ApJ...500..525S} and converted to the AB photometric system. The RGB composite was created by adapting the~\citet{2004PASP..116..133L} algorithm.}
\end{minipage}
\end{figure*}
\clearpage
\begin{figure*}
\begin{minipage}{150mm}
\centering
\includegraphics[width=15cm,height=18cm]{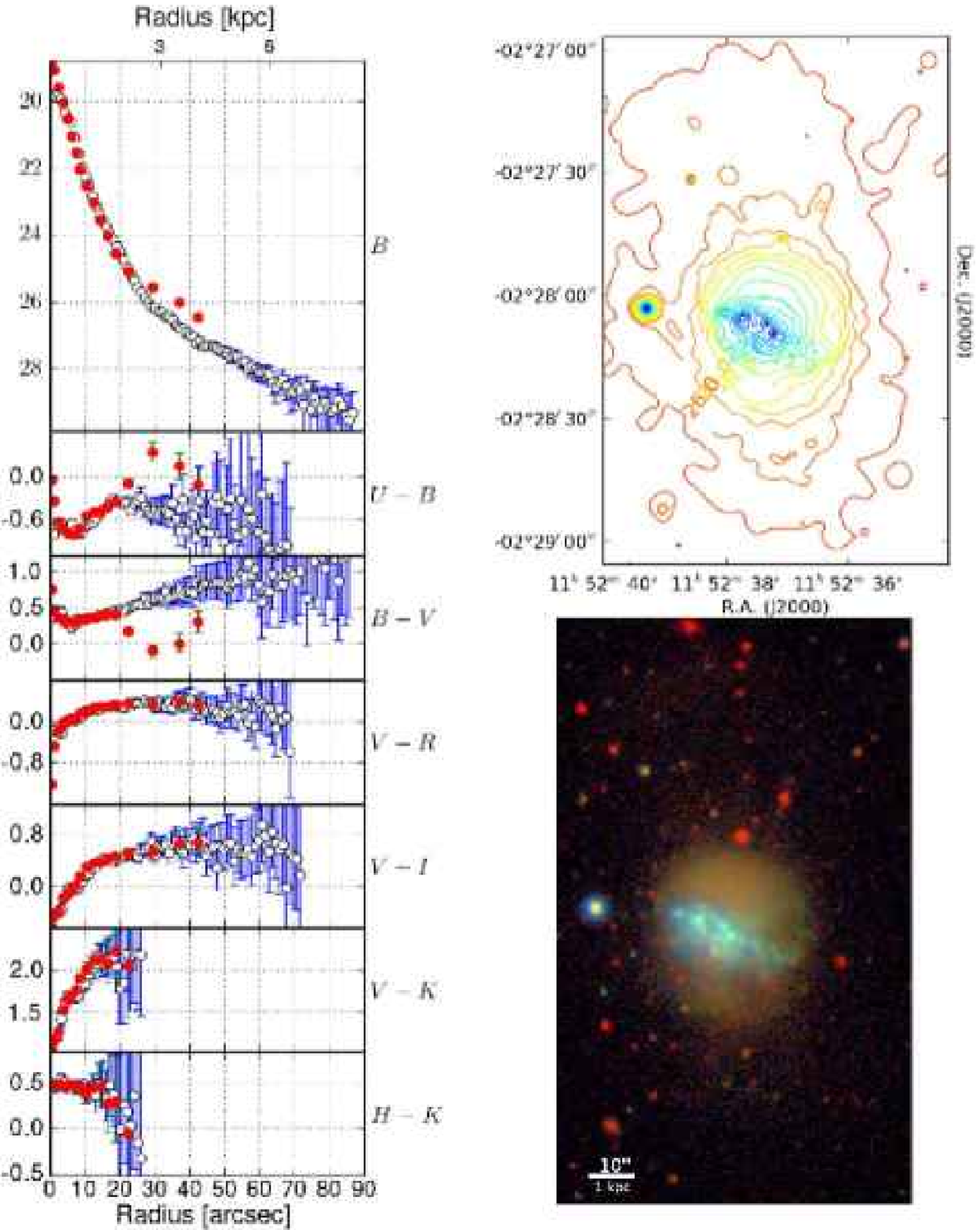}
\contcaption{\textbf{UM462}. \textit{Left panel}: Surface brightness and color radial profiles for elliptical (open circles) and isophotal (red circles) integration in the Vega photometric system. \textit{Upper right panel}: contour plot based on the $B$ band. Isophotes fainter than $22.5$, $25.5$ are iteratively smoothed with a boxcar median filter of sizes $5$, $15$ pixels respectively. \textit{Lower right panel}: A true color RGB composite image using the U,B,I filters. Each channel has been corrected for Galactic extinction following~\citet{1998ApJ...500..525S} and converted to the AB photometric system. The RGB composite was created by adapting the~\citet{2004PASP..116..133L} algorithm.}
\end{minipage}
\end{figure*}
\clearpage
\begin{figure*}
\begin{minipage}{150mm}
\centering
\includegraphics[width=15cm,height=18cm]{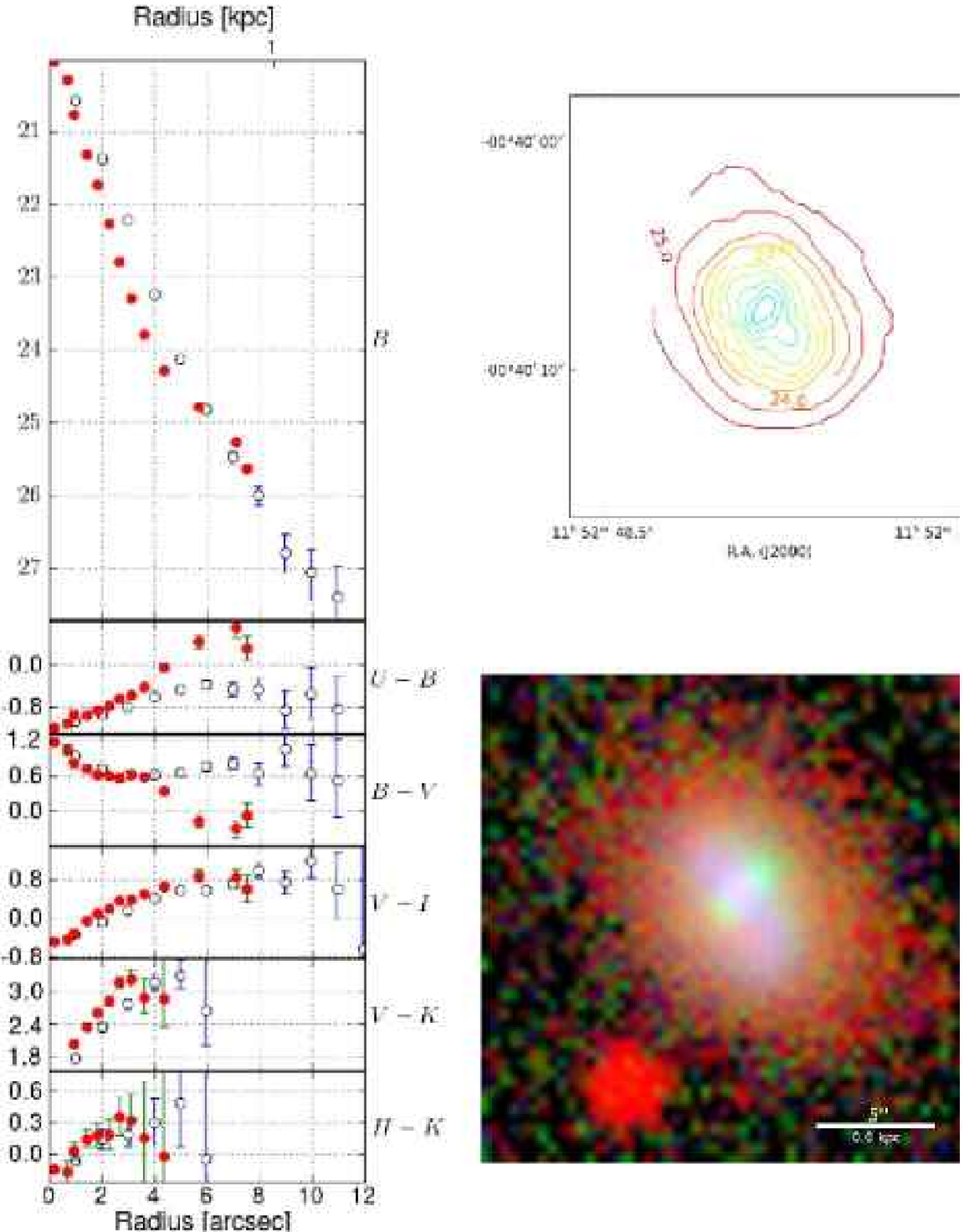}
\contcaption{\textbf{UM463}. \textit{Left panel}: Surface brightness and color radial profiles for elliptical (open circles) and isophotal (red circles) integration in the Vega photometric system. \textit{Upper right panel}: contour plot based on the $B$ band. Isophotes fainter than $23.0$ are smoothed with a boxcar median filter of size $5$ pixels. \textit{Lower right panel}: A true color RGB composite image using the U,B,I filters. Each channel has been corrected for Galactic extinction following~\citet{1998ApJ...500..525S} and converted to the AB photometric system. The RGB composite was created by adapting the~\citet{2004PASP..116..133L} algorithm.}
\end{minipage}
\end{figure*}
\clearpage
\begin{figure*}
\begin{minipage}{150mm}
\centering
\includegraphics[width=15cm,height=18cm]{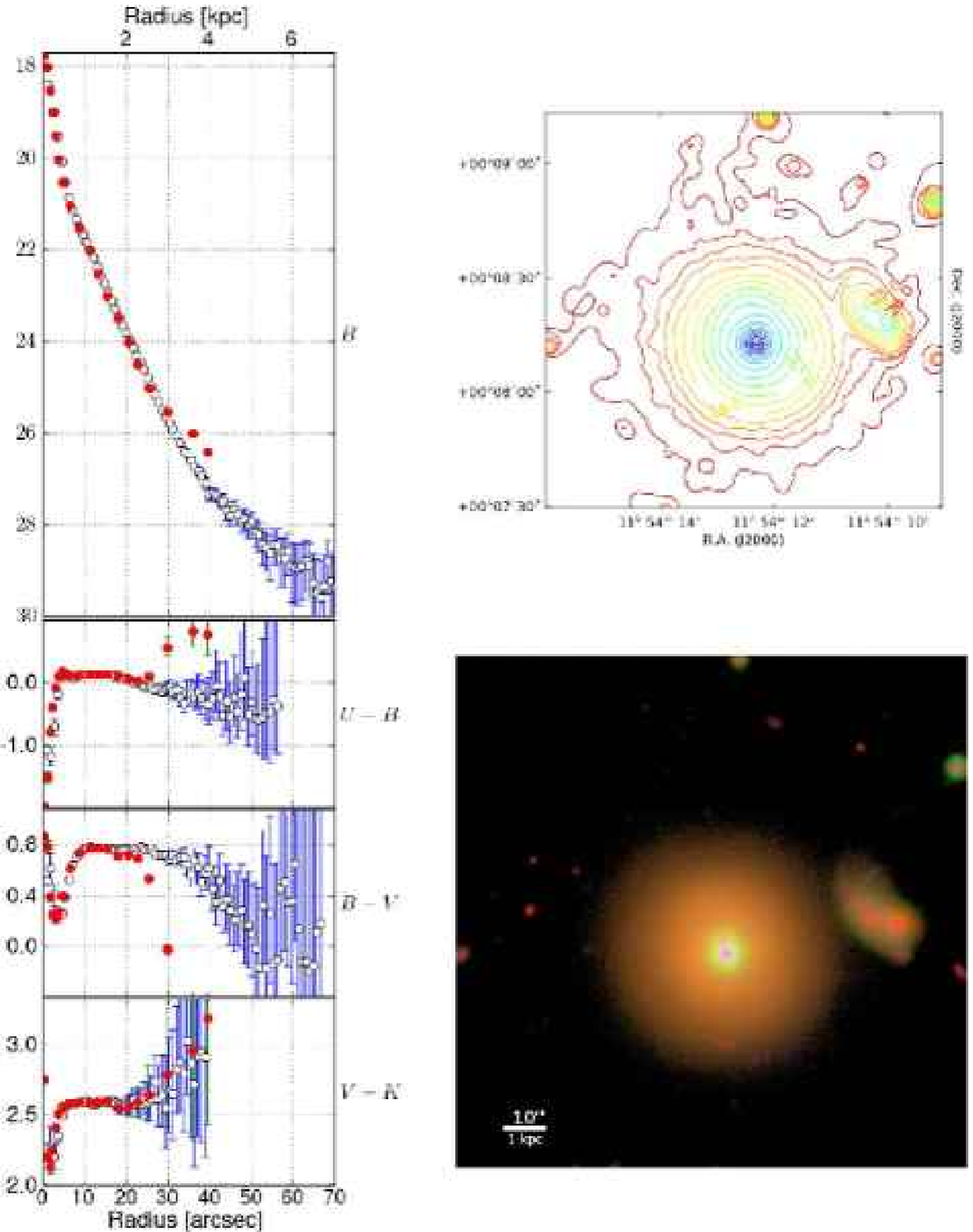}
\contcaption{\textbf{UM465}. \textit{Left panel}: Surface brightness and color radial profiles for elliptical (open circles) and isophotal (red circles) integration in the Vega photometric system. \textit{Upper right panel}: contour plot based on the $B$ band. Isophotes fainter than $23.0$, $25.5$ are iteratively smoothed with a boxcar median filter of sizes $5$, $15$ pixels respectively. \textit{Lower right panel}: A true color RGB composite image using the U,B,V filters. Each channel has been corrected for Galactic extinction following~\citet{1998ApJ...500..525S} and converted to the AB photometric system. The RGB composite was created by adapting the~\citet{2004PASP..116..133L} algorithm.}
\end{minipage}
\end{figure*}
\clearpage
\begin{figure*}
\begin{minipage}{150mm}
\centering
\includegraphics[width=15cm,height=18cm]{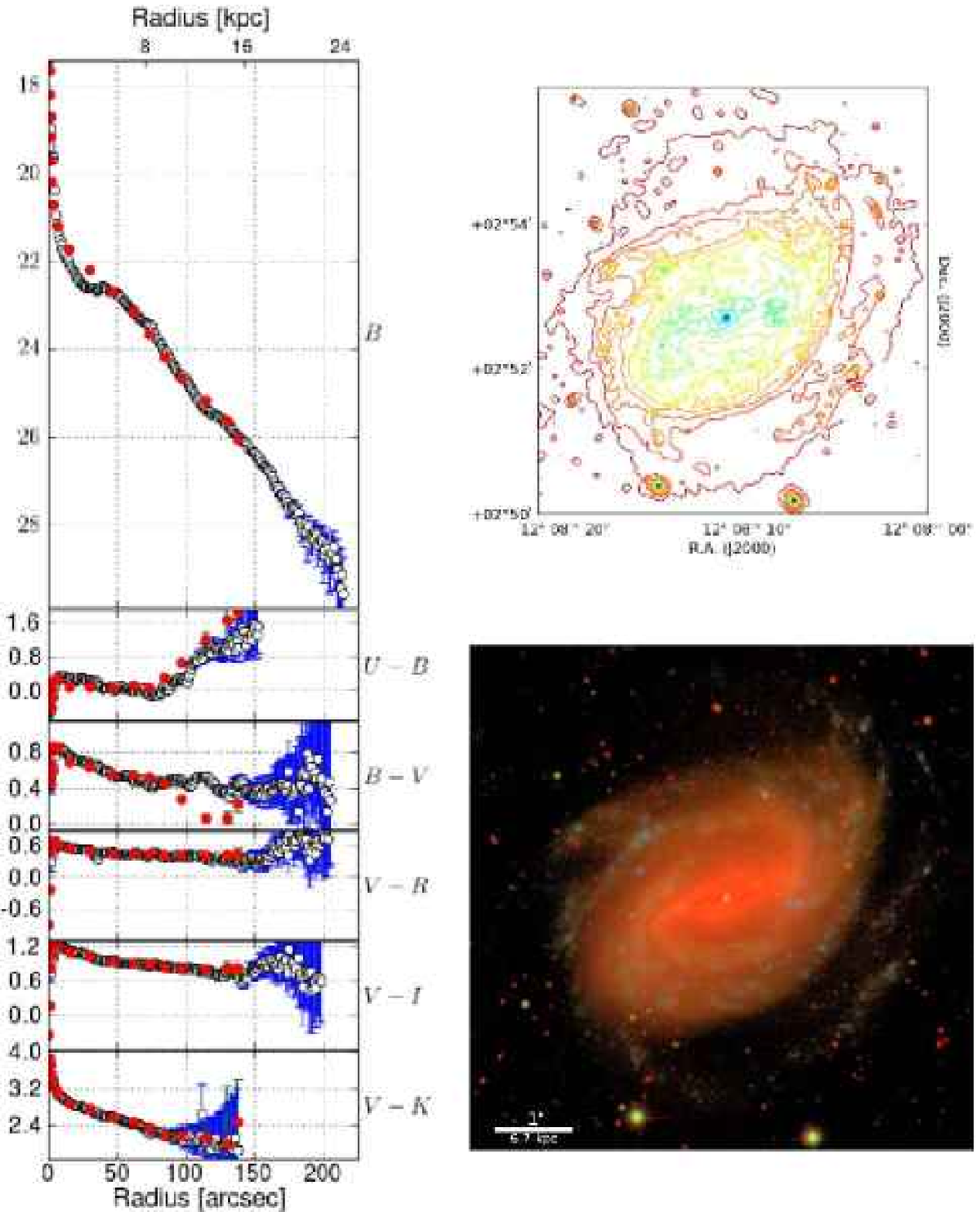}
\contcaption{\textbf{UM477}. \textit{Left panel}: Surface brightness and color radial profiles for elliptical (open circles) and isophotal (red circles) integration in the Vega photometric system. \textit{Upper right panel}: contour plot based on the $B$ band. Isophotes fainter than $22.0$, $25.0$ are iteratively smoothed with a boxcar median filter of sizes $5$, $15$ pixels respectively. \textit{Lower right panel}: A true color RGB composite image using the U,B,I filters. Each channel has been corrected for Galactic extinction following~\citet{1998ApJ...500..525S} and converted to the AB photometric system. The RGB composite was created by adapting the~\citet{2004PASP..116..133L} algorithm.}
\end{minipage}
\end{figure*}
\clearpage
\begin{figure*}
\begin{minipage}{150mm}
\centering
\includegraphics[width=15cm,height=18cm]{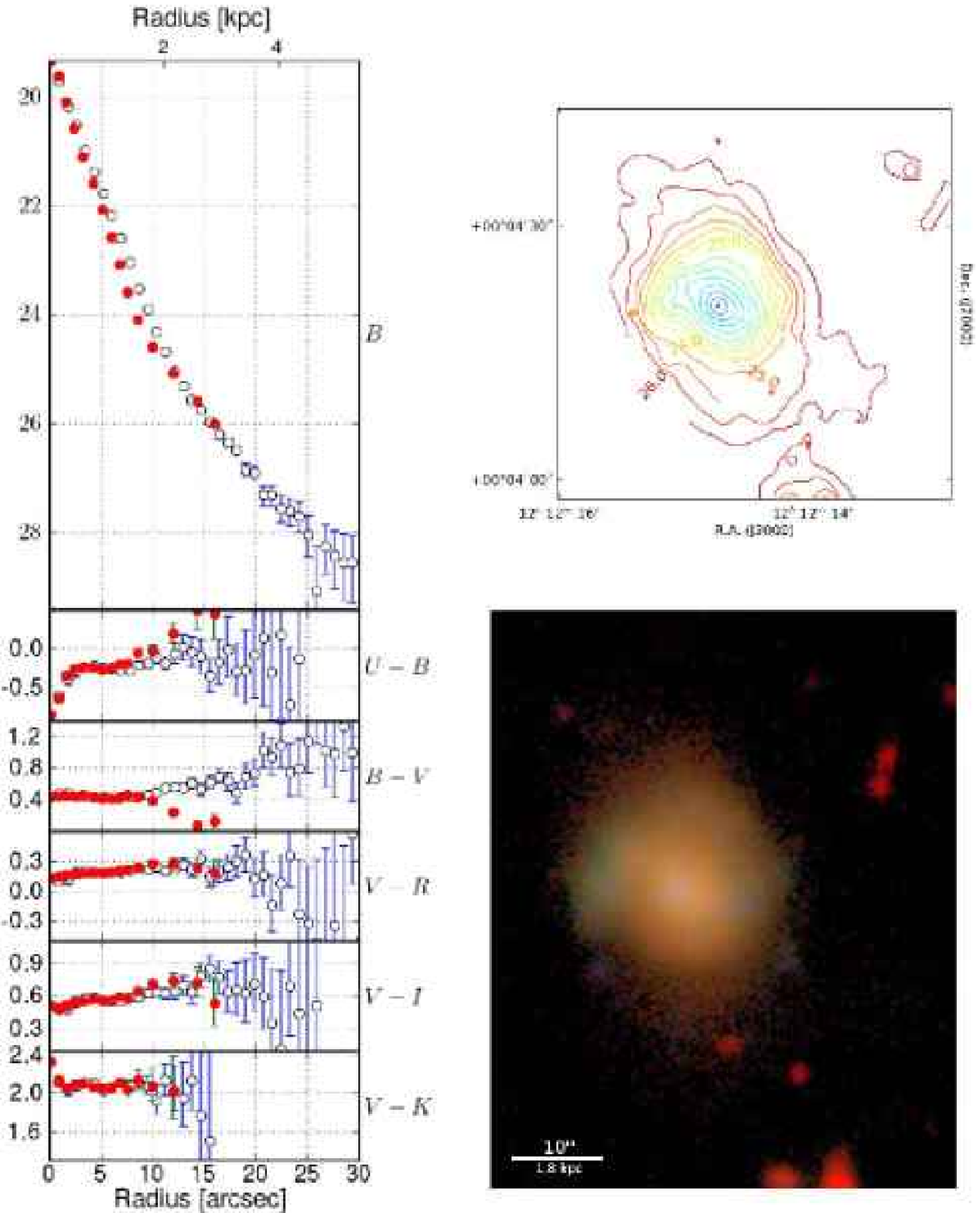}
\contcaption{\textbf{UM483}. \textit{Left panel}: Surface brightness and color radial profiles for elliptical (open circles) and isophotal (red circles) integration in the Vega photometric system. \textit{Upper right panel}: contour plot based on the $B$ band. Isophotes fainter than $23.5$, $25.5$ are iteratively smoothed with a boxcar median filter of sizes $5$, $15$ pixels respectively. \textit{Lower right panel}: A true color RGB composite image using the U,B,I filters. Each channel has been corrected for Galactic extinction following~\citet{1998ApJ...500..525S} and converted to the AB photometric system. The RGB composite was created by adapting the~\citet{2004PASP..116..133L} algorithm.}
\end{minipage}
\end{figure*}
\clearpage
\begin{figure*}
\begin{minipage}{150mm}
\centering
\includegraphics[width=15cm,height=18cm]{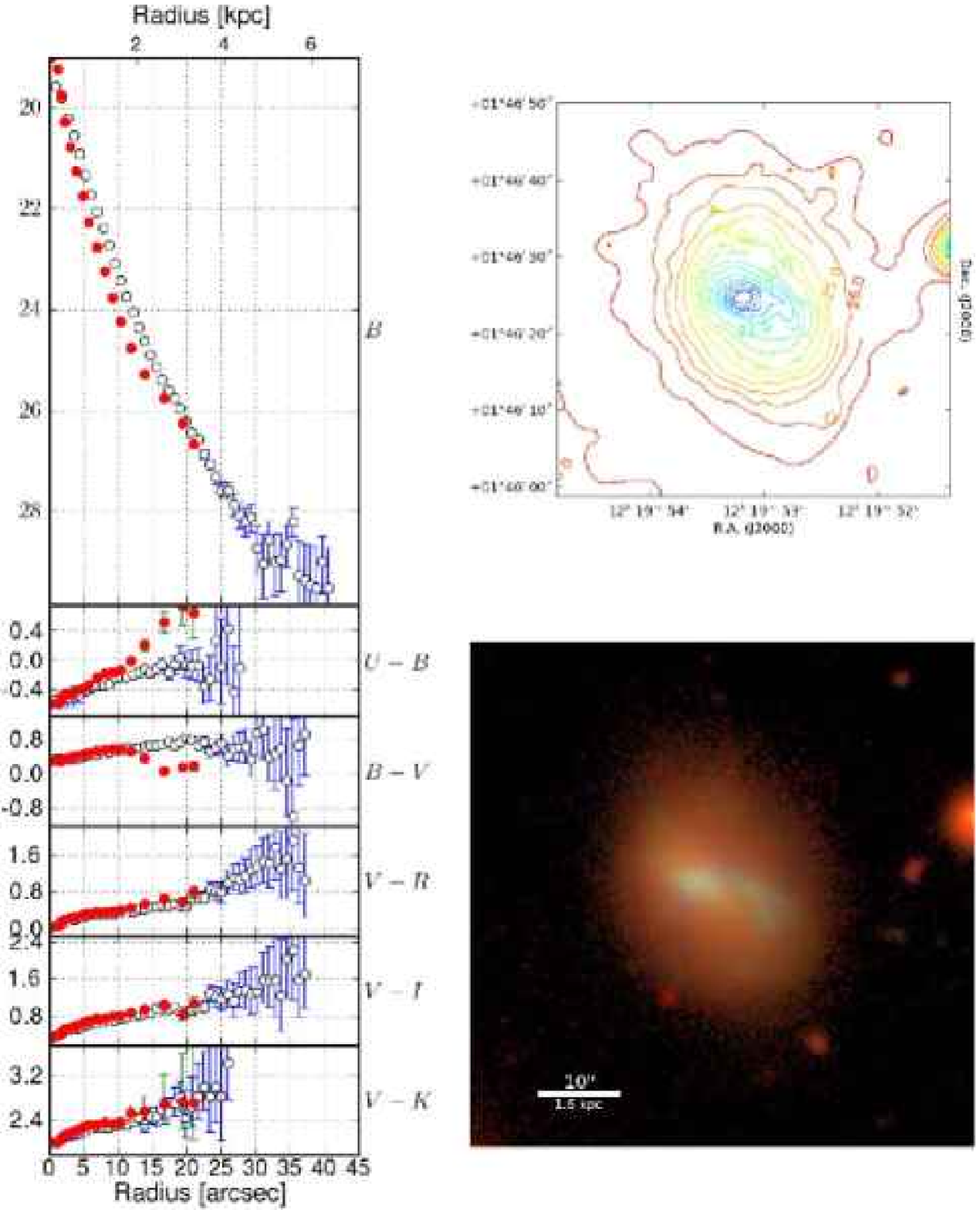}
\contcaption{\textbf{UM491}. \textit{Left panel}: Surface brightness and color radial profiles for elliptical (open circles) and isophotal (red circles) integration in the Vega photometric system. \textit{Upper right panel}: contour plot based on the $B$ band. Isophotes fainter than $23.5$, $25.5$ are iteratively smoothed with a boxcar median filter of sizes $5$, $15$ pixels respectively. \textit{Lower right panel}: A true color RGB composite image using the U,B,I filters. Each channel has been corrected for Galactic extinction following~\citet{1998ApJ...500..525S} and converted to the AB photometric system. The RGB composite was created by adapting the~\citet{2004PASP..116..133L} algorithm.}
\end{minipage}
\end{figure*}
\clearpage
\begin{figure*}
\begin{minipage}{150mm}
\centering
\includegraphics[width=15cm,height=18cm]{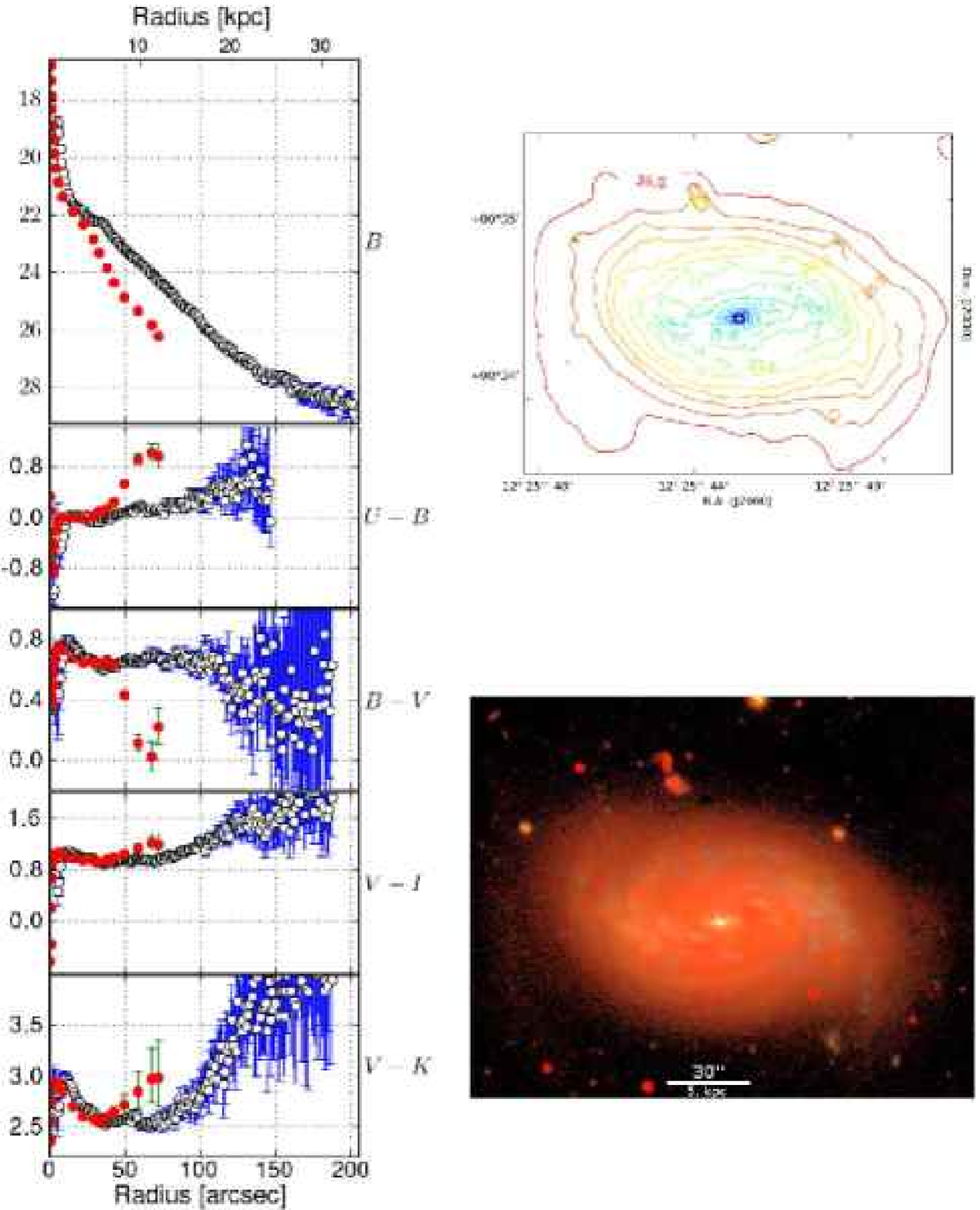}
\contcaption{\textbf{UM499}. \textit{Left panel}: Surface brightness and color radial profiles for elliptical (open circles) and isophotal (red circles) integration in the Vega photometric system. \textit{Upper right panel}: contour plot based on the $B$ band. Isophotes fainter than $22.0$, $24.5$, $25.5$ are iteratively smoothed with a boxcar median filter of sizes $5$, $15$, $25$ pixels respectively. \textit{Lower right panel}: A true color RGB composite image using the U,B,I filters. Each channel has been corrected for Galactic extinction following~\citet{1998ApJ...500..525S} and converted to the AB photometric system. The RGB composite was created by adapting the~\citet{2004PASP..116..133L} algorithm.}
\end{minipage}
\end{figure*}
\clearpage
\begin{figure*}
\begin{minipage}{150mm}
\centering
\includegraphics[width=15cm,height=18cm]{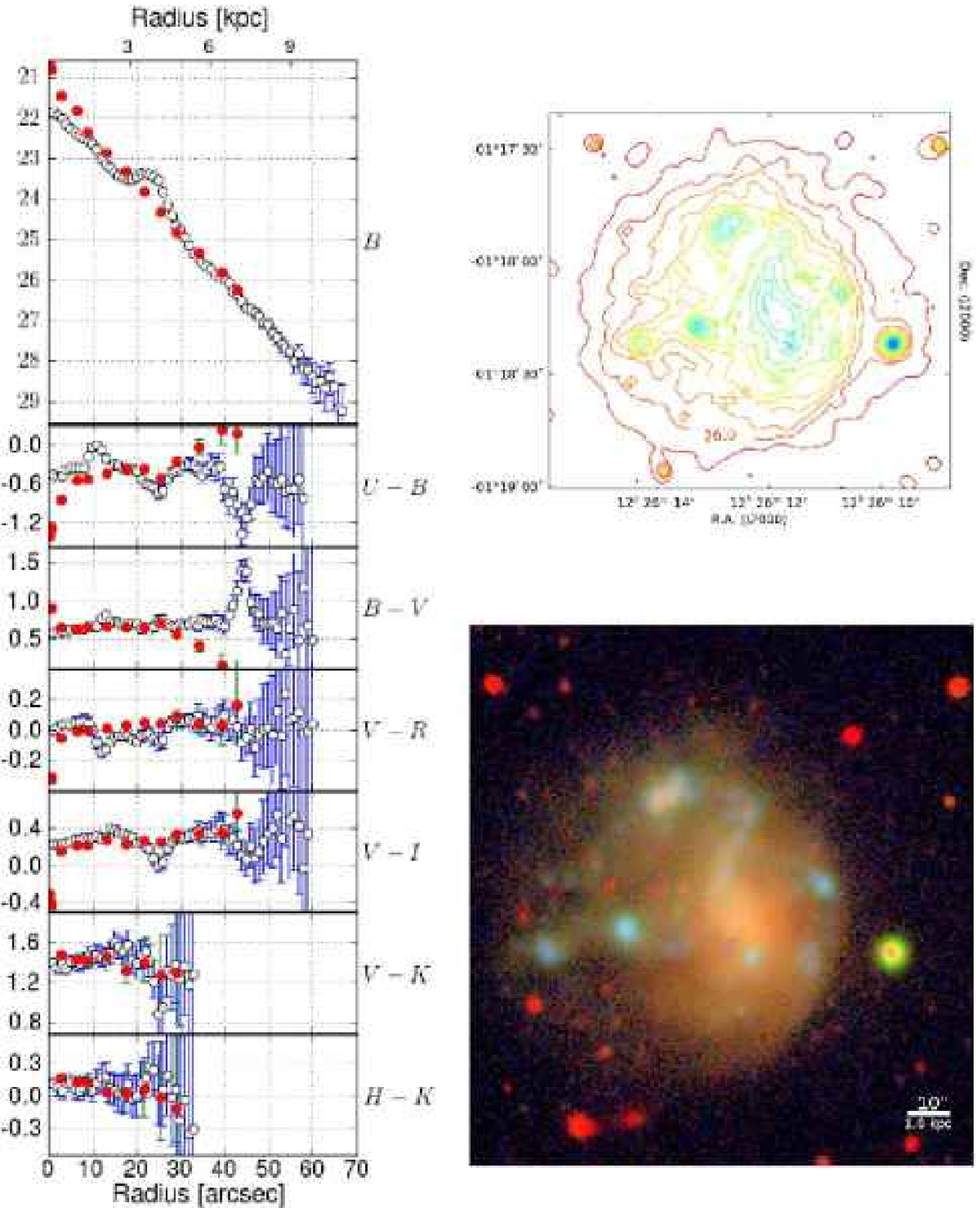}
\contcaption{\textbf{UM500}. \textit{Left panel}: Surface brightness and color radial profiles for elliptical (open circles) and isophotal (red circles) integration in the Vega photometric system. \textit{Upper right panel}: contour plot based on the $B$ band. Isophotes fainter than $23.5$, $25.5$ are iteratively smoothed with a boxcar median filter of sizes $5$, $15$ pixels respectively. \textit{Lower right panel}: A true color RGB composite image using the U,B,I filters. Each channel has been corrected for Galactic extinction following~\citet{1998ApJ...500..525S} and converted to the AB photometric system. The RGB composite was created by adapting the~\citet{2004PASP..116..133L} algorithm.}
\end{minipage}
\end{figure*}
\clearpage
\begin{figure*}
\begin{minipage}{150mm}
\centering
\includegraphics[width=15cm,height=18cm]{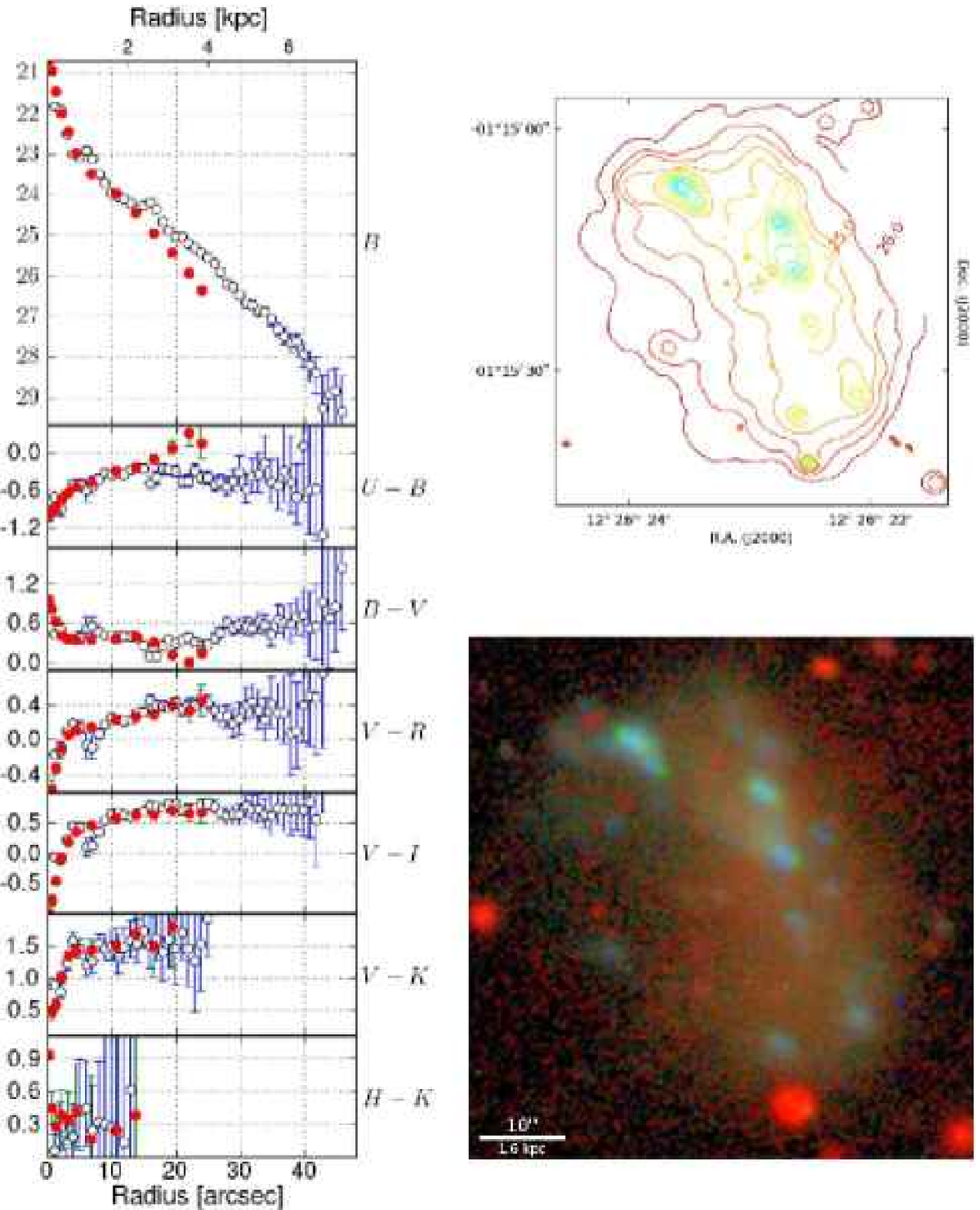}
\contcaption{\textbf{UM501}. \textit{Left panel}: Surface brightness and color radial profiles for elliptical (open circles) and isophotal (red circles) integration in the Vega photometric system. \textit{Upper right panel}: contour plot based on the $B$ band. Isophotes fainter than $23.0$, $25.5$ are iteratively smoothed with a boxcar median filter of sizes $5$, $15$ pixels respectively. \textit{Lower right panel}: A true color RGB composite image using the U,B,I filters. Each channel has been corrected for Galactic extinction following~\citet{1998ApJ...500..525S} and converted to the AB photometric system. The RGB composite was created by adapting the~\citet{2004PASP..116..133L} algorithm.}
\end{minipage}
\end{figure*}
\clearpage
\begin{figure*}
\begin{minipage}{150mm}
\centering
\includegraphics[width=15cm,height=18cm]{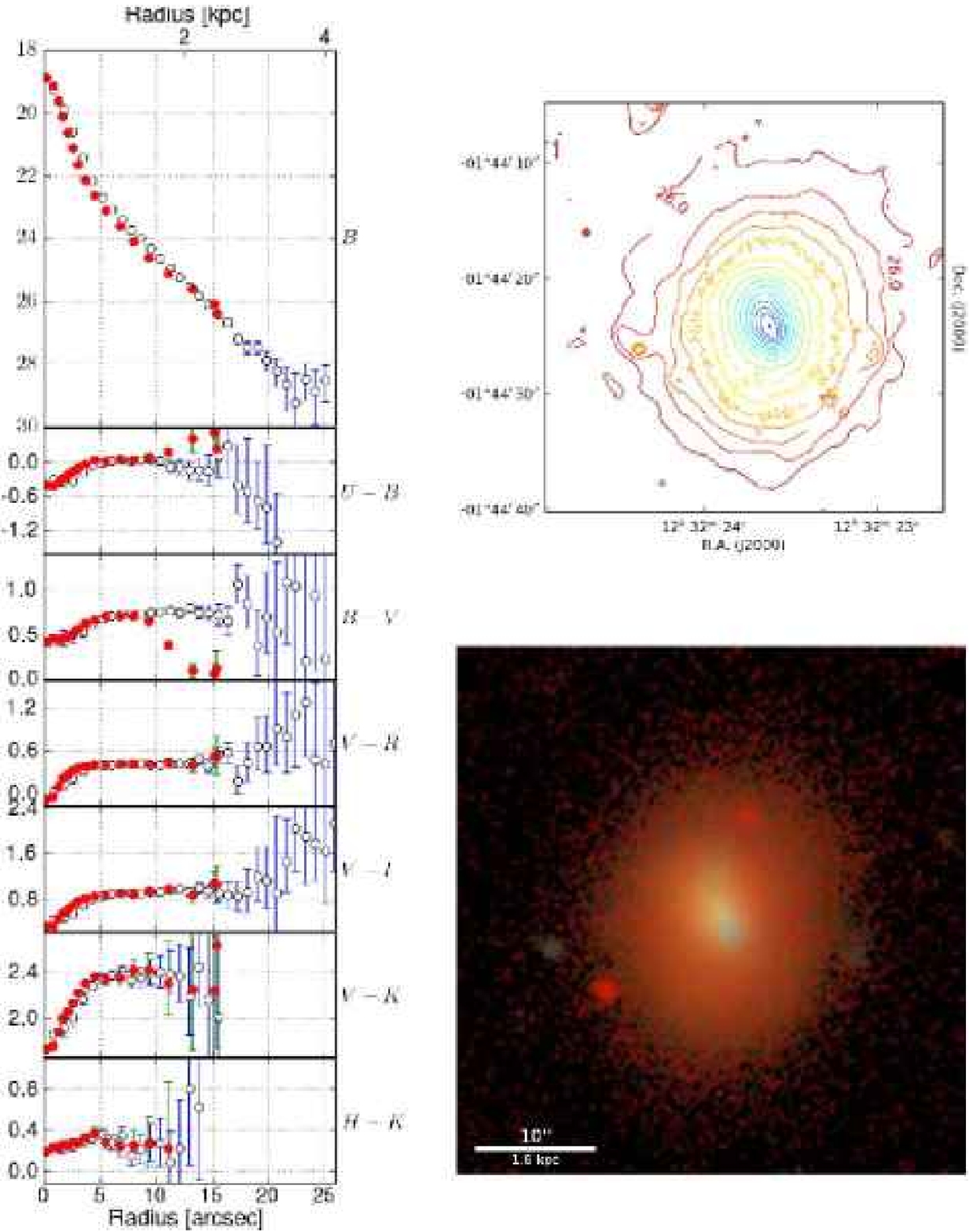}
\contcaption{\textbf{UM504}. \textit{Left panel}: Surface brightness and color radial profiles for elliptical (open circles) and isophotal (red circles) integration in the Vega photometric system. \textit{Upper right panel}: contour plot based on the $B$ band. Isophotes fainter than $24.2$, $25.5$ are iteratively smoothed with a boxcar median filter of sizes $5$, $15$ pixels respectively. \textit{Lower right panel}: A true color RGB composite image using the U,B,I filters. Each channel has been corrected for Galactic extinction following~\citet{1998ApJ...500..525S} and converted to the AB photometric system. The RGB composite was created by adapting the~\citet{2004PASP..116..133L} algorithm.}
\end{minipage}
\end{figure*}
\clearpage
\begin{figure*}
\begin{minipage}{150mm}
\centering
\includegraphics[width=15cm,height=18cm]{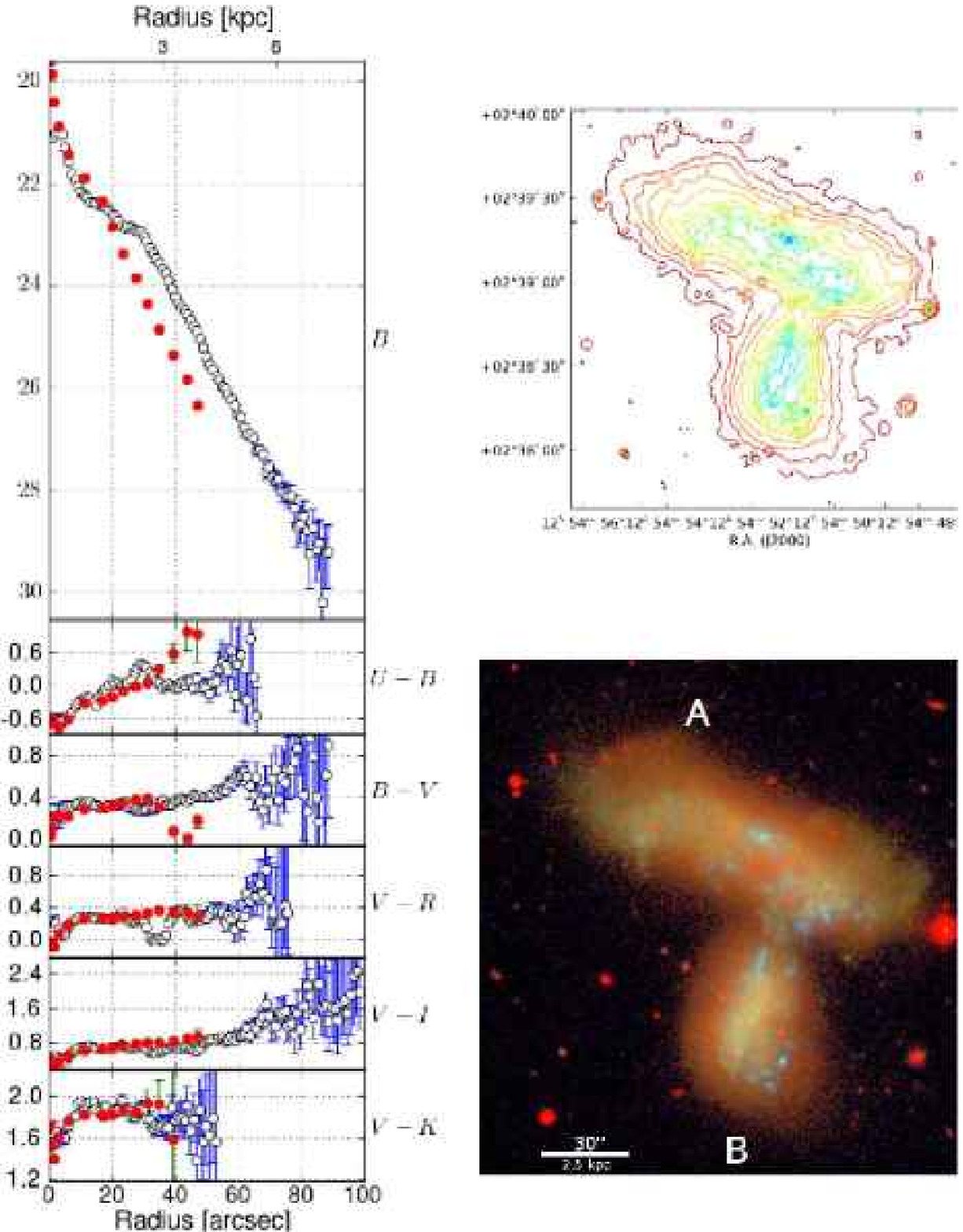}
\contcaption{\textbf{UM523A (NGC4809)}. \textit{Left panel}: Surface brightness and color radial profiles for elliptical (open circles) and isophotal (red circles) integration in the Vega photometric system. \textit{Upper right panel}: contour plot based on the $B$ band. Isophotes fainter than $23.0$, $25.5$ are iteratively smoothed with a boxcar median filter of sizes $5$, $15$ pixels respectively. \textit{Lower right panel}: A true color RGB composite image using the U,B,I filters. Each channel has been corrected for Galactic extinction following~\citet{1998ApJ...500..525S} and converted to the AB photometric system. The RGB composite was created by adapting the~\citet{2004PASP..116..133L} algorithm.}
\end{minipage}
\end{figure*}
\clearpage
\begin{figure*}
\begin{minipage}{150mm}
\centering
\includegraphics[width=15cm,height=18cm]{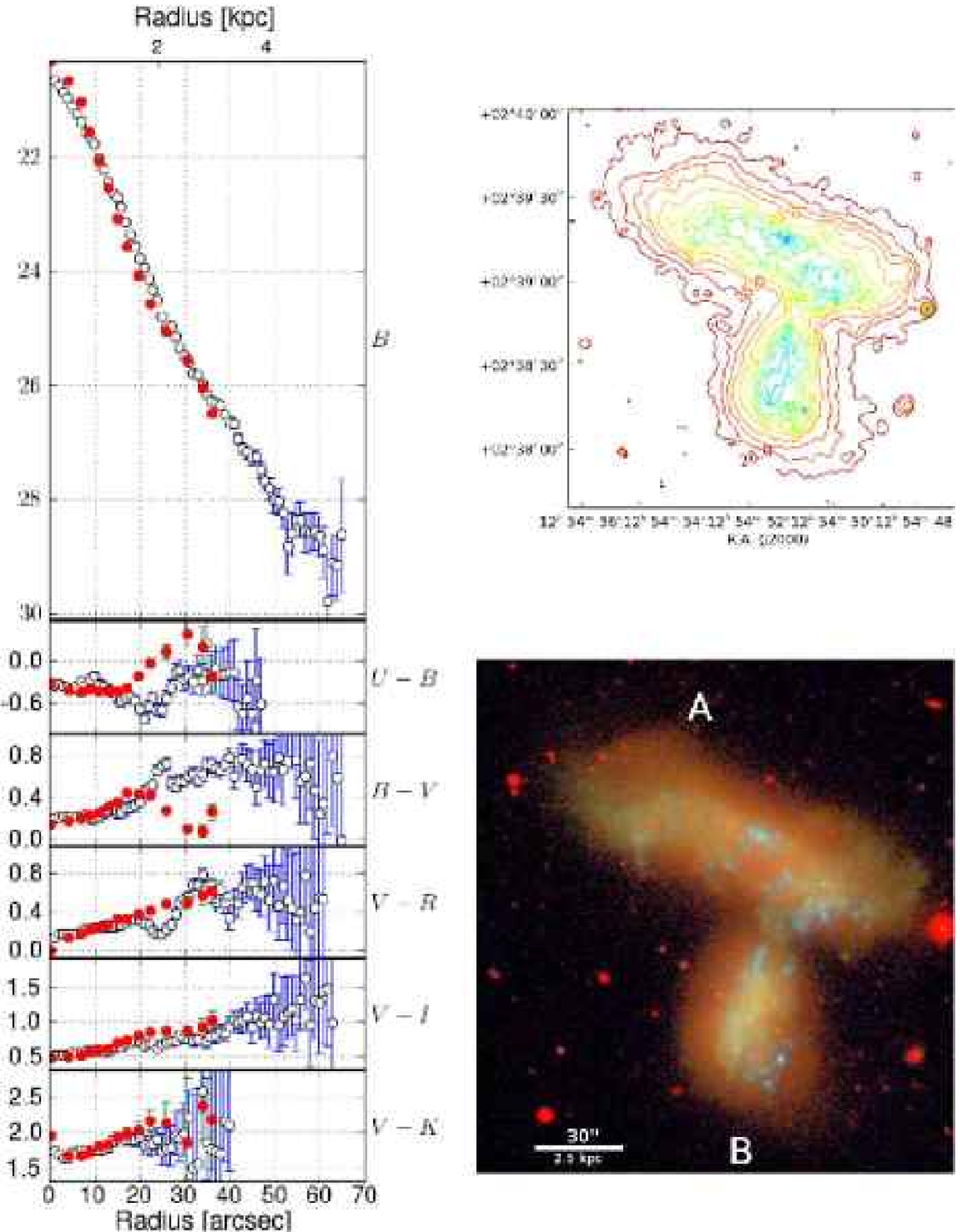}
\contcaption{\textbf{UM523B (NGC4810)}. \textit{Left panel}: Surface brightness and color radial profiles for elliptical (open circles) and isophotal (red circles) integration in the Vega photometric system. \textit{Upper right panel}: contour plot based on the $B$ band. Isophotes fainter than $23.0$, $25.5$ are iteratively smoothed with a boxcar median filter of sizes $5$, $15$ pixels respectively. \textit{Lower right panel}: A true color RGB composite image using the U,B,I filters. Each channel has been corrected for Galactic extinction following~\citet{1998ApJ...500..525S} and converted to the AB photometric system. The RGB composite was created by adapting the~\citet{2004PASP..116..133L} algorithm.}
\end{minipage}
\end{figure*}
\clearpage
\begin{figure*}
\begin{minipage}{150mm}
\centering
\includegraphics[width=15cm,height=18cm]{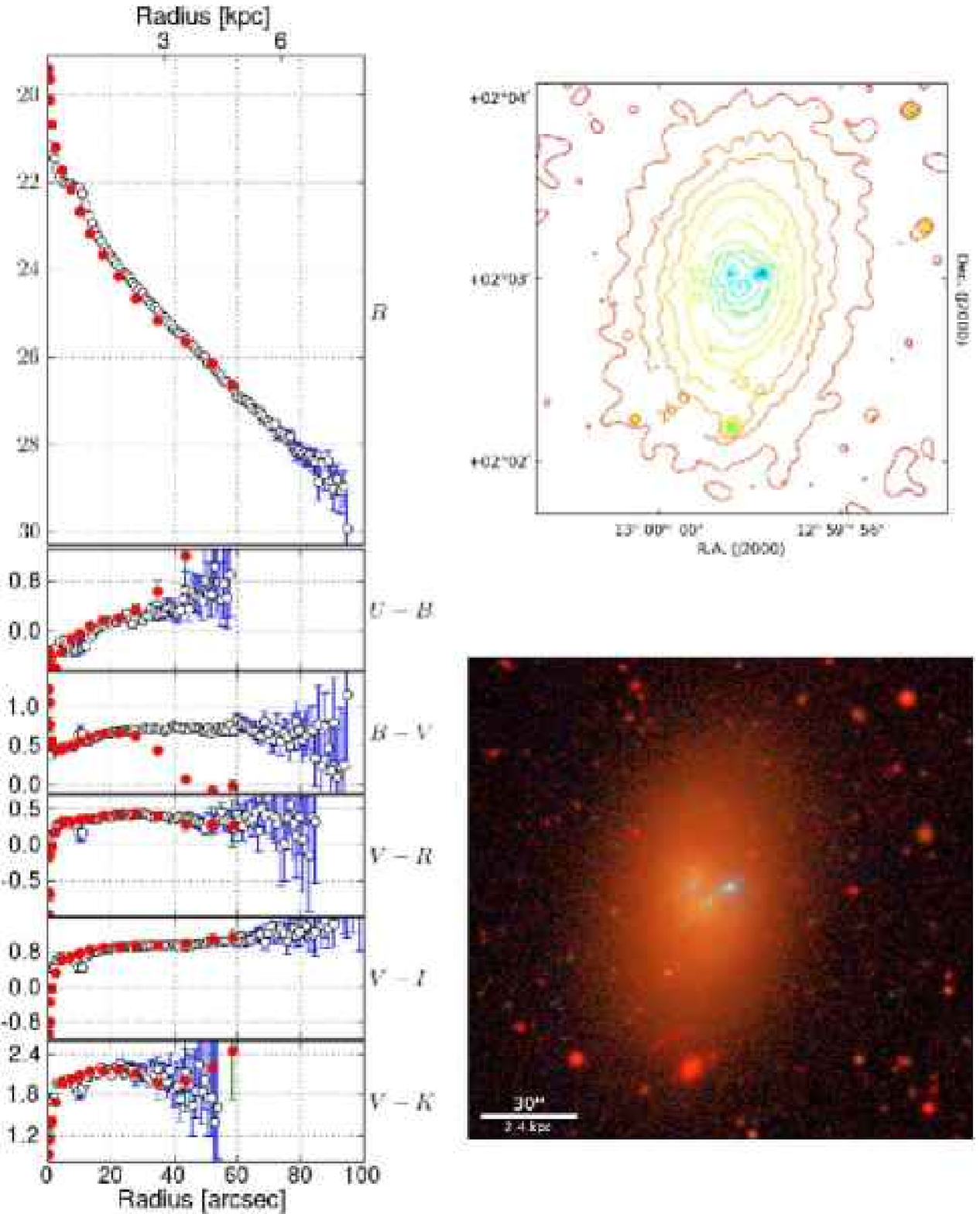}
\contcaption{\textbf{UM533}. \textit{Left panel}: Surface brightness and color radial profiles for elliptical (open circles) and isophotal (red circles) integration in the Vega photometric system. \textit{Upper right panel}: contour plot based on the $B$ band. Isophotes fainter than $23.0$, $25.5$ are iteratively smoothed with a boxcar median filter of sizes $5$, $15$ pixels respectively. \textit{Lower right panel}: A true color RGB composite image using the U,B,I filters. Each channel has been corrected for Galactic extinction following~\citet{1998ApJ...500..525S} and converted to the AB photometric system. The RGB composite was created by adapting the~\citet{2004PASP..116..133L} algorithm.}
\end{minipage}
\end{figure*}
\clearpage
\begin{figure*}
\begin{minipage}{150mm}
\centering
\includegraphics[width=15cm,height=18cm]{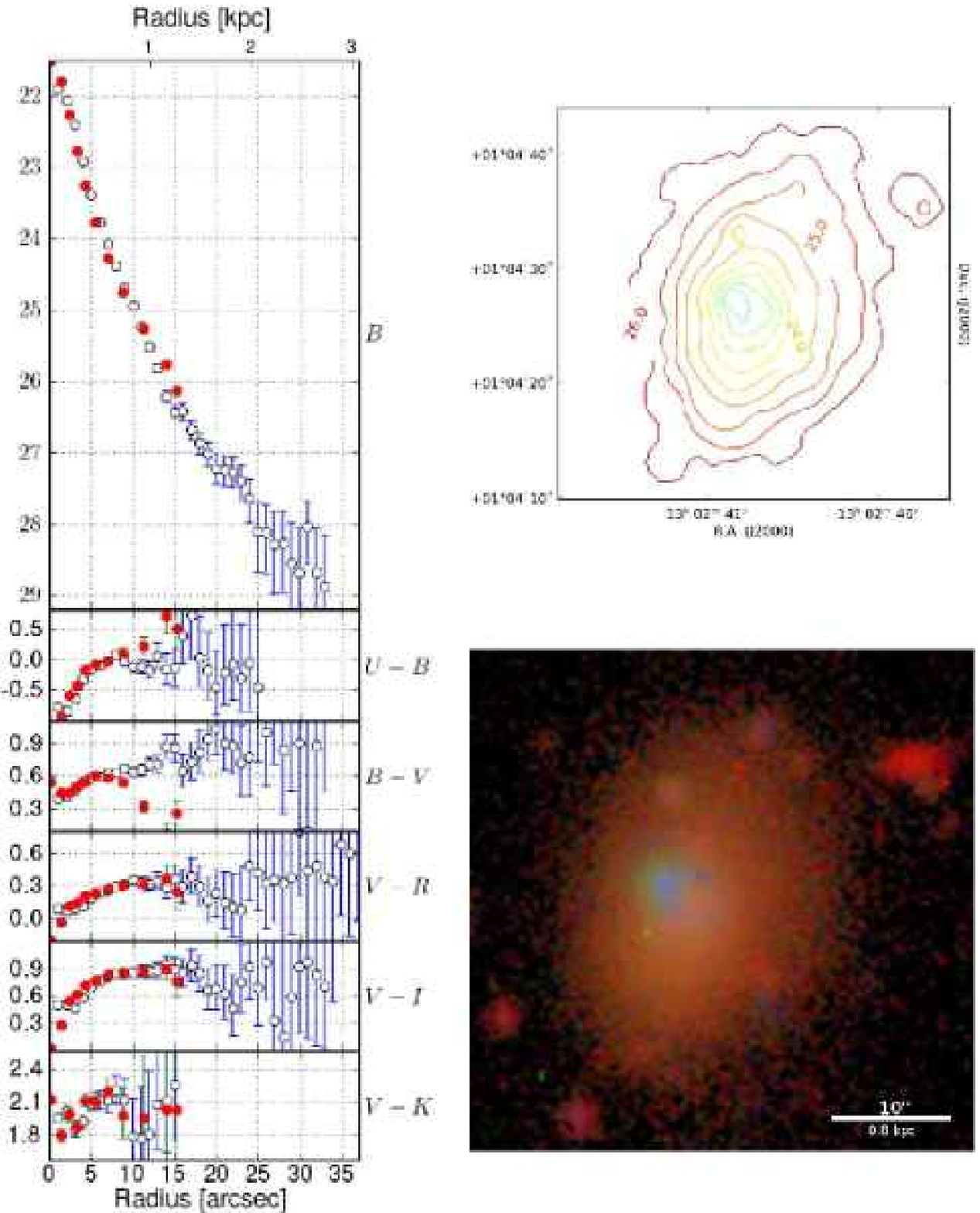}
\contcaption{\textbf{UM538}. \textit{Left panel}: Surface brightness and color radial profiles for elliptical (open circles) and isophotal (red circles) integration in the Vega photometric system. \textit{Upper right panel}: contour plot based on the $B$ band. Isophotes fainter than $23.0$, $25.5$ are iteratively smoothed with a boxcar median filter of sizes $5$, $15$ pixels respectively. \textit{Lower right panel}: A true color RGB composite image using the U,B,I filters. Each channel has been corrected for Galactic extinction following~\citet{1998ApJ...500..525S} and converted to the AB photometric system. The RGB composite was created by adapting the~\citet{2004PASP..116..133L} algorithm.}
\end{minipage}
\end{figure*}
\clearpage
\begin{figure*}
\begin{minipage}{150mm}
\centering
\includegraphics[width=15cm,height=18cm]{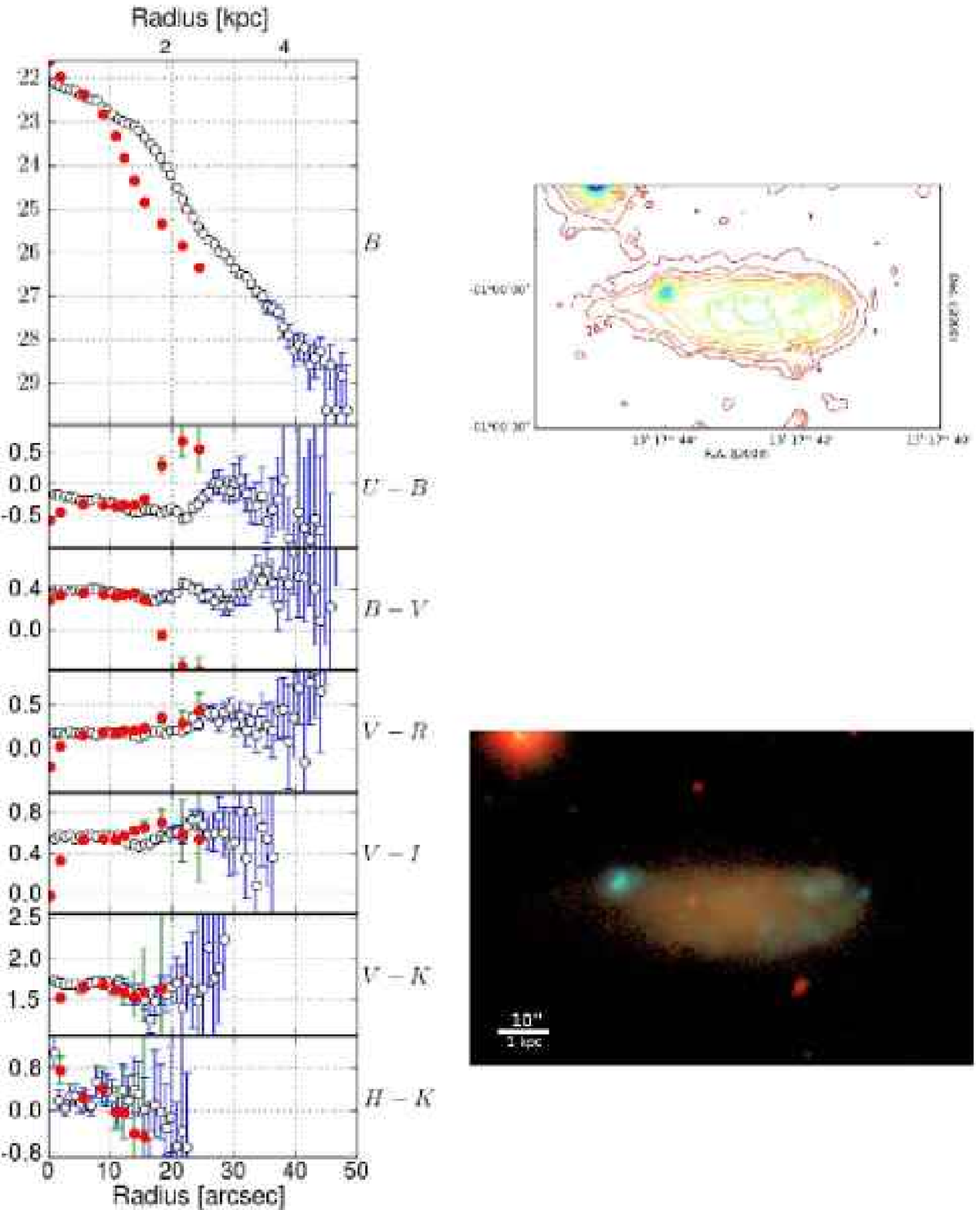}
\contcaption{\textbf{UM559}. \textit{Left panel}: Surface brightness and color radial profiles for elliptical (open circles) and isophotal (red circles) integration in the Vega photometric system. \textit{Upper right panel}: contour plot based on the $B$ band. Isophotes fainter than $23.0$, $26.5$ are iteratively smoothed with a boxcar median filter of sizes $5$, $15$ pixels respectively. \textit{Lower right panel}: A true color RGB composite image using the U,B,I filters. Each channel has been corrected for Galactic extinction following~\citet{1998ApJ...500..525S} and converted to the AB photometric system. The RGB composite was created by adapting the~\citet{2004PASP..116..133L} algorithm.}
\end{minipage}
\end{figure*}
\clearpage

\noindent However, the saturation causes charge bleeding along the columns across the galaxy (which become rows after tangential North--East projection), and to be on the cautious side we applied extensive masking, including most of the Southern part of the galaxy. The mask was heavily biased towards fainter regions beyond the central $\sim15$ arcsec, since the main star forming knot off--center to the South is bright enough to dominate the flux over the light leak from the star. Color measurements based on the \emph{I} band should nevertheless be treated with a healthy dose of suspicion, since the remaining unmasked half of the galaxy may still suffer some contamination from the saturated star at faint isophote levels. 
\subsection*{UM446}
\noindent This \emph{SS} galaxy is among the least luminous and most compact in the sample. It does not have \HI companions, even though there is an extended \HI tail to the South--East but the mass in that feature is only a few percent of the total mass of the system, and hence is not considered a legitimate candidate companion~\citep{1995ApJS...99..427T}. There is very little detected extinction based on the $H_\alpha/H_\beta$ ratio~\citep{1991A&AS...91..285T}. $B-V$ vs. $V-R$ or $H-K$ shows strong nebular emission and a very young burst of just a few Myr and intermediate metallicity ($Z\sim0.008$). $V-I$ and $V-R$ vs. $H-K$ or $V-K$ also indicate a very young age with the same $Z\sim0.008$ metallicity. $B-V$ vs. $V-I$ or $V-K$ shows the host to be very metal--poor ($Z\sim0.001$) and old ($\sim6$ Gyrs). It has a nuclear star forming knot and a second knot off-center to the North--West but regular elliptical outer isophotes, hence we classify it as an \emph{iE} BCD.
\subsection*{UM452}
\noindent This is a \emph{nE} BCD and a~\emph{DH\texttt{II}H} galaxy. It has no detected \HI companions in its vicinity~\citep{1995ApJS...99..427T}. $B-V$ vs. $V-I$, $V-K$, or $H-K$ indicates an old ($>8$ Gyrs) host with low metallicity ($Z\sim0.004$), and a burst of close--to--solar metallicity, younger than $1$ Gyr, but not by much. This is consistent with the diffuse morphology of the star forming region we observe in the RGB image, and the lack of any compact bright star forming knots. We are seeing this galaxy at the end of its most recent star formation phase, and there is largely no nebular emission contribution to any of the colors -- most colors are best fitted with the pure stellar population model.
\begin{center}
  \begin{table}
      \caption{Surface photometry parameters. The position angle ($PA^\circ$) in degrees and the ellipticity ($e$) were obtained from the $B$ band image and applied to the remaining filters. The radius where the mean surface brightness is $\mu_B\approx26.5$ mag arcsec${}^{-2}$ is the Holmberg radius, $R_{H}$, and is given in arcseconds and kpc. The absolute $B$ magnitude, $M_B$, is calculated from the area inside $R_{H}$ and has been corrected for Galactic extinction~\citep{1998ApJ...500..525S}. The $M_B$ errors are identical to the errors of the apparent $B$ magnitude in Table~\ref{totlumtbl}.}
      \protect\label{pa_holm_tbl}
      \begin{tabular}{@{}|l|r|r|r|r|r|@{}}
        \hline
        Galaxy&$PA^\circ$&$e$&$R^{''}_{H}$&$R_H^{kpc}$&$M_B$\\\hline
        UM422&$51$&$0.36$&$86.7$&$11.4$&$-18.27$\\
        UM439&$-22$&$0.37$&$34.9$&$3.4$&$-16.58$\\
        UM446&$-45$&$0.22$&$16.9$&$2.5$&$-15.67$\\
        UM452&$-64$&$0.36$&$32.9$&$4.1$&$-16.65$\\
        UM456&$35$&$0.36$&$37.8$&$5.4$&$-16.99$\\
        UM461&$90$&$0.03$&$17.9$&$1.7$&$-15.14$\\
        UM462&$67$&$0.12$&$34.9$&$3.3$&$-16.93$\\
        UM463&$27$&$0.36$&$9.0$&$1.1$&$-13.90$\\
        UM465&$20$&$0.16$&$35.4$&$3.6$&$-17.53$\\
        UM477&$-75$&$0.17$&$152.4$&$17.1$&$-19.84$\\
        UM483&$56$&$0.25$&$19.0$&$3.4$&$-16.84$\\
        UM491&$30$&$0.27$&$21.6$&$3.4$&$-16.87$\\
        UM499&$90$&$0.58$&$114.5$&$19.0$&$-19.57$\\
        UM500&$2$&$0.16$&$43.8$&$7.2$&$-17.82$\\
        UM501&$35$&$0.40$&$30.9$&$4.9$&$-16.06$\\
        UM504&$-16$&$0.24$&$16.4$&$2.6$&$-16.29$\\
        UM523A&$69$&$0.42$&$59.8$&$5.0$&$-17.09$\\
        UM523B&$-9$&$0.11$&$38.8$&$3.2$&$-16.71$\\
        UM533&$-6$&$0.25$&$55.8$&$4.5$&$-16.29$\\
        UM538&$-11$&$0.23$&$16.9$&$1.4$&$-13.66$\\
        UM559&$89$&$0.55$&$32.0$&$3.3$&$-15.60$\\
        \hline
      \end{tabular}
  \end{table}
\end{center}
\subsection*{UM456}
\noindent This Wolf--Rayet galaxy~\citep{1999A&AS..136...35S} has morphological classification~\emph{iI BCD} and a spectral classification~\emph{DH\texttt{II}H}. It is the Southern component of a system with three distinct \HI overdensities, of which only UM456 has detected emission lines, while the other two \HI clumps either have no emission lines or no optical counterpart altogether~\citep{1995ApJS...99..427T}. $B-V$ or $V-R$ vs. $V-I$ or $V-K$ shows an intermediate metallicity ($Z\sim0.008$) burst much younger than $10$ Myr, strongly affected by significant nebular emission contribution in $U-B$, $V-K$ , and $H-K$. The host is very metal--poor ($Z\sim0.001$) and older than $3$ Gyr.
\subsection*{UM461}
\noindent This Wolf--Rayet galaxy~\citep{1999A&AS..136...35S} has been classified as an \emph{SS} object. Originally thought to be tidally interacting with UM462~\citep{1995ApJS...99..427T}, this was later questioned by~\citet{1998AJ....116.1186V} who find a crossing time of $\sim700$ Myr, which is significantly longer than the age of the starburst, expected to be less than 100 Myr.~\citet{1999A&AS..138..213D} observe a double exponential profile, with the outer disk visible in the surface brightness profile beyond $\mu_B=26.75$ mag arcsec${}^{-2}$. However, our $B$ data show no evidence of a double exponential structure around or beyond this isophotal level. In fact, the surface brightness profile is very well fitted with a single disk down to $\mu_B\sim28$ mag arcsec${}^{-2}$. Since our data are deeper and have much smaller errors at fainter isophotes, the double disk structure in the surface brightness profile of~\citet{1999A&AS..138..213D} must be due to systematic sky effects, and not to the presence of a second component. Judging by $B-V$ vs. $V-R$, $V-I$ or $V-K$ the burst is only a few ($<5$) Myr old and has low metallicity ($Z\sim0.004$). This seems to be the youngest and simultaneously the most metal--poor burst in the sample, with all color--color diagrams clearly indicating a similar very young age at the same metallicity. There is strong nebular emission contribution dominating the total and burst colors in all colors, but this becomes insignificant in the two outskirt regions, which are well--fitted by a very metal--poor ($Z\sim0.001$) stellar population older than $3$ Gyrs. All integrated colors indicate that nebular emission is sufficient for a good fit with the evolutionary models, no dust is necessary. This is consistent with the extremely low extinction values measured through the Balmer decrement $H_\alpha/H_\beta$ by~\citet{1991A&AS...91..285T}. The off--center double--knot structure of the star forming regions and the irregularities in the outer isophotes (see contour plot) lead us to classify this galaxy as an \emph{iI} BCD. 
\subsection*{UM462}
\noindent This Wolf--Rayet galaxy~\citep{1999A&AS..136...35S} is classified as \emph{iE} BCD by~\citet{2001ApJS..133..321C} and as a~\emph{DH\texttt{II}H}. This galaxy is referred to as rather compact in the literature~\citep{1991A&A...241..358C,2001ApJS..133..321C,2001ApJS..136..393C}, however, we detect a very low surface brightness structure extended to the North and South of the main star forming nucleus. This structure is not only visible in the surface brightness profile beyond $\mu_B=26$ mag arcsec${}^{-2}$, but also in the contour plot. This component is previously unobserved both in $B$ band surface brightness profiles and contour plots~\citep{2001ApJS..133..321C,2001ApJS..136..393C}, however, previous observations have exposure times a factor of four shorter than ours. Together with the extended low surface brightness component this galaxy bears a striking morphological similarity to NGC 5128 (Centaurus A). In Figure~\ref{um462deepcontours} in the Appendix we show a composite figure of the RGB image superposed on these spectacular faint features. $B-V$ vs. $V-I$ or $V-K$ diagrams indicate that the age of the component previously thought to be the only host (in the region $24\lesssim\mu_B\lesssim26$ mag arcsec${}^{-2}$) is between $3$ and $4$ Gyrs and has very low metallicity ($Z\sim0.001$). Due to the large errorbars we cannot identify a specific age or metallicity of the new host component detected at $26\lesssim\mu_B\lesssim28$ in any color--color diagram, except to say that it looks to be older than $5$ Gyrs in $B-V$ vs. $V-R$ or $V-I$ but is consistent with all metallicities within the errorbars. $V-R$ vs. $H-K$ or $V-I$ indicate a very young burst of $<5$ Myr and low metallicity ($Z\sim0.004$) with significant nebular emission contribution in all filters. The presence of significant quantities of dust is suggested by the extinction effect in and around the burst regions, judging by their $U-B$ and $B-V$ colors. 
\begin{center}
  \begin{table*}
    \begin{minipage}{170mm}
      \caption{Integrated surface photometry for the sample. The integration is carried out down to the Holmberg radius $R^{''}_{H}$, which is defined from $\mu_B$ for each target and then applied to the remaining filters. All values have been corrected for Galactic extinction~\citep{1998ApJ...500..525S}.}
      \protect\label{totlumtbl}
      \begin{tabular}{@{}|l|r|r|r|r|r|r|r|r|@{}}
        \hline
        Galaxy&B&U$-$B&B$-$V&V$-$R&V$-$I&V$-$K&H$-$K\\\hline
        UM422&$13.91 \pm 0.03$&$-0.34 \pm 0.05$&$0.45 \pm 0.04$&$0.30 \pm 0.06$&$0.86 \pm 0.07$&$1.61 \pm 0.09$&$-0.30 \pm 0.10$\\
        UM439&$14.94 \pm 0.03$&$-0.48 \pm 0.05$&$0.39 \pm 0.04$&$0.28 \pm 0.03$&$0.62 \pm 0.05$&$1.77 \pm 0.06$&\\
        UM446&$16.71 \pm 0.04$&&$0.58 \pm 0.05$&$0.45 \pm 0.03$&$0.71 \pm 0.08$&$2.32 \pm 0.26$&$0.42 \pm 0.26$\\
        UM452&$15.38 \pm 0.01$&$-0.20 \pm 0.12$&$0.71 \pm 0.02$&$0.43 \pm 0.05$&$0.96 \pm 0.04$&$2.45 \pm 0.04$&$0.26 \pm 0.05$\\
        UM456&$15.37 \pm 0.01$&$-0.41 \pm 0.03$&$0.41 \pm 0.03$&$0.24 \pm 0.04$&$0.52 \pm 0.04$&$1.83 \pm 0.04$&$0.68 \pm 0.09$\\
        UM461&$16.29 \pm 0.02$&$-0.56 \pm 0.04$&$0.66 \pm 0.04$&$-0.06 \pm 0.03$&$0.10 \pm 0.04$&$1.31 \pm 0.07$&$0.53 \pm 0.14$\\
        UM462&$14.53 \pm 0.03$&$-0.66 \pm 0.15$&$0.36 \pm 0.05$&$0.08 \pm 0.25$&$0.07 \pm 0.19$&$1.77 \pm 0.07$&$0.39 \pm 0.07$\\
        UM463&$18.02 \pm 0.02$&$-0.84 \pm 0.05$&$0.74 \pm 0.04$&&$0.11 \pm 0.05$&$2.31 \pm 0.50$&$0.13 \pm 0.50$\\
        UM465&$14.05 \pm 0.01$&$-0.30 \pm 0.02$&$0.60 \pm 0.03$&&&$2.50 \pm 0.05$&\\
        UM477&$11.99 \pm 0.02$&$0.16 \pm 0.03$&$0.57 \pm 0.02$&$0.47 \pm 0.05$&$0.95 \pm 0.02$&$2.46 \pm 0.06$&\\
        UM483&$16.02 \pm 0.02$&$-0.30 \pm 0.07$&$0.45 \pm 0.06$&$0.18 \pm 0.07$&$0.56 \pm 0.06$&$2.05 \pm 0.07$&\\
        UM491&$15.69 \pm 0.01$&$-0.40 \pm 0.04$&$0.44 \pm 0.03$&$0.28 \pm 0.06$&$0.62 \pm 0.03$&$2.24 \pm 0.16$&\\
        UM499&$13.11 \pm 0.04$&$-0.08 \pm 0.04$&$0.64 \pm 0.04$&&$0.95 \pm 0.04$&$2.69 \pm 0.08$&\\
        UM500&$14.82 \pm 0.01$&$-0.42 \pm 0.02$&$0.68 \pm 0.03$&$-0.02 \pm 0.05$&$0.26 \pm 0.03$&$1.38 \pm 0.05$&$-0.01 \pm 0.13$\\
        UM501&$16.52 \pm 0.04$&$-0.45 \pm 0.04$&$0.38 \pm 0.05$&$0.22 \pm 0.04$&$0.55 \pm 0.05$&$1.43 \pm 0.16$&$0.37 \pm 0.17$\\
        UM504&$16.30 \pm 0.04$&$-0.21 \pm 0.05$&$0.56 \pm 0.05$&$0.28 \pm 0.04$&$0.69 \pm 0.10$&$2.14 \pm 0.05$&$0.28 \pm 0.06$\\
        UM523A&$14.07 \pm 0.06$&$-0.08 \pm 0.07$&$0.33 \pm 0.07$&$0.22 \pm 0.05$&$0.64 \pm 0.04$&$1.80 \pm 0.07$&\\
        UM523B&$14.46 \pm 0.06$&$-0.40 \pm 0.07$&$0.30 \pm 0.07$&$0.26 \pm 0.05$&$0.61 \pm 0.04$&$1.80 \pm 0.07$&\\
        UM533&$14.83 \pm 0.03$&$0.02 \pm 0.04$&$0.63 \pm 0.04$&$0.35 \pm 0.03$&$0.82 \pm 0.02$&$2.04 \pm 0.10$&\\
        UM538&$17.50 \pm 0.04$&$-0.34 \pm 0.06$&$0.58 \pm 0.05$&$0.22 \pm 0.07$&$0.72 \pm 0.04$&$2.01 \pm 0.06$&\\
        UM559&$16.07 \pm 0.02$&$-0.33 \pm 0.03$&$0.36 \pm 0.05$&$0.20 \pm 0.06$&$0.56 \pm 0.06$&$1.63 \pm 0.06$&$-0.01 \pm 0.07$\\
        \hline
      \end{tabular}
    \end{minipage}
  \end{table*}
\end{center}
\begin{center}
  \begin{table*}
      \caption{Oxygen--based metallicities, $H_\beta$ equivalent widths in \AA, and morphological (M class) and spectral (S class) classification compiled from the literature, or obtained in this work (Ref.=$0$). Metallicities marked with ${}^\star$ indicate that the value is estimated from Eq.~$5$ in~\citet{1989ApJS...70..479S} since tabulated values were unavailable. All spectral classifications are from~\citet{1989ApJS...70..479S}. Note that for UM422 and UM559 we see a clear inconsistency between the assigned class, $SS$, which implies the target is small and compact, and the extended morphology we observe in our contour and RGB plots for these targets. This is marked with a ``?''. }
      \protect\label{morph_tab}
      \begin{tabular}{@{}|l|l|r|r|r|r|r|r|@{}}
        \hline
        Galaxy&$12+\log{O/H}$&Ref.&$EW(H_\beta)$&Ref.&M class&Ref.&S class\\\hline
        UM422&$8.0$&1&$344$&4&$$&&$SS?$\\
        UM439&$8.0$&2&$160$&4&$iE$&7&$DH\texttt{II}H$\\
        UM446&$8.3^\star$&3&$38$&4&$iE$&0&$SS$ \\
        UM452&$8.4^\star$&3&$13.5$&6&$nE$&7&$DH\texttt{II}H$ \\
        UM456&$7.9$&1&$40$&4&$iI$&7&$DH\texttt{II}H$ \\
        UM461&$7.7$&2&$342$&4&$iI$&0&$SS$ \\
        UM462&$7.9$&2&$124$&4&$iE$&8&$DH\texttt{II}H$ \\
        UM463&$7.7$&1&$119$&4&$iE$&0&$SS$ \\
        UM465&$8.4^\star$&3&$10$&4&$nE$&0&$DANS$ \\
        UM477&$8.7$&4&$22$&4&$SB$&&$SBN$ \\
        UM483&$8.3$&1&$26$&4&$i0$&7&$DH\texttt{II}H$ \\
        UM491&$7.9$&1&$11$&4&$nE$&7&$DH\texttt{II}H$ \\
        UM499&$8.6$&4&$32$&4&$SB$&&$SBN$ \\
        UM500&$8.1$&1&$133$&4&$iI$&0&$SS$ \\
        UM501&$8.3^\star$&3&$123$&4&$iI,i0?$&0&$MI$ \\
        UM504&$8.4^\star$&3&$18$&4&$nE$&0&$DH\texttt{II}H$ \\
        UM523A&$8.1$&4&$30$&4&$iE$&0& \\
        UM523B&$8.1$&4&$30$&4&$iE$&0& \\
        UM533&$8.3^\star$&3&$101$&4&$iE$&7&$MI$ \\
        UM538&$7.8$&1&$77$&4&$iE$&0&$SS$ \\
        UM559&$7.7$&5&$325$&4&$iE/nE$&5&$SS?$ \\
        \hline
      \end{tabular}
      \medskip
      ~\\
      $1$ --~\citet{1994ApJ...420..576M}, $2$ --~\citet{2006ApJ...645.1076N},$3$ --~\citet{1989ApJS...70..479S}, $4$ --~\citet{1991A&AS...91..285T}, $5$ --~\citet{2006A&A...457...45P}, $6$ --~\citet{1989ApJS...70..447S}, $7$ --~\citet{2003ApJS..147...29G}, $8$ --~\citet{2001ApJS..133..321C}
  \end{table*}
\end{center}
\begin{center}
  \begin{table*}
    \begin{minipage}{140mm}
      \caption{Total colors for radial ranges corresponding to $\mu_B\lesssim24$, $24\lesssim\mu_B<26$, and $26\lesssim\mu_B\lesssim28$ mag arcsec${}^{-2}$. All values have been corrected for Galactic extinction~\citep{1998ApJ...500..525S}. The errors include $\sigma_{sky}$, $\sigma_{sdom}$ and $\sigma_{zp}$ added in quadrature.}
      \protect\label{totclrtbl}
      \tiny
      \begin{tabular}{@{}|l|r|r|r|r|r|r|r|r|@{}}
        \hline
        Galaxy&$\mu_B$&U$-$B&B$-$V&V$-$R&V$-$I&V$-$K&H$-$K\\\hline
        UM422&$\star$\textrm{--$24$}&$-0.33 \pm 0.05$&$0.48 \pm 0.04$&$0.31 \pm 0.07$&$0.88 \pm 0.07$&$1.76 \pm 0.30$&$-0.15 \pm 0.32$\\
        &$24$--$26$&$-0.36 \pm 0.12$&$0.49 \pm 0.07$&$0.18 \pm 0.09$&$0.79 \pm 0.15$&$1.29 \pm 2.89$&$-0.83 \pm 2.95$\\
        &$26$--$28$&$-0.70 \pm 2.54$&$1.01 \pm 1.27$&$0.20 \pm 1.07$&$1.89 \pm 1.09$&&\\\hline
        UM439&$\star$\textrm{--$24$}&$-0.51 \pm 0.06$&$0.37 \pm 0.06$&$0.14 \pm 0.05$&$0.41 \pm 0.05$&$1.78 \pm 0.07$&\\
        &$24$--$26$&$-0.21 \pm 0.07$&$0.47 \pm 0.05$&$0.70 \pm 0.05$&$1.20 \pm 0.05$&$1.89 \pm 0.26$&\\
        &$26$--$28$&$-0.29 \pm 0.40$&$0.53 \pm 0.23$&$0.91 \pm 0.27$&$1.43 \pm 0.22$&&\\\hline
        UM446&$\star$\textrm{--$24$}&&$0.55 \pm 0.07$&$0.40 \pm 0.06$&$0.63 \pm 0.08$&$2.14 \pm 0.08$&$0.38 \pm 0.10$\\
        &$24$--$26$&&$0.69 \pm 0.06$&$0.57 \pm 0.05$&$0.91 \pm 0.07$&$2.77 \pm 0.47$&$0.51 \pm 0.78$\\
        &$26$--$28$&&$0.60 \pm 0.15$&$0.73 \pm 0.16$&$1.08 \pm 0.22$&$3.71 \pm 1.17$&$0.60 \pm 2.03$\\\hline
        UM452&$\star$\textrm{--$24$}&$-0.23 \pm 0.05$&$0.69 \pm 0.04$&$0.41 \pm 0.05$&$0.93 \pm 0.05$&$2.41 \pm 0.05$&$0.29 \pm 0.08$\\
        &$24$--$26$&$0.00 \pm 0.08$&$0.83 \pm 0.06$&$0.51 \pm 0.06$&$1.06 \pm 0.06$&$2.60 \pm 0.28$&$0.21 \pm 0.52$\\
        &$26$--$28$&$-0.33 \pm 0.33$&$0.92 \pm 0.28$&$0.58 \pm 0.22$&$1.10 \pm 0.23$&$2.88 \pm 1.19$&$-0.13 \pm 1.77$\\\hline
        UM456&$\star$\textrm{--$24$}&$-0.53 \pm 0.04$&$0.36 \pm 0.04$&$0.17 \pm 0.04$&$0.40 \pm 0.05$&$1.77 \pm 0.15$&$0.34 \pm 0.22$\\
        &$24$--$26$&$-0.04 \pm 0.05$&$0.47 \pm 0.05$&$0.36 \pm 0.05$&$0.71 \pm 0.06$&$1.94 \pm 0.78$&$1.06 \pm 1.76$\\
        &$26$--$28$&$-0.10 \pm 0.34$&$0.49 \pm 0.22$&$0.48 \pm 0.26$&$0.79 \pm 0.30$&&\\\hline
        UM461&$\star$\textrm{--$24$}&$-0.58 \pm 0.09$&$0.68 \pm 0.10$&$-0.10 \pm 0.10$&$0.01 \pm 0.09$&$1.15 \pm 0.14$&$0.45 \pm 0.23$\\
        &$24$--$26$&$-0.37 \pm 0.07$&$0.54 \pm 0.05$&$0.23 \pm 0.07$&$0.67 \pm 0.07$&$2.08 \pm 0.52$&$0.77 \pm 1.31$\\
        &$26$--$28$&$-0.45 \pm 0.27$&$0.47 \pm 0.14$&$0.38 \pm 0.28$&$0.82 \pm 0.24$&&\\\hline
        UM462&$\star$\textrm{--$24$}&$-0.69 \pm 0.15$&$0.34 \pm 0.06$&$0.04 \pm 0.25$&$-0.01 \pm 0.20$&$1.72 \pm 0.09$&$0.45 \pm 0.11$\\
        &$24$--$26$&$-0.35 \pm 0.15$&$0.47 \pm 0.06$&$0.35 \pm 0.25$&$0.50 \pm 0.20$&$2.13 \pm 0.50$&$0.18 \pm 0.68$\\
        &$26$--$28$&$-0.45 \pm 0.33$&$0.75 \pm 0.20$&$0.27 \pm 0.33$&$0.53 \pm 0.33$&&\\\hline
        UM463&$\star$\textrm{--$24$}&$-0.90 \pm 0.11$&$0.74 \pm 0.11$&&$-0.01 \pm 0.11$&$2.26 \pm 0.50$&$0.15 \pm 0.50$\\
        &$24$--$26$&$-0.37 \pm 0.10$&$0.78 \pm 0.09$&&$0.63 \pm 0.09$&$2.31 \pm 1.16$&$-0.26 \pm 1.00$\\
        &$26$--$28$&$-0.76 \pm 0.44$&$0.98 \pm 0.40$&&$0.57 \pm 0.35$&&\\\hline
        UM465&$\star$\textrm{--$24$}&$-0.32 \pm 0.06$&$0.59 \pm 0.04$&&&$2.48 \pm 0.06$&\\
        &$24$--$26$&$-0.06 \pm 0.06$&$0.74 \pm 0.03$&&&$2.61 \pm 0.18$&\\
        &$26$--$28$&$-0.24 \pm 0.30$&$0.58 \pm 0.13$&&&$3.10 \pm 0.79$&\\\hline
        UM477&$\star$\textrm{--$24$}&$0.14 \pm 0.03$&$0.62 \pm 0.02$&$0.50 \pm 0.05$&$0.99 \pm 0.02$&$2.73 \pm 0.06$&\\
        &$24$--$26$&$0.56 \pm 0.11$&$0.52 \pm 0.04$&$0.43 \pm 0.07$&$0.81 \pm 0.05$&&\\
        &$26$--$28$&&$0.50 \pm 0.16$&$0.47 \pm 0.19$&$0.65 \pm 0.25$&&\\\hline
        UM483&$\star$\textrm{--$24$}&$-0.31 \pm 0.05$&$0.44 \pm 0.07$&$0.18 \pm 0.08$&$0.55 \pm 0.07$&$2.07 \pm 0.09$&\\
        &$24$--$26$&$-0.15 \pm 0.11$&$0.55 \pm 0.07$&$0.24 \pm 0.08$&$0.67 \pm 0.09$&$1.94 \pm 0.31$&\\
        &$26$--$28$&$-0.13 \pm 0.56$&$0.80 \pm 0.19$&$0.25 \pm 0.18$&$0.54 \pm 0.32$&&\\\hline
        UM491&$\star$\textrm{--$24$}&$-0.43 \pm 0.06$&$0.41 \pm 0.05$&$0.26 \pm 0.05$&$0.58 \pm 0.05$&$2.20 \pm 0.14$&\\
        &$24$--$26$&$-0.13 \pm 0.06$&$0.65 \pm 0.04$&$0.50 \pm 0.04$&$0.88 \pm 0.05$&$2.53 \pm 0.22$&\\
        &$26$--$28$&$-0.02 \pm 0.26$&$0.63 \pm 0.10$&$0.76 \pm 0.11$&$1.04 \pm 0.14$&$2.81 \pm 0.70$&\\\hline
        UM499&$\star$\textrm{--$24$}&$-0.13 \pm 0.06$&$0.63 \pm 0.05$&&$0.93 \pm 0.04$&$2.70 \pm 0.09$&\\
        &$24$--$26$&$0.18 \pm 0.05$&$0.68 \pm 0.05$&&$1.02 \pm 0.05$&$2.67 \pm 0.12$&\\
        &$26$--$28$&$0.47 \pm 0.25$&$0.62 \pm 0.18$&&$1.25 \pm 0.20$&$3.15 \pm 0.40$&\\\hline
        UM500&$\star$\textrm{--$24$}&$-0.45 \pm 0.02$&$0.62 \pm 0.04$&$0.05 \pm 0.05$&$0.25 \pm 0.04$&$1.38 \pm 0.14$&$0.07 \pm 0.20$\\
        &$24$--$26$&$-0.41 \pm 0.09$&$0.68 \pm 0.06$&$0.09 \pm 0.08$&$0.35 \pm 0.08$&$1.25 \pm 0.93$&$-0.41 \pm 1.05$\\
        &$26$--$28$&$-0.47 \pm 0.38$&$0.65 \pm 0.27$&$0.09 \pm 0.32$&$0.42 \pm 0.36$&&\\\hline
        UM501&$\star$\textrm{--$24$}&$-0.55 \pm 0.06$&$0.45 \pm 0.06$&$0.06 \pm 0.05$&$0.30 \pm 0.06$&$1.33 \pm 0.15$&$0.30 \pm 0.41$\\
        &$24$--$26$&$-0.29 \pm 0.06$&$0.40 \pm 0.05$&$0.25 \pm 0.05$&$0.61 \pm 0.06$&$1.47 \pm 0.51$&$0.37 \pm 1.67$\\
        &$26$--$28$&$-0.33 \pm 0.29$&$0.51 \pm 0.13$&$0.21 \pm 0.24$&$0.62 \pm 0.23$&$1.87 \pm 2.10$&\\\hline
        UM504&$\star$\textrm{--$24$}&$-0.23 \pm 0.09$&$0.53 \pm 0.09$&$0.25 \pm 0.07$&$0.64 \pm 0.10$&$2.10 \pm 0.08$&$0.27 \pm 0.08$\\
        &$24$--$26$&$-0.06 \pm 0.10$&$0.74 \pm 0.06$&$0.41 \pm 0.05$&$0.94 \pm 0.09$&$2.37 \pm 0.20$&$0.30 \pm 0.35$\\
        &$26$--$28$&$-0.49 \pm 0.57$&$0.78 \pm 0.24$&$0.53 \pm 0.21$&$0.87 \pm 0.33$&&\\\hline
        UM523A&$\star$\textrm{--$24$}&$-0.31 \pm 0.07$&$0.34 \pm 0.07$&$0.25 \pm 0.05$&$0.63 \pm 0.04$&$1.82 \pm 0.07$&\\
        &$24$--$26$&$0.03 \pm 0.11$&$0.38 \pm 0.07$&$0.29 \pm 0.07$&$0.76 \pm 0.06$&$1.72 \pm 0.30$&\\
        &$26$--$28$&$0.44 \pm 0.64$&$0.52 \pm 0.11$&$0.32 \pm 0.23$&$0.96 \pm 0.18$&&\\\hline
        UM523B&$\star$\textrm{--$24$}&$-0.40 \pm 0.07$&$0.25 \pm 0.07$&$0.22 \pm 0.05$&$0.59 \pm 0.04$&$1.79 \pm 0.07$&\\
        &$24$--$26$&$-0.28 \pm 0.09$&$0.44 \pm 0.07$&$0.37 \pm 0.05$&$0.78 \pm 0.05$&$2.04 \pm 0.21$&\\
        &$26$--$28$&$-0.10 \pm 0.33$&$0.58 \pm 0.09$&$0.48 \pm 0.12$&$0.94 \pm 0.13$&$1.85 \pm 1.09$&\\\hline
        UM533&$\star$\textrm{--$24$}&$-0.12 \pm 0.05$&$0.58 \pm 0.04$&$0.32 \pm 0.03$&$0.73 \pm 0.03$&$2.03 \pm 0.10$&\\
        &$24$--$26$&$0.29 \pm 0.10$&$0.71 \pm 0.04$&$0.42 \pm 0.04$&$0.96 \pm 0.03$&$2.06 \pm 0.18$&\\
        &$26$--$28$&$1.22 \pm 1.52$&$0.69 \pm 0.08$&$0.31 \pm 0.22$&$1.08 \pm 0.14$&$1.92 \pm 1.22$&\\\hline
        UM538&$\star$\textrm{--$24$}&$-0.52 \pm 0.07$&$0.51 \pm 0.06$&$0.13 \pm 0.08$&$0.59 \pm 0.05$&$1.99 \pm 0.08$&\\
        &$24$--$26$&$-0.07 \pm 0.11$&$0.67 \pm 0.07$&$0.31 \pm 0.08$&$0.88 \pm 0.06$&$1.97 \pm 0.31$&\\
        &$26$--$28$&$-0.07 \pm 0.49$&$0.74 \pm 0.20$&$0.35 \pm 0.20$&$0.92 \pm 0.20$&$2.23 \pm 1.12$&\\\hline
        UM559&$\star$\textrm{--$24$}&$-0.32 \pm 0.02$&$0.36 \pm 0.04$&$0.18 \pm 0.05$&$0.55 \pm 0.05$&$1.65 \pm 0.08$&$0.20 \pm 0.32$\\
        &$24$--$26$&$-0.31 \pm 0.07$&$0.37 \pm 0.05$&$0.26 \pm 0.06$&$0.62 \pm 0.10$&$1.81 \pm 0.45$&$-0.43 \pm 1.25$\\
        &$26$--$28$&$-0.38 \pm 0.35$&$0.49 \pm 0.16$&$0.22 \pm 0.21$&$0.58 \pm 0.44$&$1.71 \pm 2.56$&\\\hline
        \hline
      \end{tabular}
    \end{minipage}
  \end{table*}
\end{center}
\begin{figure}
  \begin{center}
    \includegraphics[width=7cm,height=6cm]{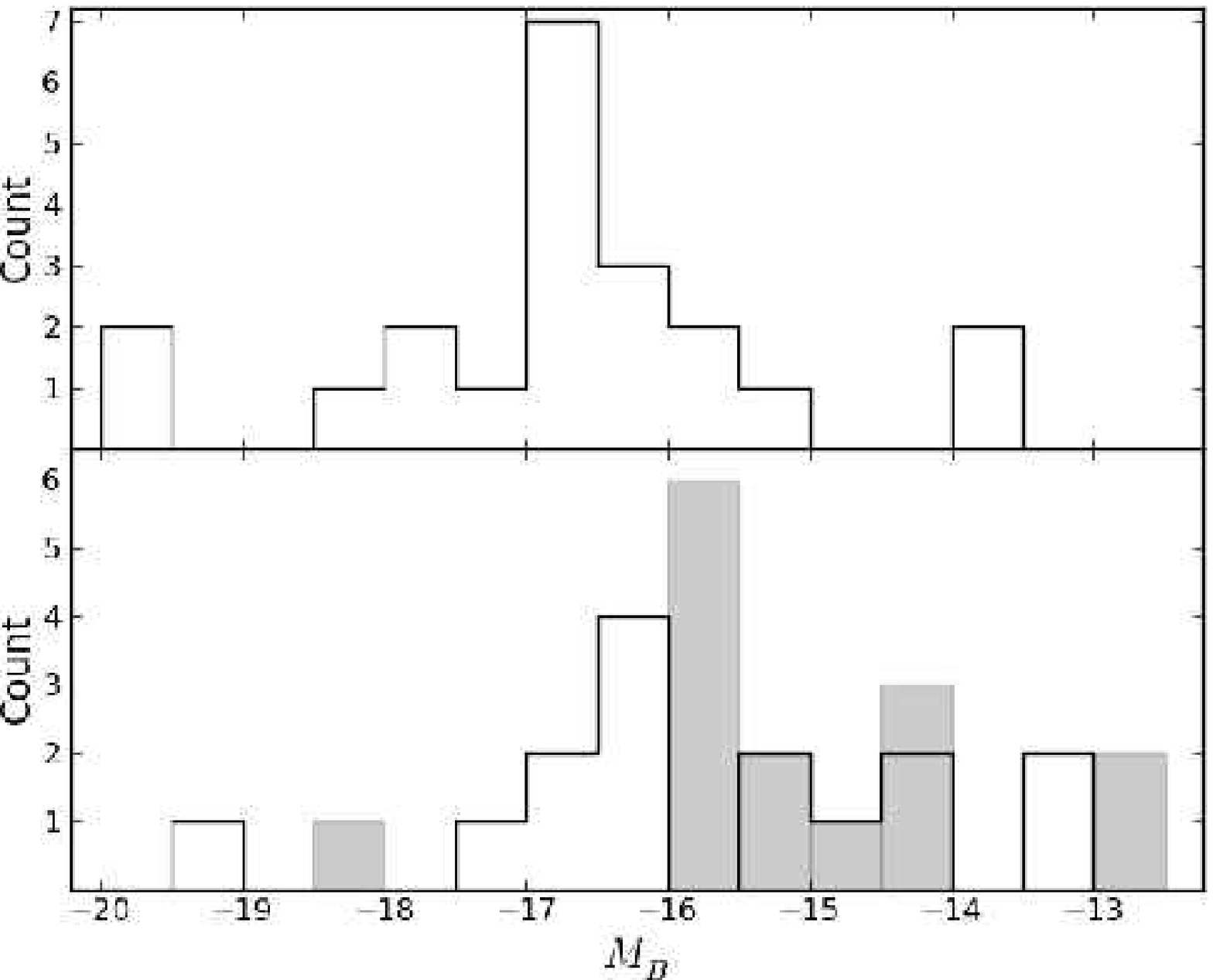}
    \caption{Absolute B magnitude distribution of the sample for the composite galaxies (upper panel), only the burst (black contour, lower panel), and the host galaxy (gray--filled steeples, lower panel). There are fewer galaxies in the lower panel due to the occasional failure in burst estimation.}\protect\label{maghist}
    \includegraphics[width=7cm,height=10cm]{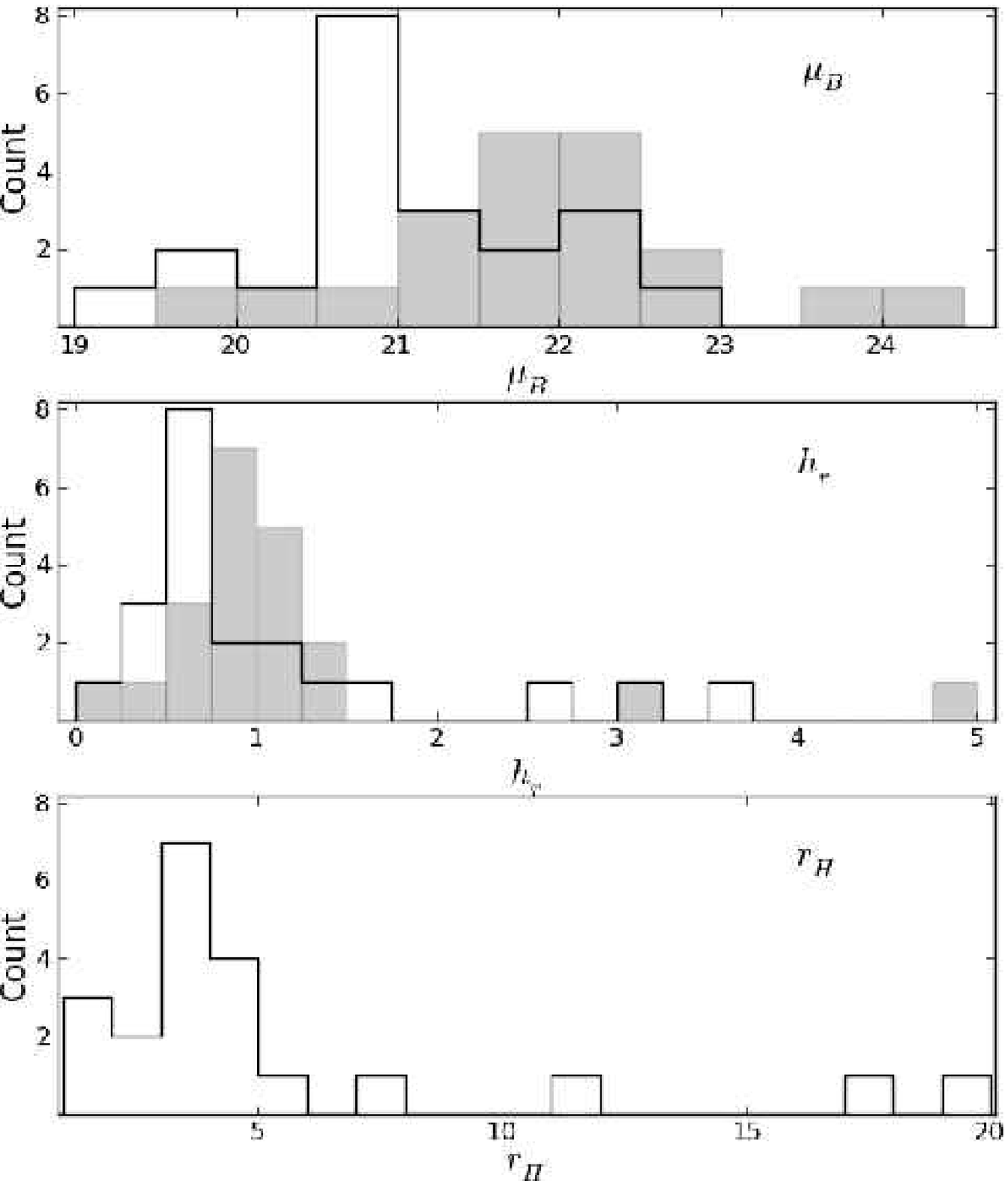}
    \caption{Top panel: Central surface brightness ($\mu_0$) distribution measured from the two regions $\mu_B=24$--$26$ (black contour) and $\mu_B=26$--$28$ mag arcsec${}^{-2}$ (gray--filled steeples). Middle panel: Same as above but for the scale length, $h_r$, in kpc. Lower panel: A distribution of the Holmberg radius, $r_H$, in kpc.}\protect\label{paramhist}
  \end{center}
\end{figure}
\begin{figure}
  \begin{center}
    \includegraphics[width=8cm,height=11cm]{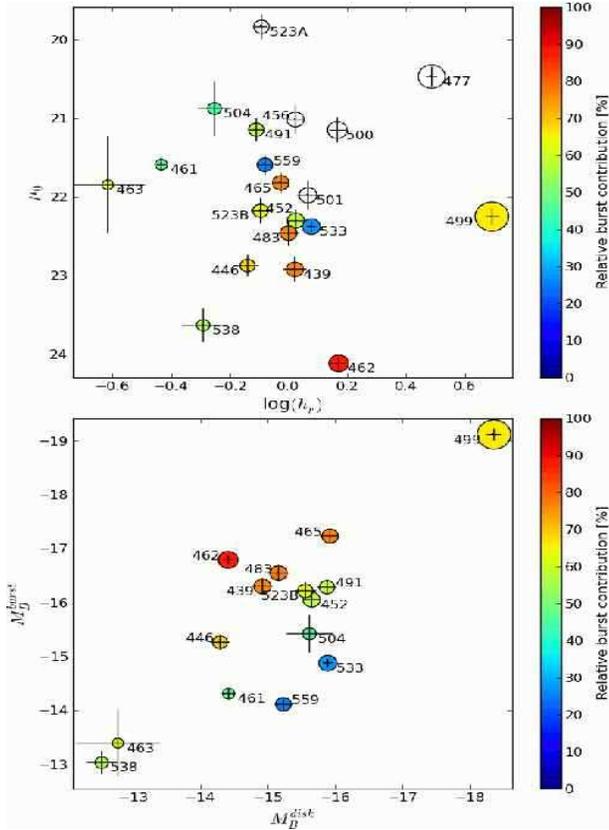}
    \caption{Upper panel: B--band $\mu_0$ in mag arcsec${}^{-2}$ vs. logged $h_r$ in kpc, both measured over the $\mu_B=26$--$28$ mag arcsec${}^{-2}$ region. Lower panel: absolute magnitude of the burst vs. absolute magnitude of the (assumed exponential) host. Marker size in both panels reflects the scale length. The colors indicate the relative burst contribution to the total (burst+host) luminosity of each target. Targets for which the burst strength estimation failed are given with open circles. Errorbars are overplotted on each marker.}\protect\label{disk_burst_hr}
  \end{center}
\end{figure}
\begin{center}
  \begin{table*}
      \caption{\textbf{Host structural parameters.} Absolute $B$ magnitude of the host ($M_B^{host}$) obtained by subtracting the burst luminosity. The scale length $h_r$ in arcseconds and kpc, and the central surface brightness $\mu_0$ based on a weighted least squares fit to the deepest image ($B$ band) for two radial ranges derived from $24\lesssim\mu_B<26$ and $26\lesssim\mu_B\lesssim28$ mag arcsec${}^{-2}$. Correction for Galactic extinction~\citep{1998ApJ...500..525S} has been applied.}
      \protect\label{scalelentbl}
      \tiny
      \begin{tabular}{@{}|l|r|r|r|r|r|r|@{}}
        \hline
        Galaxy&$\mu_B$&$M_B^{host}$&$h_r^{''}$&$h_r^{kpc}$&$\mu_0^\textrm{B}$\\\hline
        UM422&$24$--$26$&$-17.7$&$20.57\pm0.75$&$2.71\pm0.10$&$21.90\pm0.11$\\
        &$26$--$28$&&$6.17\pm0.22$&$0.81\pm0.03$&$11.35\pm0.58$\\\hline
        UM439&$24$--$26$&$-15.6$&$6.64\pm0.10$&$0.65\pm0.01$&$21.09\pm0.07$\\
        &$26$--$28$&$-14.9$&$10.71\pm0.43$&$1.05\pm0.04$&$22.92\pm0.16$\\\hline
        UM446&$24$--$26$&$-14.6$&$3.38\pm0.07$&$0.49\pm0.01$&$21.49\pm0.08$\\
        &$26$--$28$&$-14.3$&$4.98\pm0.18$&$0.72\pm0.03$&$22.88\pm0.14$\\\hline
        UM452&$24$--$26$&$-15.6$&$7.03\pm0.10$&$0.87\pm0.01$&$21.55\pm0.05$\\
        &$26$--$28$&$-15.6$&$8.58\pm0.27$&$1.06\pm0.03$&$22.31\pm0.14$\\\hline
        UM456&$24$--$26$&$-15.9$&$9.50\pm0.20$&$1.36\pm0.03$&$22.00\pm0.06$\\
        &$26$--$28$&&$7.34\pm0.23$&$1.05\pm0.03$&$21.02\pm0.18$\\\hline
        UM461&$24$--$26$&$-14.1$&$3.77\pm0.06$&$0.35\pm0.01$&$21.42\pm0.07$\\
        &$26$--$28$&$-14.4$&$3.94\pm0.06$&$0.37\pm0.01$&$21.59\pm0.09$\\\hline
        UM462&$24$--$26$&$-16.1$&$5.45\pm0.07$&$0.52\pm0.01$&$20.58\pm0.06$\\
        &$26$--$28$&$-14.4$&$15.59\pm0.40$&$1.48\pm0.04$&$24.12\pm0.08$\\\hline
        UM463&$24$--$26$&$-13.1$&$1.59\pm0.00$&$0.19\pm0.00$&$20.72\pm0.02$\\
        &$26$--$28$&$-12.8$&$2.06\pm0.26$&$0.24\pm0.03$&$21.85\pm0.62$\\\hline
        UM465&$24$--$26$&$-16.9$&$6.48\pm0.03$&$0.65\pm0.00$&$19.97\pm0.02$\\
        &$26$--$28$&$-15.9$&$9.37\pm0.25$&$0.94\pm0.03$&$21.82\pm0.13$\\\hline
        UM477&$24$--$26$&&$28.35\pm0.49$&$3.19\pm0.06$&$20.88\pm0.08$\\
        &$26$--$28$&&$27.32\pm0.52$&$3.07\pm0.06$&$20.47\pm0.12$\\\hline
        UM483&$24$--$26$&$-16.0$&$3.45\pm0.16$&$0.62\pm0.03$&$20.56\pm0.22$\\
        &$26$--$28$&$-15.2$&$5.54\pm0.20$&$1.00\pm0.04$&$22.46\pm0.16$\\\hline
        UM491&$24$--$26$&$-15.9$&$4.36\pm0.12$&$0.68\pm0.02$&$20.63\pm0.12$\\
        &$26$--$28$&$-15.9$&$4.94\pm0.13$&$0.78\pm0.02$&$21.15\pm0.15$\\\hline
        UM499&$24$--$26$&$-19.2$&$21.49\pm0.18$&$3.57\pm0.03$&$20.63\pm0.05$\\
        &$26$--$28$&$-18.4$&$29.60\pm0.64$&$4.92\pm0.11$&$22.25\pm0.11$\\\hline
        UM500&$24$--$26$&&$7.43\pm0.34$&$1.22\pm0.06$&$20.41\pm0.23$\\
        &$26$--$28$&&$8.94\pm0.24$&$1.46\pm0.04$&$21.15\pm0.16$\\\hline
        UM501&$24$--$26$&$-14.5$&$10.07\pm0.36$&$1.60\pm0.06$&$22.82\pm0.08$\\
        &$26$--$28$&&$7.32\pm0.27$&$1.16\pm0.04$&$21.98\pm0.19$\\\hline
        UM504&$24$--$26$&$-15.3$&$3.49\pm0.03$&$0.56\pm0.00$&$20.93\pm0.05$\\
        &$26$--$28$&$-15.6$&$3.51\pm0.20$&$0.56\pm0.03$&$20.88\pm0.35$\\\hline
        UM523A&$24$--$26$&&$9.14\pm0.24$&$0.76\pm0.02$&$19.43\pm0.16$\\
        &$26$--$28$&&$9.75\pm0.20$&$0.81\pm0.02$&$19.84\pm0.16$\\\hline
        UM523B&$24$--$26$&$-16.1$&$6.98\pm0.24$&$0.58\pm0.02$&$20.79\pm0.17$\\
        &$26$--$28$&$-15.6$&$9.67\pm0.31$&$0.80\pm0.03$&$22.18\pm0.16$\\\hline
        UM533&$24$--$26$&$-15.3$&$13.81\pm0.11$&$1.12\pm0.01$&$22.18\pm0.04$\\
        &$26$--$28$&$-15.9$&$14.76\pm0.22$&$1.20\pm0.02$&$22.38\pm0.07$\\\hline
        UM538&$24$--$26$&$-12.8$&$3.75\pm0.04$&$0.31\pm0.00$&$22.07\pm0.05$\\
        &$26$--$28$&$-12.5$&$6.17\pm0.44$&$0.51\pm0.04$&$23.63\pm0.22$\\\hline
        UM559&$24$--$26$&&$5.35\pm0.21$&$0.56\pm0.02$&$19.68\pm0.23$\\
        &$26$--$28$&$-15.2$&$7.97\pm0.19$&$0.83\pm0.02$&$21.59\pm0.13$\\\hline
        \hline
      \end{tabular}
  \end{table*}
\end{center}
\subsection*{UM463}
\noindent This \emph{SS} galaxy is not particularly gas--rich as it is undetected in \HI~\citep{1995ApJS...99..427T,2000A&A...361...19S} with a detection limit of $6.9\times10^6M_\odot$. Indeed, we see only mild nebular emission contribution in the color diagrams, with an old ($>5$ Gyr) very metal--poor ($Z\sim0.001$) host. We found no morphological classification in the literature, so we classify this galaxy as \emph{iE} BCD based on its contour plot and RGB image. Note that the NIR photometry of this galaxy is uncertain, since we found no stars close to the galaxy (on the final stacked image) and hence could not verify our final photometry against \emph{2MASS} (nor against \emph{SDSS} in the optical, for that matter). A correction based on offsets from Pickles stellar library values and stellar evolutionary tracks had to be applied instead. Since this correction is an estimate, we retain a $0.5$ mag error for all NIR measurements of this galaxy. The optical colors we measure for this galaxy are unusual. $U-B$ is the bluest for the entire sample, while $B-V$ is rather red. We checked our $U,~V,~I$ photometry against downloaded archived data from the {\em Hubble Space Telescope} in $F330W,~F550M,~F814W$ and found good agreement. This makes us suspicious of the $B$ band, however~\citet{1995MNRAS.275....1T} find a very similar apparent $B$ magnitude ($B=17.87$) which is consistent with ours ($B=18.02$). An additional though weaker argument is that both $B$ and $V$ band observations for this galaxy were taken during the same night and right after each other. Though we find the colors unusual, we cannot find fault in the photometry.
\subsection*{UM465}
\noindent This is \emph{DANS} galaxy with relatively low \HI mass. The small near--by galaxy visible to the North--West of UM465 is thought to be a companion, however, it must be a purely optical companion because it is not detected in \HI~\citep{1995ApJS...99..427T}. Due to its very regular isophotes at all radii we classify it as \emph{nE} BCD. $B-V$ vs $V-K$ shows a metal--poor ($Z\sim0.004$) host older than $5$ Gyrs, while the nuclear starburst is consistent with high metallicities ($Z\sim0.02$), an age $>50$ Myr, and negligible contribution from nebular emission in any of the colors.
\subsection*{UM477}
\noindent This is a gas--rich barred spiral with a central starburst region, which is a member of an interacting pair. It is one of only two galaxies classified as \emph{SBN} in our sample, the other one being UM499 -- another spiral. The burst metallicity is close to solar, but the measured extinction is moderate~\citep[][$H_\alpha/H_\beta=4.50,~12+\log{O/H}=8.69$]{1991A&AS...91..285T}. The star formation occurs along the spiral arms at large distances from the center. This is reflected in the behavior of the $B-V$ vs $V-R$, $V-I$, or $V-K$ colors in the regions $24\lesssim\mu_B\lesssim26$ and $26\lesssim\mu_B\lesssim28$ mag arcsec${}^{-2}$, which show greater age and more nebular emission contamination than the central colors. Judging by $V-I$ vs. $V-K$ the nucleus appears to be $>5$ Gyr and of low metallicity ($Z\sim0.004$), with no hint of extinction in $U-B$.
\subsection*{UM483}
\noindent UM483\protect\footnote{\footnotesize Note that UM483 is misclassified on \emph{NED} as a Seyfert 1, but the given reference catalog (V{\'e}ron--Cetty \& V{\'e}ron 2006) does not actually contain any object with similar RA and Dec.} is a Wolf--Rayet galaxy~\citep{1999A&AS..136...35S}, classified as \emph{i0} BCD by~\citet{2003ApJS..147...29G} and \emph{DHIIH} by~\citet{1989ApJS...70..479S}. The star forming regions form an ark--like structure surrounding the outskirts of a red host with regular elliptical isophotes. This galaxy has no \HI companion~\citep{1996ApJS..102..189T,2000A&A...361...19S}. The burst is not particularly recent, $U-B$ or $V-I$ vs $V-K$ colors place it somewhat younger than $1$ Gyr and consistent with $Z\gtrsim0.008$ and negligible nebular emission contribution. The host is older than $2$ Gyrs and of low metallicity $Z\lesssim0.004$.
\subsection*{UM491}
\noindent This is an \emph{nE} BCG and a~\emph{DH\texttt{II}H} galaxy. It has no \HI companions~\citep{1996ApJS..102..189T,2000A&A...361...19S} and low extinction~\citep{1991A&AS...91..285T}. $B-V$ vs $V-I$ or $V-K$ shows a metal--poor ($Z\sim0.004$) host older than $4$ Gyr and an intermediate metallicity burst ($Z\sim0.008$) older than $10$ Myr. $V-R$ vs. $V-I$ indicates that contamination from nebular emission is negligible.
\subsection*{UM499}
\noindent This Wolf--Rayet~\citep{1999A&AS..136...35S} galaxy is a normal spiral, classified as a \emph{SBN} because of its nuclear starburst activity. The metallicity of the burst is fairly high as appropriate for normal galaxies, and the measured extinction is significant~\citep[][$H_\alpha/H_\beta=8.32,~12+\log{O/H}=8.56$]{1991A&AS...91..285T}. Its $f_{25}/f_{100}$ IRAS colors suggest that the central starburst is only of moderate importance, since this flux ratio is low. Similarly to UM477, the star formation in this galaxy occurs along the spiral arms, giving the $24\lesssim\mu_B\lesssim26$ and $26\lesssim\mu_B\lesssim28$ mag arcsec${}^{-2}$ regions a high metallicity ($Z\sim0.02$) as indicated by $U-B$ or $B-V$ vs. $V-K$ or $V-I$. The colors of the star formation region are, however, dominated by an old population with an age $\gtrsim8$ Gyrs. The central colors show a similarly old population but of intermediate metallicity $Z\sim0.008$. Note that the profile break at $\mu_B\sim27.5$ mag arcsec${}^{-2}$ corresponds to a real structure as seen in Figure~\ref{deepcontours1}.
\subsection*{UM500}
\noindent Together with UM501 this \emph{SS} galaxy forms a binary pair with an extended \HI bridge between the two~\citep{1995ApJS...99..427T}. It is the more massive and extended of the two galaxies in both the optical and \HI. The measured extinction is low~\citep{1991A&AS...91..285T}. Based on its somewhat regular albeit noisy outer isophotes in the contour plot and the location of the star forming regions in the RGB image one could classify it as \emph{iE} BCG. However, the presence of what looks like spiral arm remnants make this a \emph{iI,M} candidate. This is further supported by the appearance of the galaxy in the NIR, where the circular outer envelope is not observed at all. In the $B$ band the host galaxy has a remarkably disk--like structure recognized both from the contour plot and the surface brightness profile. The multiple star forming regions are located at large radii from the center. The burst estimation fails here, but the total, central, and $24\lesssim\mu_B\lesssim26$ and $26\lesssim\mu_B\lesssim28$ mag arcsec${}^{-2}$ region colors are all very similar, require significant contribution from nebular emission to fit the tracks, and are clumped together in most color--color diagrams. The similarity in color across different physical regions reflects the well--mixed nature of the morphology and we cannot achieve enough separation between the host and the burst to estimate its metallicity or age. 
\subsection*{UM501}
\noindent This \emph{MI} galaxy is the second member of the binary pair. It is less massive in \HI than its companion~\citep{1995ApJS...99..427T}, and it also has slightly higher extinction and metallicity than UM500~\citep{1991A&AS...91..285T,1993AJ....106.1784C}.~\citet{1991A&A...241..358C} state that this galaxy has its starburst in a small companion or in an external \HII region of the galaxy, but our deeper contour plot and RGB image show that in fact the numerous star forming regions are embedded beyond the $24.5^{th}$ $B$ band isophote in a single object with vaguely elliptical isophotes. The isophotes are not regular even at very faint levels, however, indicating that this may be a merger of smaller objects. We give the classification \emph{iI} since we cannot be more specific without kinematic data on the individual starburst knots. The morphology of this galaxy is very similar to UM500, with numerous dispersed starburst knots at large and small radii from the center. The measured colors are again clumped together in the color--color diagrams and require a large contribution from nebular emission. We have a burst estimate for this target, but the surface brightness profile is remarkably flat inspite of the irregular burst morphology. This is an indication that the excess luminosity above the exponential disk must underestimate the actual burst luminosity. We see the effect of this underestimation in all color--color diagrams -- all metallicity tracks of both models fail to reproduce our burst estimate.
\subsection*{UM504}
\noindent This~\emph{DH\texttt{II}H} galaxy has low extinction~\citep{1991A&AS...91..285T}. It possibly has an \HI companion with no optical counterpart~\citep{1995ApJS...99..427T}, however, other authors do not detect any \HI companions~\citep{2000A&A...361...19S}. We classify it as \emph{nE} BCD based on its centrally located star forming region and regular outer isophotes seen in the contour plot. $B-V$ vs. $V-I$ or $V-K$ indicate a very metal--poor ($Z\sim0.001$) host older than $5$ Gyrs. Nebular emission contribution is not necessary to model most of the colors, though only the burst can be modeled by both tracks with and without nebular emission in e.g. $U-B$ and $V-I$ vs $H-K$, thus indicating that some amount of nebular emission must still be present. The burst cannot be much younger than $1$ Gyr.
\subsection*{UM523}
\noindent This is an interacting pair, $NGC4809$ (our $UM523A$) and $NGC4810$ (our $UM523B$). No individual \HI measurements exist for the members of the pair, since they are in contact (beyond the $23.5^{th}~B$ isophote) and get confused in \HI surveys. Both galaxies have very similar morphologies, with numerous compact blue star forming regions dispersed throughout the individual disks, though an increase in the abundance of the star forming knots is notable near the contact region between the two. $UM523A$ has very low extinction~\citep{1991A&AS...91..285T}, so it is reasonable to expect the same for $UM523B$. Indeed, we see no dust reddening of the central isophotal $U-B$ or $B-V$ radial profiles. Both galaxies have regular outer isophotes, so we classify each as \emph{iE} BCGs.
\subsubsection*{UM523A (NGC4809)}
\noindent Significant nebular emission is necessary to explain the colors in $B-V$ vs. $V-R$ and $V-K$. Different physical regions in the galaxy give very similar colors, indicating that they all contain a mixture of young and old stellar populations, together with gas. Due to this well--mixed nature we found no conclusive constraints on the age and metallicity of any of the different regions. The burst estimation fails here, which is just as well, since it would likely be a severe underestimate.
 \subsubsection*{UM523B (NGC4810)}
\noindent Nebular emission is similarly needed here to explain the total and burst colors. The $26\lesssim\mu_B\lesssim28$ mag arcsec${}^{-2}$ region is free from SF knots, and is consistent with a single, old ($>3$ Gyrs), very metal--poor ($Z\sim0.001$) stellar population in $V-K$ vs. $B-V$, $V-R$, and $V-I$. The total and central colors are very similar in all diagrams though many of the SF knots can be found beyond the $\mu_B=24$ isophote. This implies that our approximation underestimates the burst.
\subsection*{UM533}
\noindent This is an \emph{iE} BCD~\citep{2003ApJS..147...29G} and a \emph{MI} galaxy with non-negligible extinction~\citep[][$H_\alpha/H_\beta=5.21,~12+\log{O/H}=8.10$]{1991A&AS...91..285T}. It has no \HI companions~\citep{1995ApJS...99..427T,2000A&A...361...19S}. There are a few star forming blue knots close to the galactic center in the RGB image, but the extended regular elliptical host seems to be the dominating component of this galaxy. All regions except the burst estimate are well--fitted with a pure stellar population in $B-V$ vs. $V-I$ or $V-K$ diagrams, with an age $>5$ Gyrs and very low metallicity ($Z\sim0.001$). The total colors of the galaxy are very similar to the colors in the outer regions, which is consistent with the observation that the burst is not dominant. Nebular emission contribution is required to fit the burst in $U-B$ vs. $B-V$, and $B-V$ vs. $V-R$ and $V-I$ diagrams, both indicating a burst age younger than $10$ Myr and a moderate metallicity ($Z\sim0.008$). 
\subsection*{UM538}
\noindent This metal--poor \emph{SS} galaxy is the least luminous in the sample. It has no \HI companions and low detected \HI mass~\citep{1995ApJS...99..427T,2000A&A...361...19S}, making it one of the least massive galaxies in the sample as well. Its only star forming region is off--center, distorting the central isophotes, so we classify it as \emph{iE} BCD. The burst is not dominating, and cannot be fitted with tracks including nebular emission in any color--color diagram. A pure stellar population model fits all measurements. $B-V$ vs $V-K$ indicates a burst age $>100$ Myr with low metallicity ($Z\sim0.004$), and very low metallicity ($Z\sim0.001$) for the old ($>5$ Gyrs) host, which is also supported by $V-I$ vs $V-K$ diagrams.
\subsection*{UM559}
\noindent This \emph{SS} galaxy has an \HI companion~\citep{1995ApJS...99..427T} and low extinction~\citep{1991A&AS...91..285T}. It is very metal--poor and it is classified as \emph{iE/iI} BCD by~\citet{2006A&A...457...45P}. Compact starburst regions are visible in the outskirts, but the RGB image hints at the presence of diffuse blue regions dispersed throughout the galaxy. This is consistent with the clustering we observe in color--color diagrams, where the colors of different regions are very similar to each other. The nature of the sampled areas must obviously be well--mixed, and we cannot distinguish an age or metallicity for any separate physical component, but a significant nebular emission contribution is decidedly necessary to fit all of the measurements. 
\begin{center}
  \begin{table*}
    \begin{minipage}{140mm}
      \caption{Estimated luminosity in excess of the exponential disk defined by $h_r^{''}$ and $\mu_0$. The upper and lower numbers for each galaxy are for disk properties derived from the radial ranges corresponding to $24\lesssim\mu_B<26$ and $26\lesssim\mu_B\lesssim28$ mag arcsec${}^{-2}$, respectively. Fields are left blank where the estimation method failed. All values have been corrected for Galactic extinction~\citep{1998ApJ...500..525S}. The $\%$ column gives the $B$ band relative burst contribution to the total galaxy luminosity.}
      \protect\label{burstclrtbl}
      \tiny
      \begin{tabular}{@{}lllrrrrrrr@{}}
        \hline
        Galaxy&$\%$&$B_\star$&$(U-B)_\star$&$(B-V)_\star$&$(V-R)_\star$&$(V-I)_\star$&$(V-K)_\star$&$(H-K)_\star$\\\hline
        UM422&$2$&$18.3 \pm 0.1$&$-0.93 \pm 0.12$&$0.68 \pm 0.12$&$1.43 \pm 0.07$&$1.67 \pm 0.08$&$4.31 \pm 0.10$&$1.67 \pm 0.10$\\
        &&&&&&&&\\\hline
        UM439&$45$&$15.8 \pm 0.1$&$-0.72 \pm 0.08$&$0.29 \pm 0.07$&$-1.18 \pm 0.05$&&$1.55 \pm 0.08$&\\
        &$77$&$15.2 \pm 0.2$&$-0.52 \pm 0.17$&$0.34 \pm 0.16$&$-0.06 \pm 0.04$&$0.09 \pm 0.06$&$2.09 \pm 0.45$&\\\hline
        UM446&$42$&$17.6 \pm 0.1$&&$0.43 \pm 0.08$&$0.19 \pm 0.04$&$0.12 \pm 0.06$&$0.93 \pm 0.06$&$0.05 \pm 0.13$\\
        &$69$&$17.1 \pm 0.1$&&$0.56 \pm 0.15$&$0.29 \pm 0.04$&$0.53 \pm 0.06$&$0.92 \pm 0.09$&\\\hline
        UM452&$40$&$16.4 \pm 0.1$&$-0.46 \pm 0.06$&$0.50 \pm 0.05$&$0.29 \pm 0.05$&$0.71 \pm 0.04$&$2.12 \pm 0.03$&$0.31 \pm 0.06$\\
        &$58$&$16.0 \pm 0.1$&$-0.19 \pm 0.15$&$0.64 \pm 0.15$&$0.35 \pm 0.05$&$0.84 \pm 0.05$&$1.92 \pm 0.07$&$3.34 \pm 0.11$\\\hline
        UM456&$32$&$16.6 \pm 0.1$&$-0.90 \pm 0.07$&$0.15 \pm 0.07$&$-0.32 \pm 0.04$&$-0.54 \pm 0.05$&$1.04 \pm 0.06$&$-0.25 \pm 0.10$\\
        &&&&&&&&\\\hline
        UM461&$42$&$17.2 \pm 0.1$&$-0.78 \pm 0.08$&$0.84 \pm 0.07$&$-0.46 \pm 0.04$&$-1.13 \pm 0.04$&$-0.57 \pm 0.12$&$-0.60 \pm 0.16$\\
        &$47$&$17.1 \pm 0.1$&$-0.63 \pm 0.10$&$0.81 \pm 0.09$&$-0.35 \pm 0.05$&$-0.86 \pm 0.05$&&\\\hline
        UM462&$43$&$15.5 \pm 0.1$&$-0.97 \pm 0.16$&$0.18 \pm 0.07$&$-0.52 \pm 0.25$&$-1.39 \pm 0.20$&$0.76 \pm 0.08$&$3.00 \pm 0.11$\\
        &$88$&$14.7 \pm 0.1$&$-0.68 \pm 0.16$&$0.31 \pm 0.09$&$0.02 \pm 0.25$&$-0.08 \pm 0.20$&$1.65 \pm 0.12$&$0.68 \pm 0.12$\\\hline
        UM463&$37$&$19.1 \pm 0.1$&$-1.30 \pm 0.06$&$0.81 \pm 0.06$&&$-2.53 \pm 0.07$&&\\
        &$63$&$18.5 \pm 0.6$&$-0.97 \pm 0.62$&$0.75 \pm 0.62$&&$-1.11 \pm 0.09$&$2.74 \pm 0.50$&\\\hline
        UM465&$34$&$15.2 \pm 0.1$&$-0.68 \pm 0.04$&$0.19 \pm 0.04$&&&$2.05 \pm 0.06$&\\
        &$76$&$14.4 \pm 0.1$&$-0.33 \pm 0.14$&$0.58 \pm 0.14$&&&$2.33 \pm 0.07$&\\\hline
        UM477&&&&&&&&\\
        &&&&&&&&\\\hline
        UM483&$44$&$16.9 \pm 0.2$&$-0.44 \pm 0.22$&$0.34 \pm 0.22$&$0.12 \pm 0.08$&$0.42 \pm 0.08$&$2.12 \pm 0.09$&\\
        &$77$&$16.3 \pm 0.2$&$-0.31 \pm 0.17$&$0.37 \pm 0.17$&$0.18 \pm 0.08$&$0.49 \pm 0.08$&$2.24 \pm 0.09$&\\\hline
        UM491&$47$&$16.5 \pm 0.1$&$-0.63 \pm 0.13$&$0.14 \pm 0.13$&$-0.11 \pm 0.05$&$0.03 \pm 0.05$&$1.72 \pm 0.14$&\\
        &$59$&$16.3 \pm 0.2$&$-0.59 \pm 0.16$&$0.15 \pm 0.15$&$-0.32 \pm 0.05$&$-0.00 \pm 0.04$&$0.63 \pm 0.14$&\\\hline
        UM499&$15$&$15.2 \pm 0.1$&$-0.97 \pm 0.06$&$0.43 \pm 0.06$&&$0.12 \pm 0.03$&$3.14 \pm 0.08$&\\
        &$66$&$13.6 \pm 0.1$&$-0.27 \pm 0.11$&$0.66 \pm 0.11$&&$0.67 \pm 0.04$&$2.25 \pm 0.09$&\\\hline
        UM500&&&&&&&&\\
        &&&&&&&&\\\hline
        UM501&$12$&$18.8 \pm 0.1$&$-0.92 \pm 0.08$&$0.61 \pm 0.08$&$-0.65 \pm 0.04$&$-1.56 \pm 0.05$&$0.76 \pm 0.06$&$0.80 \pm 0.09$\\
        &&&&&&&&\\\hline
        UM504&$48$&$17.1 \pm 0.1$&$-0.39 \pm 0.06$&$0.31 \pm 0.06$&$0.01 \pm 0.04$&$0.17 \pm 0.08$&$1.68 \pm 0.05$&$0.40 \pm 0.07$\\
        &$45$&$17.2 \pm 0.4$&$-0.22 \pm 0.37$&$0.30 \pm 0.35$&$-0.15 \pm 0.06$&$0.08 \pm 0.08$&$2.15 \pm 0.06$&$-0.63 \pm 0.25$\\\hline
        UM523A&&&&&&&&\\
        &&&&&&&&\\\hline
        UM523B&$28$&$15.9 \pm 0.2$&$0.16 \pm 0.18$&$-1.34 \pm 0.17$&&&&\\
        &$64$&$14.9 \pm 0.2$&$-0.46 \pm 0.17$&$0.00 \pm 0.16$&$-0.36 \pm 0.05$&$0.25 \pm 0.05$&$0.60 \pm 0.10$&\\\hline
        UM533&$22$&$16.5 \pm 0.1$&$-0.61 \pm 0.05$&$0.27 \pm 0.04$&$-0.01 \pm 0.03$&$-0.44 \pm 0.02$&$1.60 \pm 0.09$&\\
        &$27$&$16.2 \pm 0.1$&$-0.83 \pm 0.09$&$0.38 \pm 0.08$&$0.39 \pm 0.03$&$-1.43 \pm 0.02$&$1.48 \pm 0.12$&\\\hline
        UM538&$16$&$19.5 \pm 0.1$&$-1.26 \pm 0.06$&$0.01 \pm 0.06$&$-1.68 \pm 0.07$&&$1.10 \pm 0.07$&\\
        &$56$&$18.1 \pm 0.2$&$-0.53 \pm 0.22$&$0.38 \pm 0.22$&$0.10 \pm 0.08$&$0.52 \pm 0.05$&$1.82 \pm 0.09$&\\\hline
        UM559&&&&&&&&\\
        &$26$&$17.6 \pm 0.1$&$-0.74 \pm 0.14$&$0.35 \pm 0.14$&$-0.37 \pm 0.06$&$0.24 \pm 0.07$&&\\\hline
        \hline
      \end{tabular}
    \end{minipage}
  \end{table*}
\end{center}
\section[]{Results and discussion}\protect\label{discuss}
\noindent In Figures~\ref{maghist} and~\ref{paramhist} we present the distribution of the total absolute $B$ band luminosity, the burst and the host luminosities separately, the central surface brightness, the scale length and the Holmberg radius for all galaxies in the sample. The sample contains mostly dwarfs, strongly peaked at $M_B\sim-16.7$ mag, and two bright spirals with $M_B\sim-20$ mag. The luminosity of the hosts is naturally fainter but already here we can conclude that the burst, although dominating the total galaxy luminosity, only increases the light output by about a magnitude for most galaxies. Most galaxies in this sample also seem to be very compact, with very similar scale lengths and Holmberg radii. There are of course outliers in both luminosity and size, with some extremely faint and very extended galaxies or vice versa, however, on the large there appears to be a distinct subgroup of objects with nearly identical sizes and luminosities. We will come back to this group later on in the discussion.\\

\noindent Contrary to \M, in this sample we see no correlation between the central surface brightness and the scale length (upper panel, Figure~\ref{disk_burst_hr}). There is no trend for the more extended hosts to be of fainter $\mu_0$. That is not surprising since the trend seems to appear most strongly for hosts which qualify as true LSB galaxies, with $\mu_0\gtrsim23$ mag arcsec${}^{-2}$, and the scatter in the relation increases significantly for brighter $\mu_0$ (compare with Figure $9$ in \M). In this sample most objects fall above the $\mu_0\sim23$ mag arcsec${}^{-2}$ line, although the bulk of the objects around $h_r\sim1$ kpc and $\mu_0\sim22$ mag arcsec${}^{-2}$ would fall on the correlation line defined by Figure $9$ in \M. There is also no clear trend with increasing relative burst contribution -- most galaxies have moderate burst strengths ($\sim50\%$) and those seem fairly independent of $\mu_0$ and $h_r$. Similarly to \M, we find a correlation between the luminosity of the burst and the luminosity of the host (lower panel, Figure~\ref{disk_burst_hr}). The brightest burst in absolute terms, the spiral galaxy UM499, corresponds to the brightest host as expected, but its relative burst contribution is moderate, only $\sim65\%$ (yellow). Much smaller and intrinsically fainter galaxies in the sample, like UM439 and UM483, have larger relative burst contributions of $\sim80\%$ (orange).\\
\begin{figure}
  \begin{center}
    \includegraphics[width=7.5cm,height=6cm]{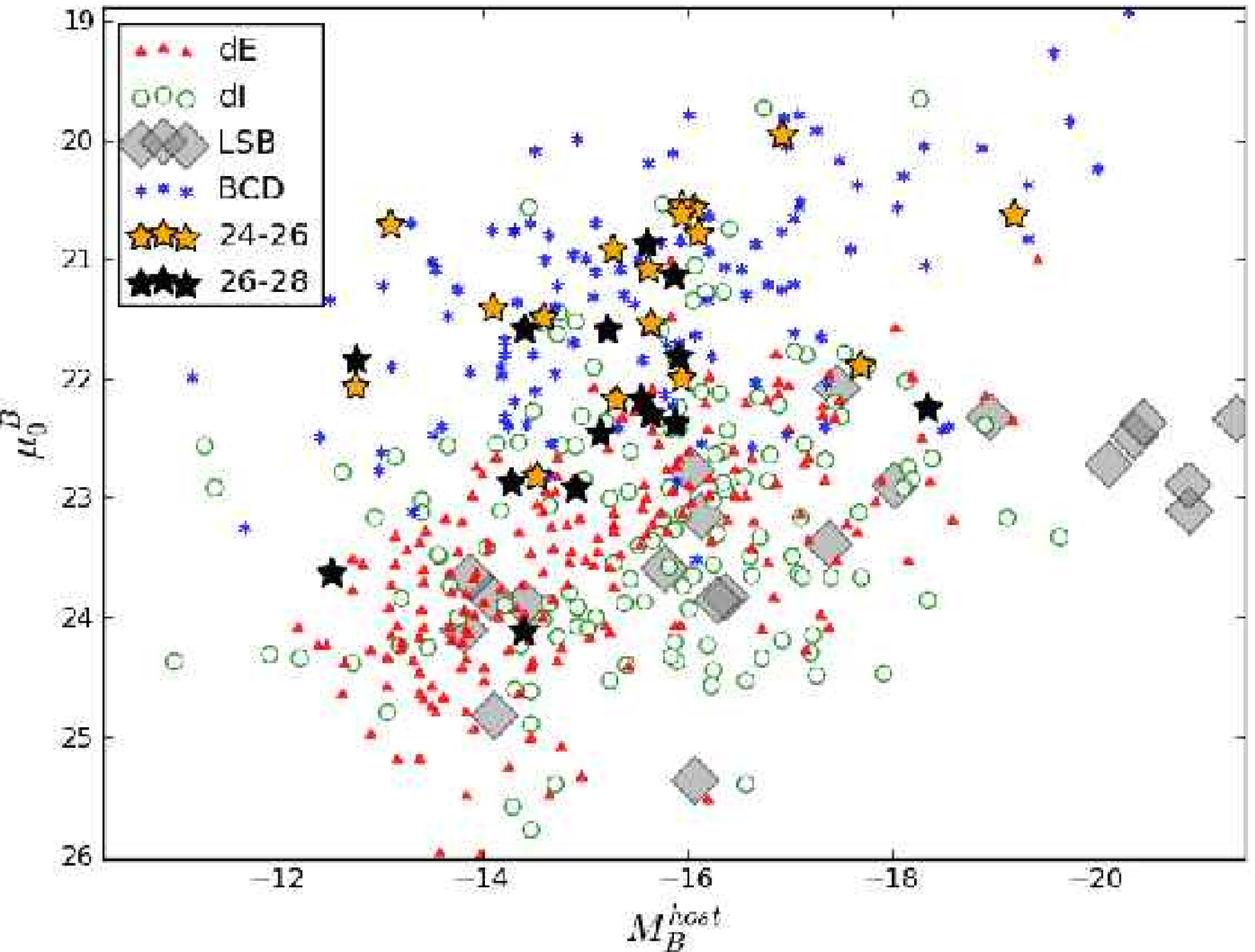}
    \includegraphics[width=7.5cm,height=6cm]{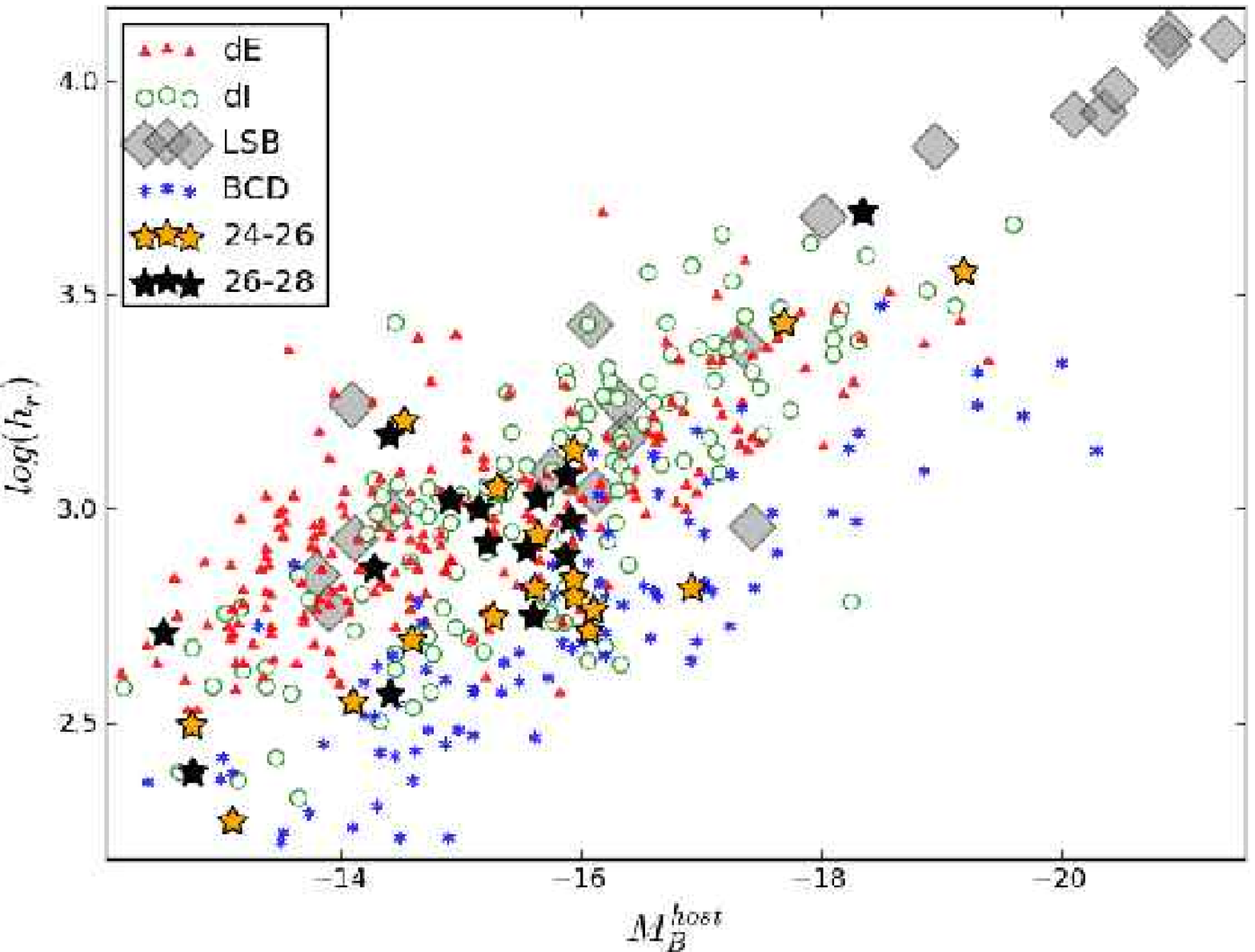}
    \caption{Comparing the structural properties of the host galaxies in our BCG sample (orange and black stars) to those of dEs, dIs, LSB and BCD galaxies from the literature. dE, dI, BCD and some LSB data were obtained from~\citet{2008A&A...491..113P} and references therein, while the giant LSB galaxies were taken from~\citet{1995AJ....109..558S} and references therein. The orange stars indicate structural parameters derived from the physical region corresponding to $\mu_B=24-26$ mag arcsec${}^{-2}$, while the black stars are from the fainter $\mu_B=26-28$ region. Note that $h_r$ is here in units of pc, not kpc.}\protect\label{papaderos}
  \end{center}
\end{figure}

\noindent Examining the surface brightness profiles in Figure~\ref{datafig} we see that most profiles are very well fitted in the outskirts with an exponential disk. In fact, there are no observed profile breaks in the region $24\lesssim\mu_B\lesssim28$ mag arcsec${}^{-2}$ for the vast majority of galaxies. UM462 is an exception, with a previously undetected extended second LSB component beyond $\mu_B=26$ mag arcsec${}^{-2}$. Two other galaxies show profile breaks at faint isophotes -- UM465 at $\mu_B\sim27.0$ mag arcsec${}^{-2}$, and UM499 at $\mu_B\sim27.5$ mag arcsec${}^{-2}$. For the latter two galaxies the structural parameters obtained from an exponential disk fit over the range $26-28$ mag arcsec${}^{-2}$ will be influenced by this change in profile slope, but they are not an accurate measure of the faint components beyond $\mu_B\sim27.0$ and $\mu_B\sim27.5$ mag arcsec${}^{-2}$ for UM465 and UM499 respectively. These structures are real, as evidenced by the extremely faint contours in Figure~\ref{deepcontours1}.\\

\noindent There are a significant number of galaxies for which the host is well approximated by a single exponential disk in the outskirts all the way down to $\mu_B\sim28$ mag arcsec${}^{-2}$. For such targets, this implies that nebular emission cannot be dominating the brighter $\mu_B=24$--$26$ mag arcsec${}^{-2}$ regions. Even though BCG literature often does not present surface brightness profiles probing much fainter than the Holmberg radius, targets with no profile change at faint ($\mu_B>26$ mag arcsec${}^{-2}$) isophotes should agree well with BCG structural parameters in the literature for both $\mu_B=24$--$26$ mag arcsec${}^{-2}$ and $\mu_B=26$--$28$ mag arcsec${}^{-2}$ regions. In Figure~\ref{papaderos} we compare the host structural parameters of our sample to those compiled by~\citet{2008A&A...491..113P} for dE, dI, and blue compact dwarfs (BCD), and to the giant low surface brightness (LSB) spiral galaxies of~\citet{1995AJ....109..558S}. In contrast to the luminous BCGs of \M~many galaxies in this sample have $\mu_0$ and $h_r$ consistent with BCD data. The rest occupy the same parameter space as dE and dI. This is suggestive of a separation of BCGs into two groups, consistent with that proposed by~\citet{1997MNRAS.288...78T}. As evidenced by the structural parameter space, the irregular extended BCGs in \M~must have different progenitors than the more compact and regular ones in this sample. We will get back to this in~\citet{Paper3}.\\

\subsection[]{Color trends}\protect\label{colortrends}
\noindent We first acknowledge that with a maximum of $21$ measurements of any photometric or structural quantity we are in the regime of small number statistics, and we should not overinterpret the observed distributions and trends in these measurements.\\

\begin{figure*}
\begin{center}
  \includegraphics[width=16.5cm]{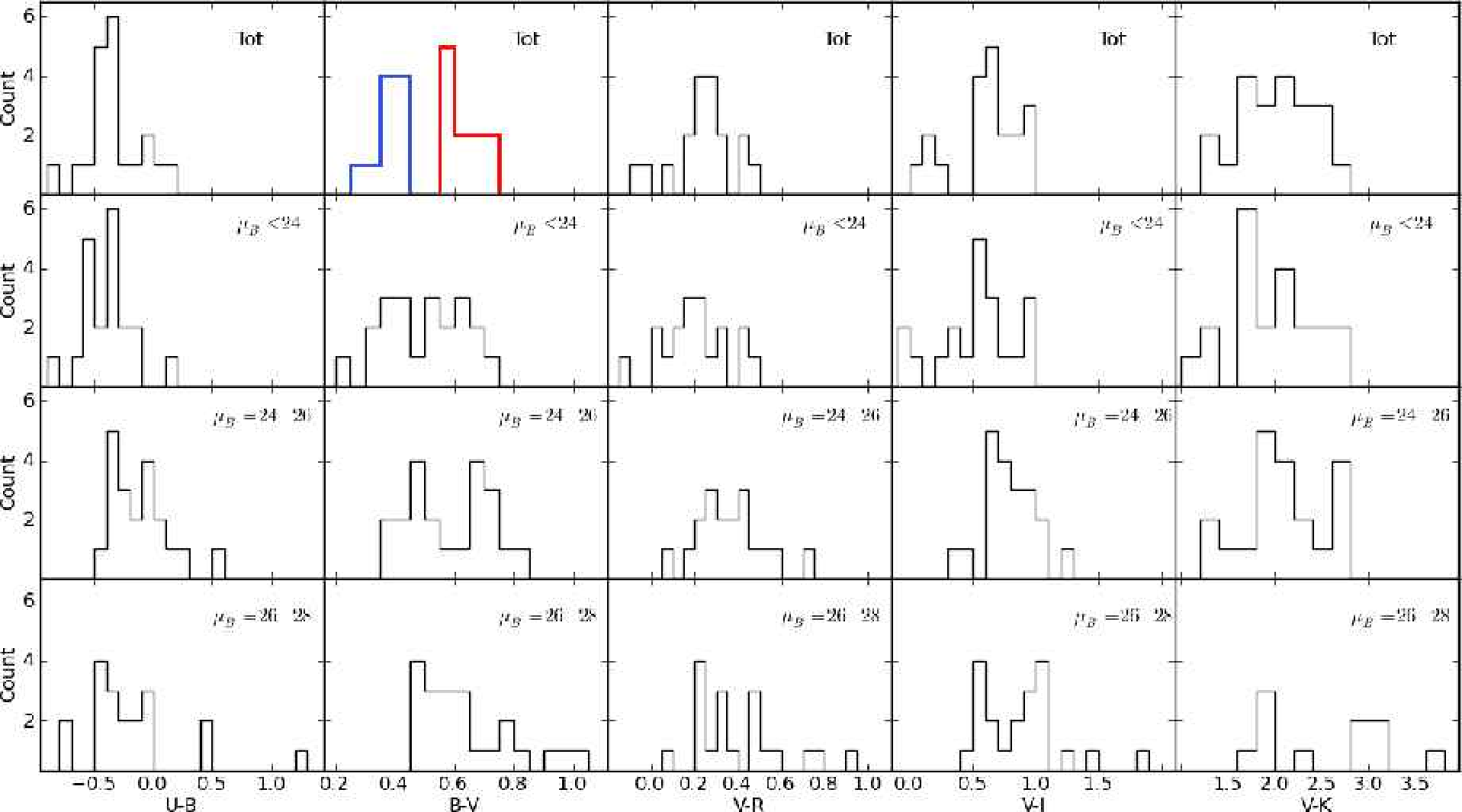}
\caption{The distribution of colors measured over different regions of the galaxies -- total color down to the Holmberg radius $r_H$ (top row); central colors down to $\mu_B=24$ mag arcsec${}^{-2}$ (second row); color in the region $24\lesssim\mu_B\lesssim26$ mag arcsec${}^{-2}$ (third row); color in the region $26\lesssim\mu_B\lesssim28$ mag arcsec${}^{-2}$ (last row). The bin sizes are $0.1$, $0.05$, $0.05$, $0.1$, $0.2$ for $U-B$, $B-V$, $V-R$, $V-I$, and $V-K$ respectively, making the bin sizes greater than or equal to the average errors in the colors in the first row (Table~\ref{totlumtbl}). The $G^B$ and $G^R$ groups in the text are marked with blue and red.}\protect\label{clrhistmosaic}
  \includegraphics[width=16.5cm]{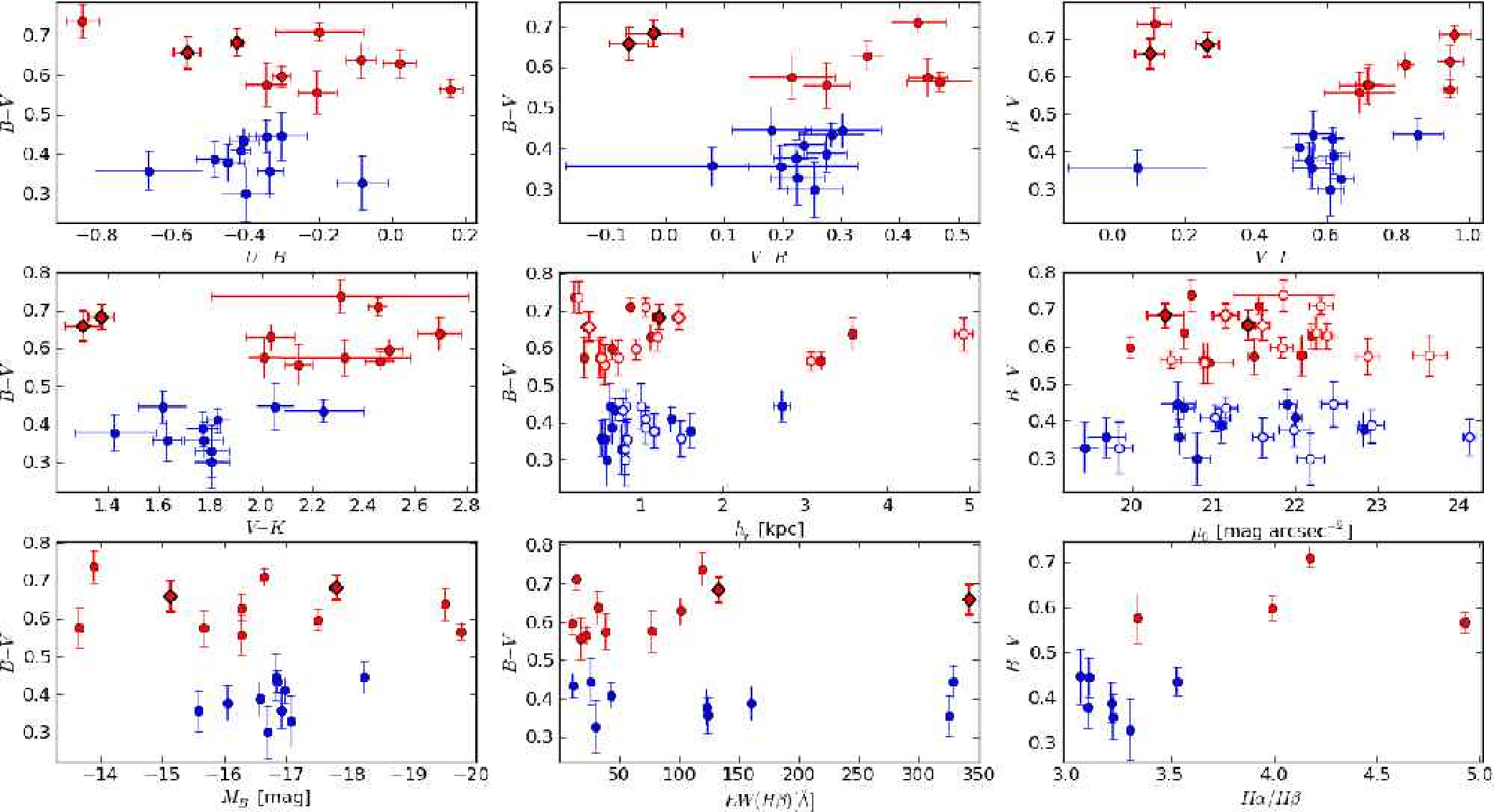}
\caption{A follow--up on the different behavior of the blue and red ($G^B$ and $G^R$ in the text) groups defined from the total $B-V$ color histogram in Figure~\ref{clrhistmosaic}. The marker coloring is self-evident, except in the scale length ($h_r$) and central surface brightness ($\mu_0$) scatter plots, where the filled and open circles correspond to disk parameters defined over $24\lesssim\mu_B\lesssim26$, and $26\lesssim\mu_B\lesssim28$ mag arcsec${}^{-2}$, respectively. The black/red diamonds are UM461 and UM500, which belong to $G^R$ but deviate in many of the plots. $M_B$ is the absolute $B$ band luminosity integrated down to $r_H$ in Vega magnitudes. The $H\beta$ equivalent width is taken from~\citet{1991A&AS...91..285T}. The $H\alpha/H\beta$ ratio, corrected for Galactic extinction, is estimated from the SDSS spectra where such existed. }\protect\label{comlementhist}
\end{center}
\end{figure*}

\noindent The color histograms in Fig.~\ref{clrhistmosaic} have bin sizes $0.1$, $0.05$, $0.05$, $0.1$, $0.2$ for $U-B$, $B-V$, $V-R$, $V-I$, and $V-K$ respectively. In the first row of the figure, where the histograms of the total colors down to the Holmberg radius are presented, the chosen bin sizes are much larger than or equal to the average errors of the colors in Table~\ref{totlumtbl}. The same bin sizes are also appropriate for the second and third rows, where the central colors down to $\mu_B=24$ mag arcsec${}^{-2}$ and the colors between $24\lesssim\mu_B\lesssim26$ mag arcsec${}^{-2}$ are presented (from Table~\ref{totclrtbl}). In the latter case increasing the bin size where necessary to exactly correspond to the average error, e.g. bin size $a=0.07$ from $\bar{\sigma}(V-R)=0.07$ instead of the used value $a=0.05$, or bin size $a=0.3$ from $\bar{\sigma}(V-K)=0.3$ instead of the used $a=0.2$, does not significantly alter the shape of the histogram, so we maintain the same bin size even here. The colors between $26\lesssim\mu_B\lesssim28$ mag arcsec${}^{-2}$, presented in the last row of Fig.~\ref{clrhistmosaic}, however, have much higher average errors than the selected bin sizes. In some extreme cases like $V-K$ taking the bin size to be the average error of $\bar{\sigma}(V-K)=1.2$ will smooth all features in the histogram. We have therefore chosen to keep the same bin size in all histograms for a specific color but in the $26\lesssim\mu_B\lesssim28$ mag arcsec${}^{-2}$ region the only meaningful statistic we can measure is the average color.\\
\begin{table}
    \caption{List of galaxies in $G^B$ and $G^R$}
    \protect\label{bimod}
    \begin{tabular}{|cc|}
      \hline
      $G^B$&$G^R$\\\hline
      UM422&UM446\\
      UM439&UM452\\
      UM456&UM461\\
      UM462&UM463\\
      UM483&UM465\\
      UM491&UM477\\
      UM501&UM499\\
      UM523A&UM500\\
      UM523B&UM504\\
      UM559 &UM533\\
      &UM538\\\hline
      \hline
    \end{tabular}
\end{table}
\begin{figure*}
\begin{center}
  \includegraphics[width=14.0cm]{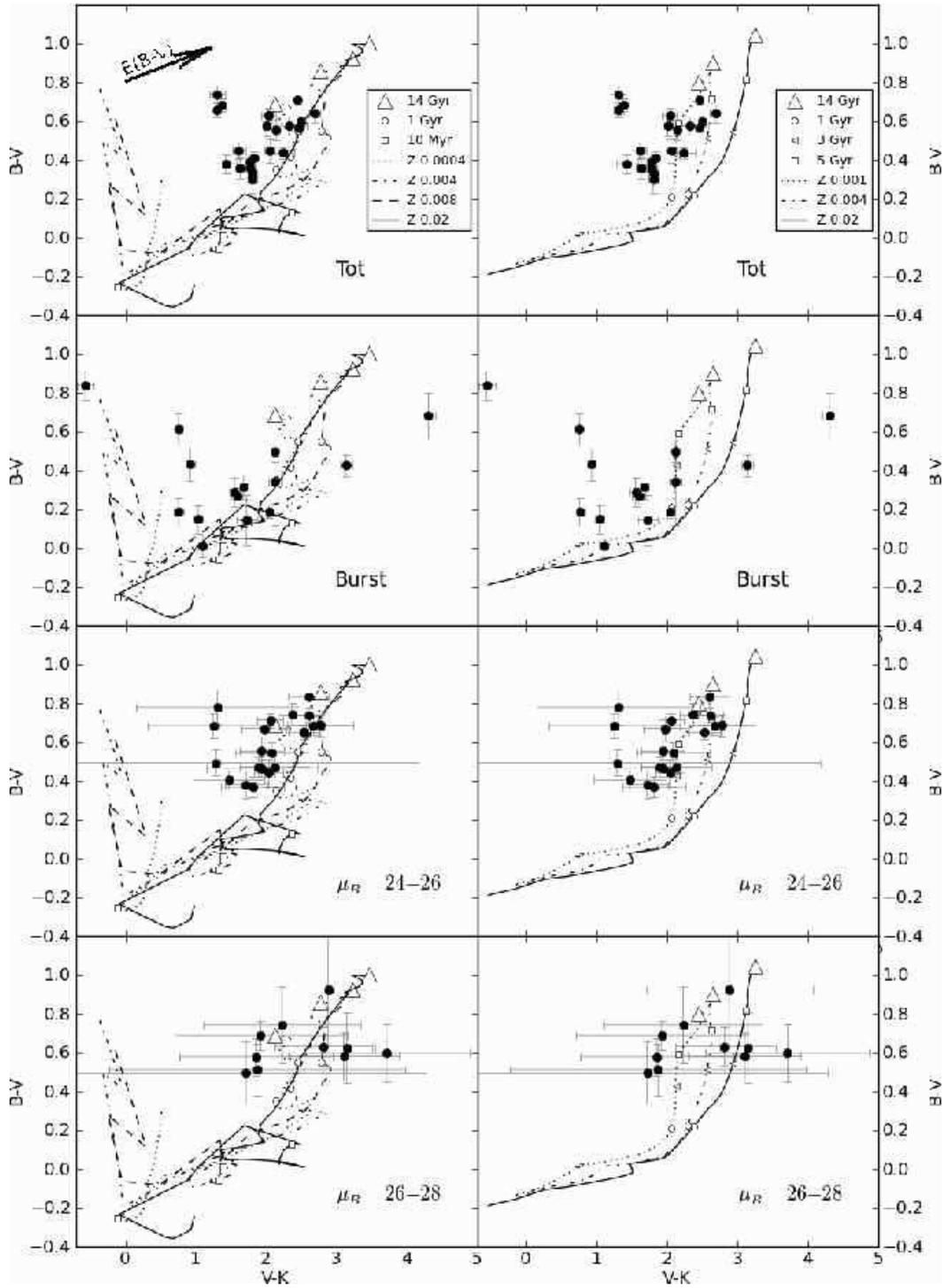}
\caption{Integrated optical/NIR colors with SEMs tracks with (left panel) and without (right panel) nebular emission. From top to bottom in both panels the plotted data points are the total colors over the entire area of the galaxy, the estimated burst colors down to $\mu_B=24$, the colors of the $\mu_B\sim24-26$ halo region, and the colors of the $\mu_B\sim26-28$ halo region. The data are corrected for Galactic extinction. The errorbars include different uncertainties depending on the case (see \S~\ref{integsurfphot}), and the burst uncertainties may be underestimated. The number of targets in the plots is not equal since the burst/host separation fails completely for a number of objects with irregular morphology, or the integrated flux inside $\mu_B\sim26-28$ is negative for the $K$ band. The left panel shows \emph{Yggdrasil} tracks with nebular emission, with metallicities $Z=0.0004$ (dotted), $Z=0.004$ (dash--dotted), $Z=0.008$ (dashed), and $Z=0.02$ (line). The ages of $10$ Myr (open square), $1$ Gyr (open circle), and $14$ Gyrs (open triangle) are marked for convenience. In the right panel the tracks are for a pure stellar population with a Salpeter IMF, $M_{min}=0.08M_\odot$, $M_{max}=120M_\odot$, an $e$--folding time of $10^9$ yr, and with metallicity $Z=0.001$ (dotted), $Z=0.004$ (dash--dotted), and $Z=0.02$ (line). The ages of $1$ (open circle), $3$ (open left triangle), $5$ (open square), and $14$ Gyrs (open triangle) are marked for convenience. }\protect\label{SEMs}
\end{center}
\end{figure*}

\noindent In the $B-V$ column of Fig.~\ref{clrhistmosaic} the total colors show a seemingly bi--modal distribution, which separates the sample essentially in two equal halves. For the sake of simplicity we will refer to these ``blue'' and ``red'' galaxies as $G^B$ and $G^R$, respectively. These are listed in Table~\ref{bimod}. This apparent separation is curious, because even though the counts are small, the bin size is larger than or equal to the relevant color errors for both groups. We do not know what the actual distribution would look like, it may be flat or even Gaussian -- a Shapiro-Wilk normality test gives a p--value $=0.18$ which does not reject the null hypothesis at a $\alpha=0.1$ significance level for our sample size. Additionally, we do not see this ``bimodality'' in any other filter and hence $G^B$ and $G^R$ are unlikely to belong to a true bimodal distribution. Nevertheless, $G^B$ and $G^R$ are clearly separated in our data and we investigate their behavior further. The separation between blue and red BGCs is preserved in the histogram of the central colors down to $\mu_B=24$ mag arcsec${}^{-2}$ (Figure~\ref{clrhistmosaic}, second row), albeit with more smoothed peaks. Such behavior is expected if the separation is due to differences in the young population. Both $G^B$ and $G^R$ shift towards bluer colors in this region due to the diminished host contribution. The same is true for the color in the $24\lesssim\mu_B\lesssim26$ mag arcsec${}^{-2}$ region, showing that the burst contamination to that region is still significant. In the $26\lesssim\mu_B\lesssim28$ mag arcsec${}^{-2}$ region we expect that we are sampling exclusively host--dominated regions, and indeed, the trend is destroyed and we observe a redwards skewed distribution instead.\\

\noindent In Figure~\ref{comlementhist} we examine the behavior of the $G^B$ and $G^R$ galaxies in terms of various other properties. For the remainder of this subsection we exclude the two spiral galaxies, UM477 and UM499, since they are not BCGs, or BCG--like. With the exception of a few outliers, $B-V$ shows a positive correlation with all other colors in the sense that $G^B$ is on average bluer than $G^R$ in every color. In terms of total galaxy luminosity, scale length, and central surface brightness both $G^B$ and $G^R$ behave in a similar fashion. From the literature we have further investigated the metallicity, \HI mass, and $H\beta$ equivalent widths ($EW(H\beta)$) for the galaxies in $G^B$ and $G^R$, and find them similar, with both groups containing objects on either end of the extremes. We also considered a possible difference in inclination, with the ellipticity as proxy but, again, saw no significant differences. One defining difference between the blue and red BCGs is found when analyzing their typical morphology. All $G^B$ galaxies have multiple star forming regions, highly irregular inner isophotes and only vaguely elliptical outer isophotes. The $G^R$ galaxies, with the exception of UM461 and UM500, have predominantly nuclear star forming regions or a single off--centered compact star forming knot. The outliers UM461 and UM500 both have multiple star forming knots and have been marked with black--red diamonds in Figure~\ref{comlementhist}. A look at Table~\ref{totlumtbl} reveals that these two galaxies are red only in $B-V$, while in every other color they fall bluewards of the sample median. This is likely due to a nebular emission contribution, which would make $B-V$ redder due to the much stronger $[OIII]~\lambda 5007$ \AA~line in the $V$ band (typical filter transmission $T>70\%$), compared to $H_\beta~\lambda 4861$ \AA~in the $B$ band ($T\sim25\%$). High nebular emission in both of these galaxies is further indicated by their very high $EW(H\beta)$ (Figure~\ref{comlementhist}, bottom row, middle panel).\\

\noindent Other than morphology of the star formation regions differences in the strength of the star formation itself should contribute to the separation between blue and red BCGs. We would naturally expect the blue BCGs to have stronger star formation. However, the estimate of the relative starburst strength fails with our method more often for $G^B$ than for $G^R$, as is to be expected from the easier--to--fit surface brightness profiles of the nuclear starbursts. We therefore do not have enough galaxies populating a histogram of the relative starburst contribution for $G^B$ to be able to tell if this group is also defined by more dominating star formation than $G^R$. Differences in extinction in blue and red BCGs may also enhance the separation between them. A quick check of $H\alpha/H\beta$ ratios from SDSS spectra (bottom row, right panel of Figure~\ref{comlementhist}) does indeed suggest a larger extinction for the $G^R$ group, albeit this is based on very few $G^R$ data points. The $U-B$ color is very sensitive to extinction, and should reflect extinction differences between the red and blue BCGs. Aside from a few outliers, the $B-V$ vs. $U-B$ plot in Figure~\ref{comlementhist} is consistent with $G^R$ having larger extinction. $G^B$ galaxies, on the other hand, all have similar total $U-B$ colors, which implies a similar age of the young population for these targets. Since this is a {\em total} color, i.e. integrated over the entire galaxy, it is a mixture of young and old stars, as well as gas. It is curious that the different components for the individual $G^B$ galaxies all conspire to produce similar total $U-B$ colors. We will come back to $G^B$ and $G^R$ when we examine the asymmetry in \S~\ref{asymdiscuss}. \\
\begin{figure}
\begin{center}
  \includegraphics[width=8cm]{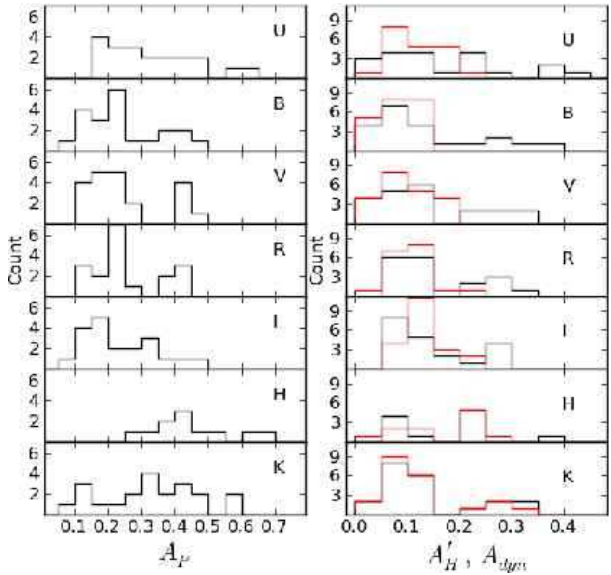}
  \caption{\textbf{Left:} distribution of $A_{P}$ asymmetry obtained over the area enclosed by the Petrosian radius $r[\eta(0.2)]$ for each galaxy and filter. \textbf{Right:} $A_{H}^\prime$ (black) over the area enclosed by the Holmberg radius, smoothed with $1\times1$ kpc box. Overplotted in red is $A_{dyn}$ asymmetry, obtained over the area enclosed by $R_{27}$ in the optical and $R_{23}$ in the NIR, with star forming knots set to a constant flux value. The bin size is $0.05$ in both panels.}\protect\label{asymhist}
\end{center}
\end{figure}
\subsection{Stellar evolutionary models}\protect\label{semsdiscuss}
\noindent In Figure~\ref{SEMs} we compare the $B-V$ vs $V-K$ colors, measured over different physical regions of the galaxies, with the predictions from stellar evolutionary models. The model tracks with nebular emission (left column) assume zero redshift and instant burst, and are based on the \emph{Yggdrasil} spectral synthesis code \citep{2011ApJ...740...13Z}, whereas the pure stellar population tracks (right column) are based on \citet{2008A&A...482..883M} isochrones, are also at zero redshift, and have an exponential SFR decay of $1$ Gyr. The total colors (first row) for about half the sample are redwards of $V-K\sim2$, having a low to moderate contribution from nebular emission coming from the star forming regions. The rest of the galaxies bluewards of $V-K\sim2$ have total $V-K$ colors clearly incompatible with those of a pure stellar population and may suffer the effects of nebular emission contamination and dust extinction. Examining only the burst color estimate down to $\mu_B=24$ mag arcsec${}^{-2}$ we see that they are better fitted with nebular emission contribution, as expected. Most data points fall on or close to the model tracks, though some clearly are incompatible with this instant burst model, and would be better modeled with an extended burst instead. The fit can only get better when we account for the difference in used filters and the non-zero redshift of the sample. The very red $V-K$ outlier in that plot (Figure~\ref{SEMs}, left column, second row) is UM422, or rather the composite galaxy together with the very extended neighbor. There are no visible blue knots of star formation inside the region $\mu_B\lesssim24$ mag arcsec${}^{-2}$ for that galaxy, so we expect the burst estimate to be very inaccurate for this target due to its morphology and profile shape, and the fact that we are not actually measuring a star forming region. The region $24\lesssim\mu_B\lesssim26$ mag arcsec${}^{-2}$ shows colors similar to the total colors in the first row of Figure~\ref{SEMs}, which implies that either the starburst is not dominating the total galaxy colors, or that the $24\lesssim\mu_B\lesssim26$ region is still very much strongly contaminated by the burst contribution. The latter is not likely since we observe continuous smooth profile slopes along the entire $24\lesssim\mu_B\lesssim28$ region, and also because we have established that the relative burst contribution is on average moderate. Further, from the predominantly flat color profiles we observe in Figure~\ref{datafig} we would not expect a big change in the integrated color at larger radii. This is unfortunately difficult to see, since the $26\lesssim\mu_B\lesssim28$ mag arcsec${}^{-2}$ region contains fewer measurements due to the limiting effect of the $K$ band. For the same reason, the errorbars here are large. Nevertheless, the location of the data points is here better fitted with the pure stellar population model, with some very metal--poor ($Z\sim0.001$) and some very metal--rich $Z\sim0.02$ hosts older than a few Gyr. The three outliers close to the solar--metallicity track beyond $V-K\sim3$ are UM446, UM465, and UM499. Solar metallicity is not unusual for spiral galaxies, hence UM499 is not truly a deviating data point. With the help of $H\alpha$ data in~\citet{Paper3} we will be able to discern whether there is any nebular emission contribution at such radii for these three galaxies, and hence whether one should compare them to the left or right panel tracks in Figure~\ref{SEMs}. For now we can conclude that the case for a so called ``red halo''~\citep{2002A&A...390..891B,2005mmgf.conf..355B,2006ApJ...650..812Z} in this sample of emission line galaxies is weak, though if taken at face value the stellar evolutionary model tracks imply unusually high metallicities for some hosts in the sample. \\
\begin{table*}
  \begin{minipage}{150mm}
    \caption{$A_{P}$ asymmetry in each filter measured inside the Petrosian radius $r[\eta(0.2)]$, given here in kpc. In the optical the $S/N$ within the Petrosian radius is significantly high so that the typical errors are $\ll0.02$ (rms), while in the NIR the $S/N$ often drops to low values of $\lesssim400$ within $r[\eta(0.2)]$, which gives typical errors of $\sim0.05$. These typical errors have been estimated by~\citet{2000ApJ...529..886C}.}
    \protect\label{asymtbl}
    \begin{tabular}{|lccccccccccccccc|}
      \hline
      Galaxy&$A_U$&$r_U$&$A_B$&$r_B$&$A_V$&$r_V$&$A_R$&$r_R$&$A_I$&$r_I$&$A_H$&$r_H$&$A_K$&$r_K$\\\hline
      UM422&$0.58$&$7.7$&$0.43$&$7.7$&$0.44$&$7.6$&$0.39$&$7.2$&$0.40$&$7.4$&$0.43$&$7.2$&$0.55$&$5.4$\\
      UM439&$0.26$&$1.2$&$0.26$&$1.3$&$0.27$&$1.3$&$0.22$&$1.5$&$0.21$&$1.7$&&&$0.28$&$1.6$\\
      UM446&&&$0.11$&$0.7$&$0.11$&$0.7$&$0.13$&$0.9$&$0.11$&$1.0$&$0.40$&$1.0$&$0.32$&$1.0$\\
      UM452&$0.27$&$1.5$&$0.24$&$1.7$&$0.23$&$1.8$&$0.24$&$1.8$&$0.23$&$2.0$&$0.31$&$2.0$&$0.27$&$2.0$\\
      UM456&$0.36$&$1.9$&$0.39$&$2.1$&$0.43$&$2.4$&$0.40$&$3.0$&$0.35$&$3.4$&$0.55$&$2.4$&$0.50$&$2.6$\\
      UM461&$0.45$&$0.7$&$0.42$&$0.7$&$0.47$&$0.7$&$0.44$&$0.7$&$0.35$&$0.9$&$0.48$&$0.9$&$0.49$&$0.8$\\
      UM462&$0.16$&$0.9$&$0.22$&$0.9$&$0.23$&$0.9$&$0.23$&$1.1$&$0.28$&$1.1$&$0.42$&$1.2$&$0.34$&$1.2$\\
      UM463&$0.17$&$0.5$&$0.15$&$0.5$&$0.20$&$0.5$&&&$0.13$&$0.6$&$0.35$&$0.6$&$0.37$&$0.7$\\
      UM465&$0.19$&$0.3$&$0.09$&$0.6$&$0.16$&$0.8$&&&&&&&$0.14$&$1.2$\\
      UM477&$0.37$&$0.6$&$0.22$&$9.3$&$0.19$&$8.0$&$0.21$&$7.4$&$0.15$&$7.3$&&&$0.30$&$0.8$\\
      UM483&$0.17$&$1.1$&$0.14$&$1.2$&$0.12$&$1.2$&$0.12$&$1.2$&$0.13$&$1.2$&&&$0.15$&$1.2$\\
      UM491&$0.23$&$1.1$&$0.22$&$1.2$&$0.21$&$1.2$&$0.20$&$1.4$&$0.19$&$1.4$&&&$0.20$&$1.4$\\
      UM499&$0.20$&$1.7$&$0.18$&$2.2$&$0.14$&$6.1$&&&$0.08$&$6.3$&&&$0.09$&$3.5$\\
      UM500&$0.63$&$4.7$&$0.36$&$4.7$&$0.42$&$4.7$&$0.41$&$4.7$&$0.31$&$4.9$&$0.45$&$4.7$&$0.42$&$4.7$\\
      UM501&$0.49$&$3.0$&$0.46$&$3.3$&$0.43$&$3.0$&$0.44$&$3.8$&$0.47$&$4.1$&$0.65$&$2.4$&$0.59$&$3.5$\\
      UM504&$0.20$&$0.6$&$0.13$&$0.7$&$0.14$&$0.7$&$0.11$&$0.8$&$0.10$&$0.8$&$0.30$&$0.8$&$0.13$&$0.8$\\
      UM523A&$0.48$&$2.6$&$0.33$&$3.5$&$0.29$&$3.5$&$0.27$&$3.6$&$0.26$&$3.5$&&&$0.40$&$3.5$\\
      UM523B&$0.30$&$1.7$&$0.18$&$1.7$&$0.17$&$1.7$&$0.15$&$1.7$&$0.15$&$1.7$&&&$0.24$&$1.8$\\
      UM533&$0.41$&$1.5$&$0.22$&$2.1$&$0.21$&$2.4$&$0.22$&$2.6$&$0.16$&$2.9$&&&$0.41$&$2.6$\\
      UM538&$0.28$&$0.5$&$0.23$&$0.7$&$0.22$&$0.8$&$0.19$&$0.9$&$0.19$&$0.9$&&&$0.37$&$0.8$\\
      UM559&$0.31$&$2.5$&$0.17$&$2.4$&$0.15$&$2.4$&$0.23$&$2.5$&$0.34$&$2.5$&$0.60$&$2.7$&$0.33$&$2.4$\\\hline
      \hline
    \end{tabular}
  \end{minipage}
\end{table*}
\subsection{Asymmetries}\protect\label{asymdiscuss}
\noindent In Table~\ref{asymtbl} and Figure~\ref{asymhist} (left panel) we present the distribution of the Petrosian $A_{P}$ asymmetries measured in each filter down to the Petrosian radius $r[\eta(0.2)]$. Since the sample consists exclusively of emission line galaxies, the composite asymmetry of a galaxy is usually dominated by the flocculent asymmetry, where we use the distinction ``flocculent'' and ``dynamical'' asymmetry as defined by~\citet{2000ApJ...529..886C}. This domination is further enhanced by our choice of the area over which the asymmetry is measured -- Table~\ref{asymtbl} shows that the Petrosian $r[\eta(0.2)]$ radius is usually quite small, and hence the enclosed area is limited to fairly bright surface brightness levels. This implies that for many galaxies the dominating component to the total asymmetry (flocculent plus dynamical) will be the asymmetry due to individual star forming knots, i.e. the flocculent asymmetry, best estimated by $A_P$. The dynamical asymmetry contribution is underestimated in this way, because the tidal tails and plumes usually associated with mergers and strong tidal interactions, can be very faint and will thus lie beyond the $r[\eta(0.2)]$ radius. Even if faint features were included their contribution would be negligible since the asymmetry is luminosity weighted. In an attempt to obtain a better estimate of the dynamical asymmetry component in \M~we examined alternative asymmetry measurements, such as the ``Holmberg'' $A_H^\prime$ asymmetry and the purely dynamical $A_{dyn}$ asymmetry. We will come back to these later on but first let us examine in detail the behavior of the Petrosian $A_P$ asymmetry.\\
\begin{figure*}
\begin{center}
  \includegraphics[width=18.0cm]{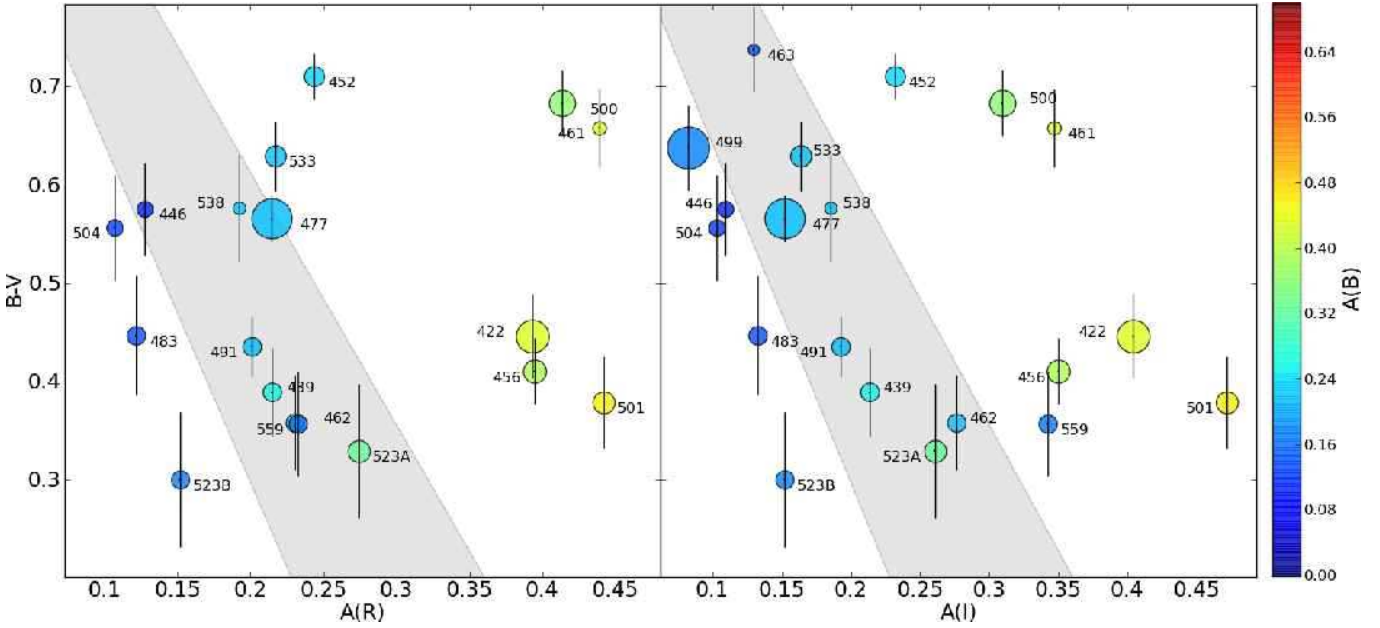}
  \caption{$B-V$ total color (Table~\ref{totlumtbl}) vs. the $A_{P}$ asymmetry in the $R$ (left panel), and the $I$ bands (right panel). The markers are color--coded by their $B$ band asymmetry, while their size reflects their Holmberg radius $r_H$. The gray area is the location of the fiducial galaxy color--asymmetry sequence as defined in~\citet{2000ApJ...529..886C}.}\protect\label{conselice}
\end{center}
\end{figure*}
\begin{table}
  \caption{Minimum asymmetries measured in each filter. The two numbers per filter per galaxy are Holmberg $A_H^\prime$ asymmetry measured over the area enclosed by the Holmberg radius $r(\mu=26.5)$ in the optical and by $r(\mu=23)$ in the NIR (top value), and the dynamical $A_{dyn}$ asymmetry, with regions $\mu<25$ ($\mu<21$) set to $25$ ($21$) mag arcsec${}^{-2}$ in the optical (NIR) (bottom value). The images are pre--processed by a boxcar average of size $1\times1$ kpc.}
  \protect\label{tab:altasym}
  \begin{tabular}{|l|r|r|r|r|r|r|r|}
    \hline
    Galaxy&$A_U$&$A_B$&$A_V$&$A_R$&$A_I$&$A_H$&$A_K$\\
    \hline
    UM422 &$ 0.41$ &$ 0.36$ &$ 0.34$ &$ 0.30$ &$ 0.28$ &$ 0.25$ &$ 0.32$\\
    &$ 0.18$ &$ 0.15$ &$ 0.15$ &$ 0.13$ &$ 0.12$ &$ 0.23$ &$ 0.29$\\
    UM439 &$ 0.09$ &$ 0.09$ &$ 0.10$ &$ 0.15$ &$ 0.18$ &  &$ 0.10$\\
    &$ 0.13$ &$ 0.12$ &$ 0.16$ &$ 0.21$ &$ 0.21$ &  &$ 0.11$\\
    UM446 &  &$ 0.05$ &$ 0.05$ &$ 0.05$ &$ 0.06$ &$ 0.08$ &$ 0.10$\\
    &  &$ 0.03$ &$ 0.05$ &$ 0.08$ &$ 0.11$ &$ 0.10$ &$ 0.12$\\
    UM452 &$ 0.10$ &$ 0.11$ &$ 0.11$ &$ 0.11$ &$ 0.11$ &$ 0.21$ &$ 0.12$\\
    &$ 0.10$ &$ 0.05$ &$ 0.08$ &$ 0.08$ &$ 0.10$ &$ 0.24$ &$ 0.11$\\
    UM456 &$ 0.23$ &$ 0.28$ &$ 0.30$ &$ 0.28$ &$ 0.26$ &$ 0.35$ &$ 0.35$\\
    &$ 0.11$ &$ 0.09$ &$ 0.08$ &$ 0.10$ &$ 0.13$ &$ 0.27$ &$ 0.32$\\
    UM461 &$ 0.20$ &$ 0.20$ &$ 0.23$ &$ 0.20$ &$ 0.16$ &$ 0.09$ &$ 0.11$\\
    &$ 0.09$ &$ 0.08$ &$ 0.08$ &$ 0.10$ &$ 0.10$ &$ 0.09$ &$ 0.10$\\
    UM462 &$ 0.07$ &$ 0.09$ &$ 0.10$ &$ 0.10$ &$ 0.13$ &$ 0.10$ &$ 0.14$\\
    &$ 0.10$ &$ 0.06$ &$ 0.08$ &$ 0.10$ &$ 0.12$ &$ 0.08$ &$ 0.08$\\
    UM463 &$ 0.04$ &$ 0.02$ &$ 0.04$ &  &$ 0.09$ &$ 0.11$ &$ 0.08$\\
    &$ 0.06$ &$ 0.02$ &$ 0.05$ &  &$ 0.06$ &$ 0.14$ &$ 0.06$\\
    UM465 &$ 0.04$ &$ 0.04$ &$ 0.04$ &  &  &  &$ 0.04$\\
    &$ 0.06$ &$ 0.05$ &$ 0.05$ &  &  &  &$ 0.05$\\
    UM477 &$ 0.20$ &$ 0.14$ &$ 0.12$ &$ 0.11$ &$ 0.10$ &  &$ 0.08$\\
    &$ 0.17$ &$ 0.11$ &$ 0.10$ &$ 0.12$ &$ 0.11$ &  &$ 0.09$\\
    UM483 &$ 0.08$ &$ 0.07$ &$ 0.07$ &$ 0.08$ &$ 0.07$ &  &$ 0.08$\\
    &$ 0.08$ &$ 0.07$ &$ 0.16$ &$ 0.12$ &$ 0.15$ &  &$ 0.07$\\
    UM491 &$ 0.12$ &$ 0.12$ &$ 0.12$ &$ 0.11$ &$ 0.11$ &  &$ 0.10$\\
    &$ 0.08$ &$ 0.06$ &$ 0.07$ &$ 0.11$ &$ 0.12$ &  &$ 0.06$\\
    UM499 &$ 0.12$ &$ 0.09$ &$ 0.08$ &  &$ 0.07$ &  &$ 0.08$\\
    &$ 0.15$ &$ 0.11$ &$ 0.18$ &  &$ 0.14$ &  &$ 0.14$\\
    UM500 &$ 0.36$ &$ 0.28$ &$ 0.29$ &$ 0.27$ &$ 0.25$ &$ 0.25$ &$ 0.26$\\
    &$ 0.11$ &$ 0.11$ &$ 0.11$ &$ 0.11$ &$ 0.12$ &$ 0.21$ &$ 0.22$\\
    UM501 &$ 0.37$ &$ 0.34$ &$ 0.31$ &$ 0.30$ &$ 0.29$ &$ 0.25$ &$ 0.26$\\
    &$ 0.13$ &$ 0.11$ &$ 0.11$ &$ 0.19$ &$ 0.24$ &$ 0.25$ &$ 0.26$\\
    UM504 &$ 0.04$ &$ 0.05$ &$ 0.04$ &$ 0.05$ &$ 0.06$ &$ 0.06$ &$ 0.07$\\
    &$ 0.08$ &$ 0.03$ &$ 0.03$ &$ 0.05$ &$ 0.19$ &$ 0.05$ &$ 0.06$\\
    UM523A &$ 0.27$ &$ 0.24$ &$ 0.22$ &$ 0.21$ &$ 0.21$ &  &$ 0.22$\\
    &$ 0.21$ &$ 0.10$ &$ 0.09$ &$ 0.13$ &$ 0.15$ &  &$ 0.13$\\
    UM523B &$ 0.21$ &$ 0.10$ &$ 0.12$ &$ 0.10$ &$ 0.12$ &  &$ 0.10$\\
    &$ 0.18$ &$ 0.14$ &$ 0.13$ &$ 0.13$ &$ 0.11$ &  &$ 0.12$\\
    UM533 &$ 0.13$ &$ 0.08$ &$ 0.08$ &$ 0.08$ &$ 0.06$ &  &$ 0.09$\\
    &$ 0.06$ &$ 0.03$ &$ 0.03$ &$ 0.06$ &$ 0.08$ &  &$ 0.06$\\
    UM538 &$ 0.08$ &$ 0.06$ &$ 0.08$ &$ 0.06$ &$ 0.05$ &  &$ 0.04$\\
    &$ 0.03$ &$ 0.04$ &$ 0.09$ &$ 0.07$ &$ 0.07$ &  &$ 0.04$\\
    UM559 &$ 0.20$ &$ 0.12$ &$ 0.12$ &$ 0.11$ &$ 0.12$ &$ 0.21$ &$ 0.09$\\
    &$ 0.16$ &$ 0.12$ &$ 0.12$ &$ 0.13$ &$ 0.20$ &$ 0.21$ &$ 0.09$\\
    \hline
  \end{tabular}
\end{table}
\begin{table}
     \caption{Concentration parameter for each filter}
    \protect\label{tab:conc}
    \begin{tabular}{@{}|l|r|r|r|r|r|r|r|r|@{}}
      \hline
      Galaxy&$C_U$&$C_B$&$C_V$&$C_R$&$C_I$&$C_H$&$C_K$\\\hline
      UM422&$2.2$&$2.5$&$2.6$&$2.6$&$2.7$&$2.8$&$2.4$\\
      UM439&$2.2$&$2.2$&$2.4$&$2.7$&$2.7$&&$2.7$\\
      UM446&&$2.4$&$2.4$&$2.7$&$3.0$&$3.0$&$2.1$\\
      UM452&$2.2$&$2.6$&$2.7$&$2.7$&$2.9$&$2.9$&$2.9$\\
      UM456&$2.4$&$2.9$&$3.0$&$3.2$&$3.3$&$2.7$&$3.2$\\
      UM461&$1.8$&$1.8$&$1.5$&$2.1$&$2.4$&$2.4$&$2.2$\\
      UM462&$2.4$&$2.6$&$2.6$&$2.4$&$2.4$&$2.6$&$2.6$\\
      UM463&$1.5$&$1.5$&$1.5$&&$2.0$&$2.0$&$2.0$\\
      UM465&$1.5$&$2.1$&$2.8$&&&&$3.5$\\
      UM477&$2.4$&$2.7$&$3.0$&$3.2$&$3.2$&&$3.3$\\
      UM483&$1.8$&$2.1$&$2.1$&$2.1$&$2.1$&&$2.1$\\
      UM491&$1.8$&$2.1$&$2.4$&$2.4$&$2.4$&&$2.4$\\
      UM499&$2.4$&$2.6$&$4.0$&&$3.7$&&$3.2$\\
      UM500&$2.4$&$2.6$&$2.4$&$2.7$&$2.4$&$2.6$&$2.3$\\
      UM501&$3.0$&$2.7$&$2.7$&$2.9$&$2.7$&$2.5$&$2.8$\\
      UM504&$1.5$&$2.0$&$2.4$&$2.7$&$2.7$&$2.7$&$2.7$\\
      UM523A&$3.3$&$2.7$&$2.6$&$2.6$&$2.6$&&$2.5$\\
      UM523B&$2.8$&$2.7$&$2.8$&$2.4$&$2.5$&&$2.5$\\
      UM533&$2.4$&$2.8$&$2.8$&$2.9$&$2.8$&&$2.8$\\
      UM538&$2.4$&$2.4$&$2.6$&$2.8$&$2.8$&&$2.6$\\
      UM559&$2.2$&$2.1$&$2.4$&$2.5$&$2.5$&$2.7$&$2.5$\\
      \hline
    \end{tabular}
\end{table}

\noindent This is a volume limited sample with fairly similar redshifts (Table~\ref{nedtable}). The difference in distance to the galaxies varies at most by a factor of $2$, and hence differences in measured asymmetry values are not due to simple resolution effects. All optical filters show a strongly peaked distribution at small $A_{P}\sim0.2$, which we expect since the minimum asymmetry is usually found in the starbursting knots, regardless of their location in the galaxy. There are, of course, exceptions to this as is the case for galaxies with starforming regions of comparable brightness located symmetrically on either side of the geometric center, e.g. $UM559$, $UM483$ in this sample. For most galaxies, the physical location of the minimum asymmetry does not change with filter, however the value of the asymmetry does so, with the most drastic change observed when going from optical to NIR filters. The $H$ and $K$ histograms both show a distinct shift of the majority of the targets to much higher asymmetry values, $A_{P}\sim0.4$, than in the optical. This change in asymmetry from optical to NIR reflects the change between flocculent to dynamical domination in the integrated asymmetry value. In the NIR the contribution to the light output of the galaxy from old stars is significant, while the burst is of diminished importance. Hence, we expect the NIR asymmetry to be a reflection of the departure from the symmetric ground state due to dynamical effects, such as merging or tidal interactions. We can distinguish three groups of galaxies based on their $A_{P}$ asymmetry behavior in the optical and NIR.
\begin{itemize}
\item \textbf{Small optical and small NIR $A_{P}$ asymmetry:} These are predominantly members from the \emph{nE} BCG class, have regular isophotes, and include all the nuclear starbursts. Thus, they have small flocculent and small dynamical asymmetries. They are $UM439$, $UM452$, $UM465$, $UM477$, $UM483$, $UM491$, $UM499$, and $UM504$.
\item \textbf{Small optical and large NIR $A_{P}$ asymmetry:} This group is characterized by having a spatially extended burst region, or multiple SF knots off--center. The host is otherwise regular at faint isophotes, i.e. the dominant morphological classes here \emph{iE} and \emph{SS} BCGs. These are UM446, UM462, UM463, UM523A, UM523B, UM533, UM538, and UM559. This group has small flocculent and large dynamical asymmetries. 
\item \textbf{Large optical and large NIR $A_{P}$ asymmetry:} This group contains all galaxies with highly irregular morphologies and/or numerous SF knots. These are UM422, UM456, UM461, UM500, and UM501. Note that all targets classified as mergers (\emph{iI}) are found here. This group has large flocculent and large dynamical asymmetries.
\end{itemize}
\noindent Interestingly, we found no galaxies with large optical and small NIR $A_P$ asymmetries in this sample. If we use optical asymmetry as a proxy for the flocculent asymmetry component, and NIR asymmetry as a proxy for the dynamical asymmetry component, then such a combination would imply a galaxy in or close to the symmetric ground state but with a non-nuclear starforming region. This is difficult to achieve because any compact localized off-center star formation presumably would require some sort of tidal interaction or merger to trigger, which in turn would raise the dynamical asymmetry value. While possible, such morphological setup is obviously rare. \\
\begin{figure}
  \begin{center}
    \includegraphics[width=8cm]{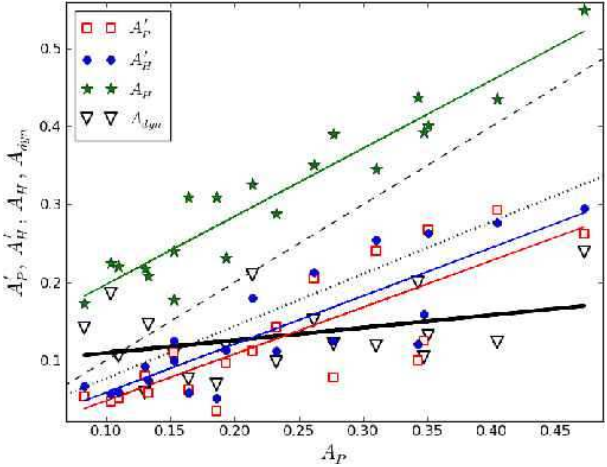}
    \caption{Petrosian asymmetry ($A_P$) vs. all alternative asymmetry measurements. $A_H$ is the Holmberg asymmetry without any smoothing of the images, while a $1\times1$ kpc smoothing box has been applied to $A_P^\prime$ (Petrosian), $A_H^\prime$ (Holmberg), and dynamical $A_{dyn}$ asymmetry. The uninterrupted lines are least square fits to the similarly colored data points. The dashed line has a slope of $1$. The dotted line is an extrapolation from the correlation of~\citet{2003ApJS..147....1C} for E, S0, Sa--b, Sc--d and Irr, which lie beyond $A_P\sim0.6$. All asymmetries here are from the $I$ band, in order to facilitate comparison with~\citet{2003ApJS..147....1C}.}\protect\label{allasymmetries}
  \end{center}
\end{figure}

\noindent There is a known correlation between the (blue) color and the (red) $A_P$ asymmetry for spheroids, disks and irregulars~\citep{2000ApJ...529..886C}, and in Figure~\ref{conselice} we compare the total $B-V$ color of the galaxies in the sample to the $R$ and $I$ asymmetries. We have estimated the region of the fiducial galaxy color-asymmetry sequence by~\citet{2000ApJ...529..886C}, and plotted it for comparison. Galaxies deviating from this sequence are too asymmetric for their observed color, which is an indication of a merger or ongoing interaction. All of the potential mergers in the group with large optical and NIR asymmetries are indeed located to the right of the fiducial line. Star formation alone cannot account for their measured asymmetries, and a boost to the asymmetry by dynamical processes is needed. We note that our previous suspicions about the merger nature of UM500 are reinforced in this figure, with UM500 falling clearly to the far right of the fiducial line. We find this convincing and now firmly classify UM500 as a \emph{iI,M} BCG. This means that the galaxy group with large optical and large NIR $A_P$ asymmetry now exclusively captures all classified mergers in our sample.\\

\noindent The fact that the Petrosian $A_P$ asymmetry varies with wavelength regime is a strong indication that it is not completely flocculent dominated as was the case in \M. If this is the case it should be also evident from the behavior of the alternative asymmetry measurements shown in Table~\ref{tab:altasym} and described in \S~\ref{casparam}. The right panel of Figure~\ref{asymhist} shows the distribution of the Holmberg $A_H^\prime$ and the dynamical $A_{dyn}$ asymmetries. The $A_H^\prime$ distribution differs significantly from the $A_P$ values. The grouping by optical/NIR asymmetry we presented above is now destroyed, and only two major groups emerge -- one with small ($\lesssim0.2$) optical and small NIR $A_H^\prime$ which contains the majority of the galaxies, and one with large ($\gtrsim0.2$) optical and large NIR $A_H^\prime$, which contains UM422, UM456, UM500, UM501, and UM523A. The fact that $A_H^\prime$ does not follow the same distribution as $A_P$ implies that, contrary to \M, the latter is not completely flocculent dominated and the contribution of the dynamical component to $A_P$ is not insignificant. We also note that the $K$ band $A_H^\prime$ asymmetry is nearly identical to the dynamical $A_{dyn}$ asymmetry distribution. The same is true for the $H$ band but is better seen comparing the $A_H^\prime$ and $A_{dyn}$ values in Table~\ref{tab:altasym}. This confirms our assumption that the NIR asymmetry is a very good proxy of the dynamical asymmetry component. Further, in the $A_{dyn}$ asymmetry distribution there are essentially no galaxies with large ($\gtrsim0.2$) optical asymmetries. This is consistent with the optical asymmetry being flocculent dominated, a dominance which is effectively neutralized through our method of obtaining $A_{dyn}$. In terms of morphological class we would expect $A_{dyn}$ to only be able to distinguish mergers from the rest. Indeed, the galaxies here fall into two groups, the majority having small ($\lesssim0.2$) optical and NIR $A_{dyn}$ asymmetries. The exceptions are UM422, UM456, UM500, and UM501, all of which are mergers, and have optical $A_{dyn}\sim0.1$ and a much larger NIR $A_{dyn}\sim0.3$ (averaged values). In other words, our measure of the purely dynamical component, $A_{dyn}$, successfully represents the effect mergers have on the morphology of a galaxy. Note that $A_H^\prime$ performs nearly as well in distinguishing mergers from non-mergers (one false positive notwithstanding), and is therefore an acceptable measure of the dynamical asymmetry for the low luminosity BCGs in our sample.\\

\noindent In Figure~\ref{allasymmetries} we plot these alternative asymmetry measurements versus the Petrosian $A_P$ asymmetry. Similar to \M, we find a strong correlation between the smoothed and unsmoothed Holmberg ($A_H^\prime,~A_H$) and Petrosian ($A_P^\prime,~A_P$) asymmetries. The correlation coefficients for $A_P$ vs $A_P^\prime,~A_H^\prime,~A_H$ are $R=0.8,~0.9,~0.9$, with line slopes $\mathfrak{m}=0.60,~0.62,~0.87$ respectively. Contrary to \M~we also find a medium strength correlation between $A_P$ (predominantly flocculent) and $A_{dyn}$ (dynamical) asymmetries. Pearson's $R$ for this correlation is $\sim0.4$, which is only marginally significant for a sample of this size. This is consistent with our findings for the BCG sample of \M.\\
\begin{figure}
\begin{center}
  \includegraphics[width=8cm]{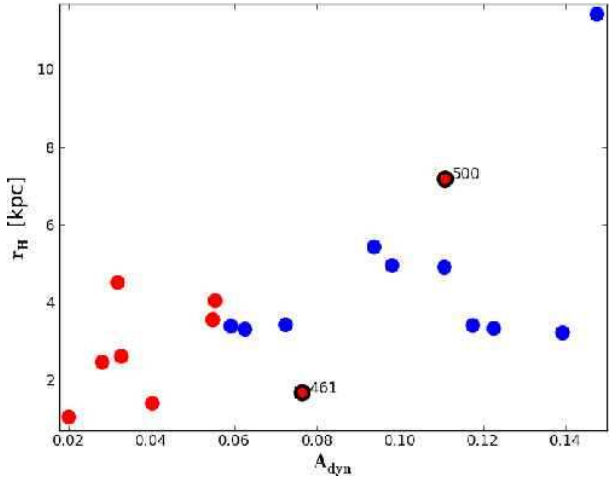}
  \caption{$G^B$ (blue circles) and $G^R$ (red circles) in a Holmberg radius ($r_H$) vs. dynamical $B$ band asymmetry ($A_{dyn}$) parameter space. The black/red circles are the deviating $G^R$ galaxies UM461 and UM500. The two spiral galaxies are not included in the plot.}\protect\label{holmdynA}
\end{center}
\end{figure}

\noindent Figure~\ref{holmdynA} shows the behavior of the blue $G^B$ and red $G^R$ BCGs (defined in \S~\ref{colortrends}) in terms of the dynamical asymmetry component $A_{dyn}$ and Holmberg radius, $r_H$. There is only a weak trend for the $G^B$ galaxies to have larger Holmberg radii and hence to be more extended but clearly they have higher dynamical asymmetries than the $G^R$ galaxies. While we established in \S~\ref{colortrends} that the star formation morphologies between blue and red BCGs clearly vary, the dynamical asymmetries suggest that also the morphologies of the underlying hosts of blue and red BCGs are different. As we already saw in the preceding discussion some of the $G^B$ galaxies are clearly mergers or show signs of strong tidal interaction. Other members of the $G^B$ group, e.g. UM491 or UM483, seem unlikely merger candidates yet they, too, display high dynamical asymmetries and blue colors. It is possible that the dynamical asymmetry component allows us to identify not only the obvious major mergers but also the more ``mature'' or minor ones. For example, the location and morphology of the star forming regions in UM491 or UM483 are suggestive of a long passed dynamical disturbance. We will investigate the ability of $A_{dyn}$ to detect minor mergers in detail in~\citet{Paper3}. \\

\begin{figure*}
  \begin{center}
    \includegraphics[width=17cm,height=8cm]{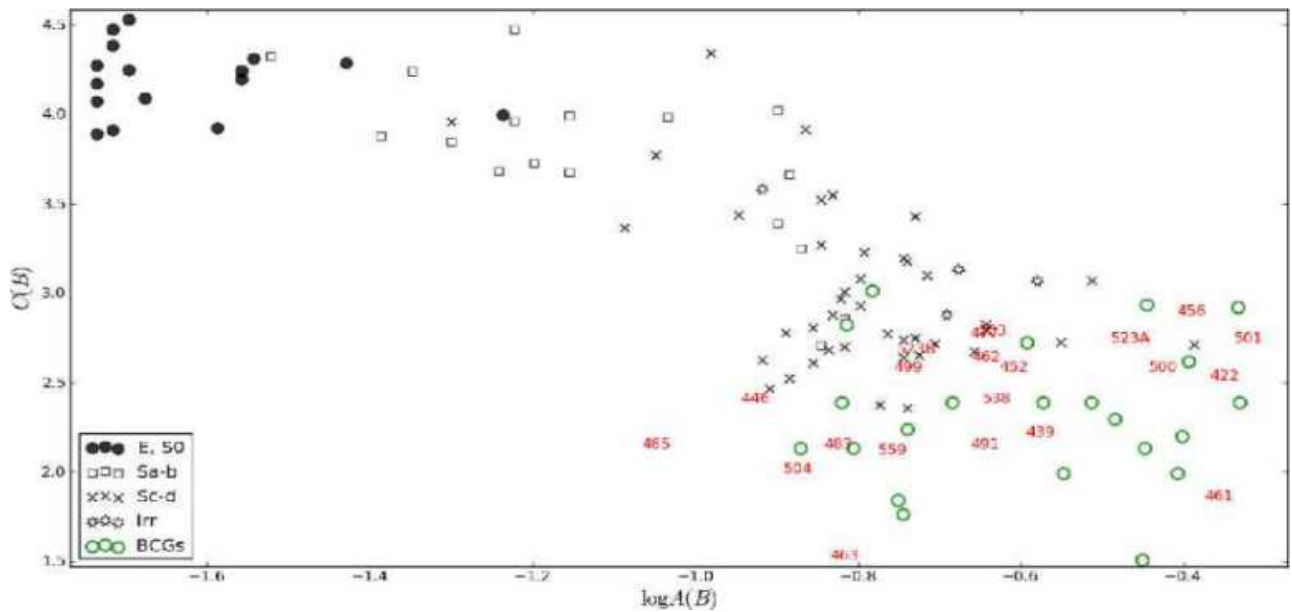}
    \caption{Concentration vs Petrosian $A_P$ Asymmetry for our sample (red text) compared to spheroids, early and late type disks, and irregulars taken from~\citet{2000ApJ...529..886C} and the BCG galaxies from \M. To make all labels as readable as possible we omitted the SBS0335--052E measurement from \M. The red text indicates the UM number of the respective galaxy.}\protect\label{concVSasym}
  \end{center}
\end{figure*}

\noindent The galaxies in our sample are clearly separated from spheroids and early and late type disks in the concentration--asymmetry parameter space (Figure~\ref{concVSasym}). They occupy the same region as the BCGs in \M, with large asymmetries and small concentration indices. Note that BCGs show significant scatter in the concentration--asymmetry parameter space compared to normal galaxies.
\section[]{Conclusions}\protect\label{conclude}
\noindent We have presented deep broadband imaging data in $UBVRIHKs$ for a volume limited ($11\le Ra\le14$h, $v\le2100$ km s${}^{-1}$) sample of $21$ emission line galaxies, comprising $19$ BCGs and two spirals. We have analyzed the surface brightness and radial color profiles, contour maps and RGB images of the galaxies, and provided a morphological classification based on~\citet{1986sfdg.conf...73L} where such was missing in the literature. Separating each galaxy into different regions, we were able to obtain a central surface brightness and scale length from two radial ranges, estimate the burst luminosity and its relative contribution to the total light, as well as investigate the behavior of the regions in terms of color with respect to other galaxy properties. \\

\noindent Most of the galaxies have no break in the surface brightness profiles in the outer parts, and are well fitted with a single exponential throughout the range $\mu_B=24-28$ mag arcsec${}^{-2}$. For UM462 we observe a previously undetected second exponential disk component which dominates the profile beyond $\mu_B\sim26$ mag arcsec${}^{-2}$. This component is symmetrically extended to the North and South of the central galaxy regions, and is also clearly visible in the contour plot. The presence of this component makes the UM462 host a true low surface brightness galaxy, with a central surface brightness of $\mu_B=24.1$ mag arcsec${}^{-2}$ and a scale length $h_r=1.48$ kpc.\\

\noindent Comparing the integrated colors for different components of the galaxy, i.e. burst, host, and composite total, to stellar evolutionary models both with and without nebular emission contribution we are able to give indications of the metallicity and age of both the young burst and the host populations in quite a few cases. The models indicate that the typical host galaxy is metal poor ($Z\lesssim0.004$), though some hosts require an unexpectedly high metallicity ($Z\sim0.02$). A careful investigation of such cases  will be provided in a future paper~\citep{Paper3}. \\

\noindent We have derived Petrosian $A_{P}$ asymmetries based on the region inside the Petrosian radius $r[\eta(0.2)]$ for each galaxy in each filter. The center of asymmetry minimum stays constant in all filters for most galaxies. We detect a strong peak of the asymmetry distribution of the galaxies in the optical around $A_{P}\sim0.2$, which shifts to $A_{P}\sim0.4$ in the NIR for the majority of galaxies, though some retain their small $0.2$ asymmetry value. This separates the galaxies into three different groups based on the behavior of the $A_{P}$ asymmetry in the different filters. These groups are correlated with the morphological class of a galaxy, \emph{nE}s having small optical and NIR asymmetries, \emph{iE}s having small optical and large NIR asymmetries, and \emph{iI}s and \emph{iI,M}s having both large optical and NIR asymmetries. The latter group is clearly deviating from the fiducial galaxy color--asymmetry sequence of~\citet{2000ApJ...529..886C} for spheroidals, disks and irregulars, which is consistent with the group's morphological classification as tidally interacting/merging galaxies.\\

\noindent The Petrosian $A_{P}$ asymmetry for this sample is dominated by the flocculent component (i.e. due to the star formation), though to a lesser extent than for the luminous BCGs in \M. The alternative asymmetry measurements we have used carry valuable additional information which is unavailable if one solely considers $A_{P}$. Our ``Holmberg'' ($A_H^\prime$) and ``dynamical'' ($A_{dyn}$) asymmetries confirm that NIR asymmetry is a good proxy for the dynamical asymmetry component (i.e. due to the galaxy morphology). Similar to \M~we find that the dynamical asymmetry component is weakly, if at all, correlated with $A_{P}$. We find a strong correlation between $A_P$ and $A_H^\prime$ in the sense that $A_H^\prime\approx0.62\times A_P-0.003$. This should be compared to~\citet{2003ApJS..147....1C} who finds $A_{Global}^\prime\approx0.67\times A_P+0.01$. \\

\noindent The BCGs in the sample, i.e. excluding the two spiral galaxies, seem to divide into a blue and red groups in all colors. This division is due to differences in star formation and nebular emission contribution, with the blue BCGs having brighter, more extended, and more irregular star forming regions compared to the red ones. The hosts of the blue BCGs also show higher dynamical asymmetries ($A_{dyn}$).\\

\noindent In a concentration--asymmetry plot emission line galaxies occupy the region with low concentration and high asymmetry, i.e. it is not possible to distinguish between luminous blue compact galaxies and the less vigorously star forming galaxies of this sample in this parameter space. This is interesting considering the otherwise very different behavior and structural parameters of the galaxies in this paper and the luminous blue compacts of~\citet{Paper1}.

\section*{Acknowledgments}

\noindent G.\"O. is a Royal Swedish Academy of Sciences Research Fellow supported by a grant from the Knut and Alice Wallenberg Foundation. G.\"O. acknowledges support from the Swedish research council (VR) and the Swedish National Space Board. E.Z. acknowledges research grants from the Swedish Research Council and the Swedish National Space Board. J.M. and I.M. acknowledge financial support from the Spanish grant AYA2010-15169 and from the Junta de Andalucia through TIC-114 and the Excellence Project P08-TIC-03531.

\noindent ALFOSC is provided by the Instituto de Astrofisica de Andalucia (IAA) under a joint agreement with the University of Copenhagen and NOTSA. \\

\noindent This work made use of the NASA/IPAC Extragalactic Database (NED) which is operated by the Jet Propulsion Laboratory, California Institute of Technology, under contract with the National Aeronautics and Space Administration.

%
\appendix

\section{Extremely faint contours}\protect\label{extremecontours}
\noindent In Figure~\ref{deepcontours1} we present extremely faint contours of the galaxies in the sample. Both contours and images have been heavily smoothed to reduce noise. The diagonal streak features in the UM422 image are satellite tracks.
\begin{figure*}
\begin{center}
  \includegraphics[width=18cm]{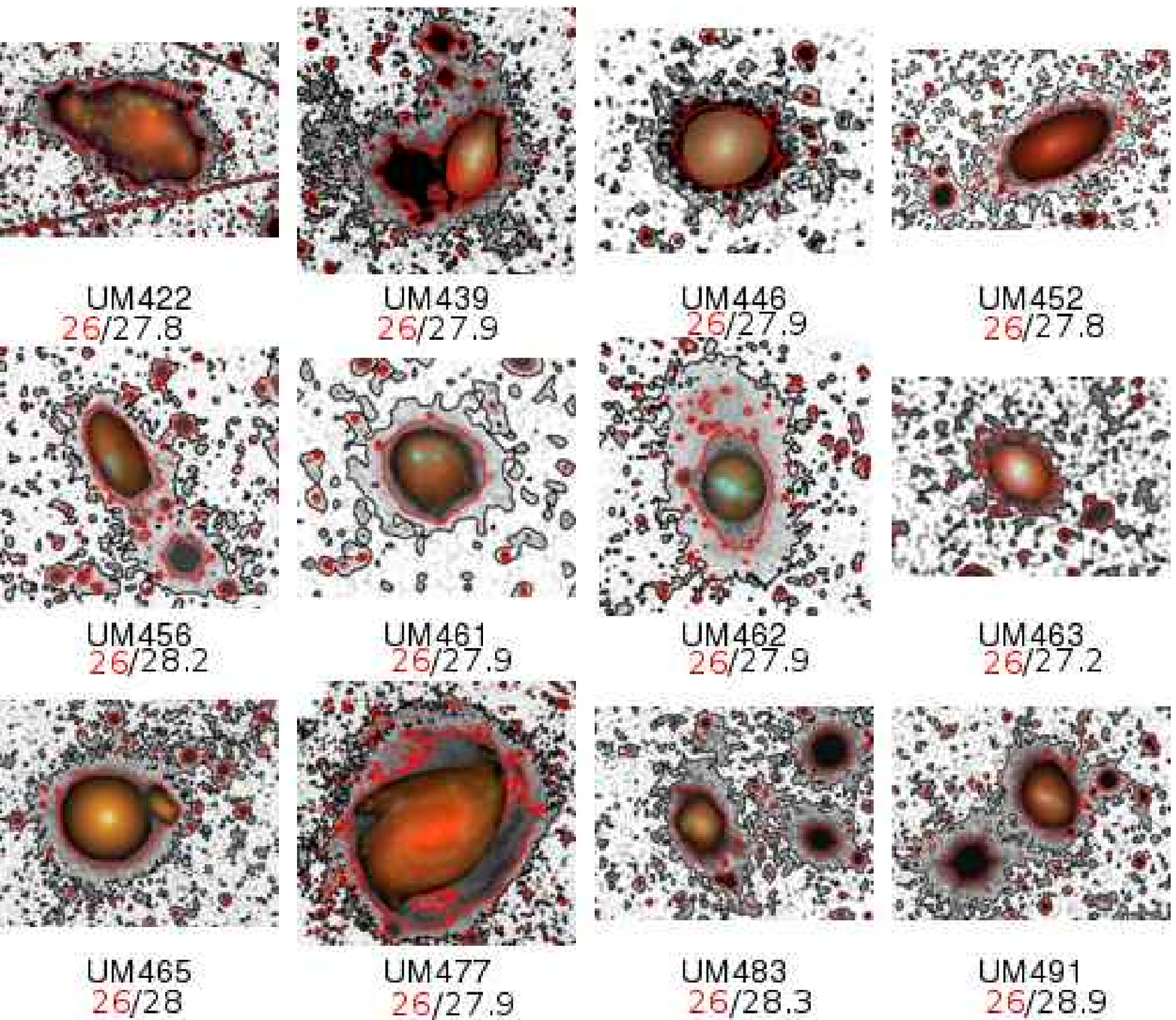}
\caption{Extreme contours. Numbers in red/black indicate the isophotal level of the red/black contour in mag arcsec${}^{-2}$.}\protect\label{deepcontours1}
\end{center}
\end{figure*}
\begin{figure*}
\begin{center}
  \includegraphics[width=18cm]{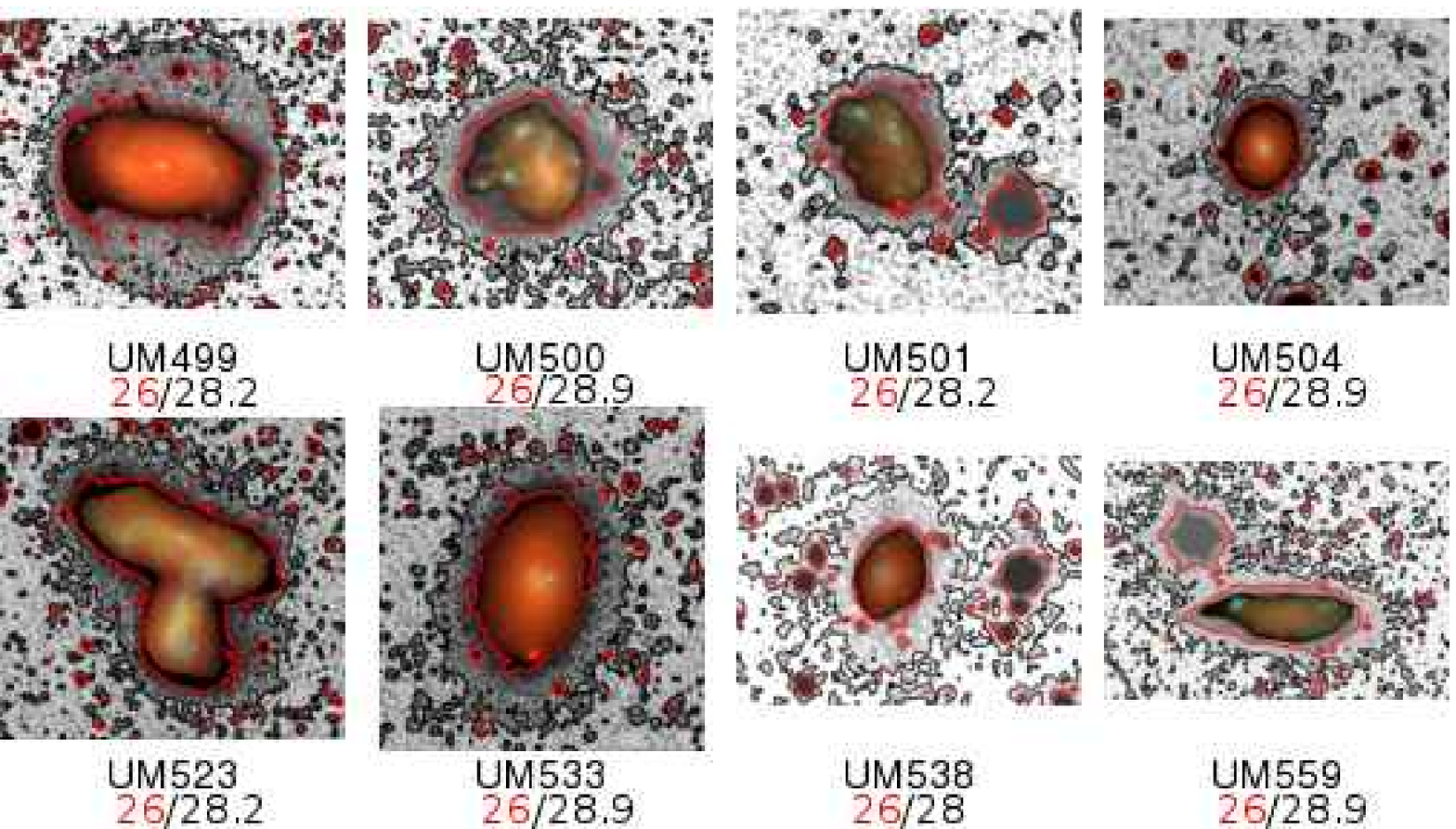}
\contcaption{ }
\end{center}
\end{figure*}
\subsection*{UM462}
\begin{figure*}
\begin{center}
  \includegraphics[width=17cm]{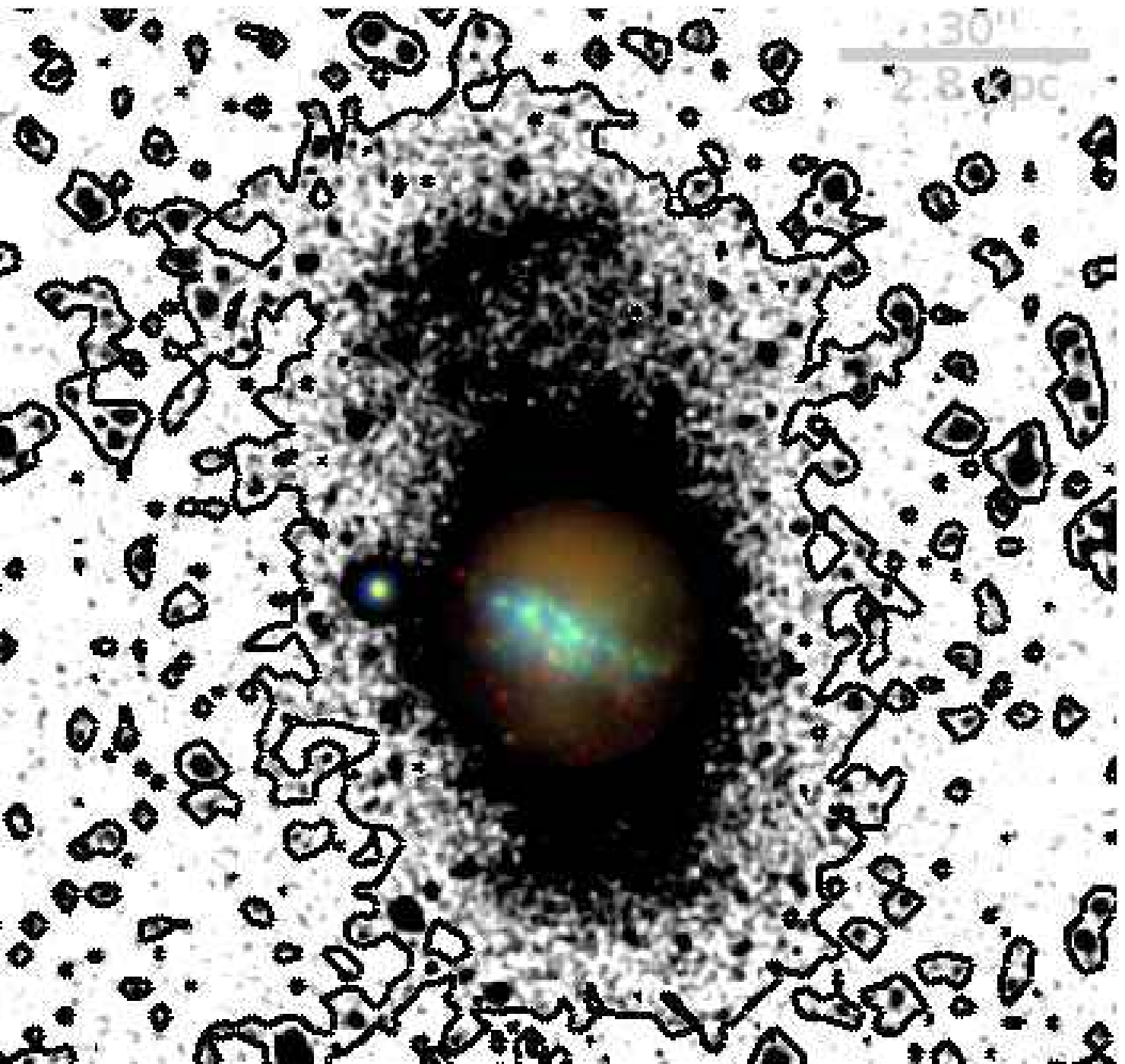}
\caption{A zoom on UM462. Note that these spectacular faint features bear a striking morphological similarity to NGC 5128 (Centaurus A). The black contours are at $\mu_B=27.9$ mag arcsec${}^{-2}$.}\protect\label{um462deepcontours}
\end{center}
\end{figure*}

\protect\label{um462appendix}
\noindent Figure~\ref{um462deepcontours} shows a zoom of the spectacular low surface brightness features of UM462.

\bibliographystyle{mn2e}
\bibliography{micheva2}

\bsp

\label{lastpage}

\end{document}